\documentclass[traditabstract]{aa}

\usepackage{natbib}
\usepackage[stable]{footmisc}
\usepackage{amsmath}
\usepackage{amssymb}
\usepackage[varg]{txfonts}
\usepackage{placeins}
\usepackage{natbib}
\usepackage{graphicx}
\usepackage{epstopdf}
\usepackage{ifthen}
\usepackage[breaklinks, colorlinks, citecolor=blue]{hyperref}
\usepackage{overpic}
\usepackage{fixltx2e}
\usepackage{rotating}
\usepackage{lscape}
\usepackage{threeparttable}
\usepackage{arydshln}
\usepackage[]{xcolor}

\bibpunct{(}{)}{;}{a}{}{,} 

\begin{document}

\title{Using ALMA to resolve the nature of the early star-forming large-scale
 structure PLCK\,G073.4$-$57.5}
\titlerunning{Resolving the \textit{Planck}/\textit{Herschel} source
 G073.4$-$57.5 with ALMA}

\author{R\"udiger~Kneissl\inst{\ref{inst1},\ref{inst2}}
\and Maria~del~Carmen~Polletta\inst{\ref{inst3},\ref{inst4}}
\and Clement~Martinache\inst{\ref{inst5},\ref{inst6}}
\and Ryley~Hill\inst{\ref{inst7}}
\and Benjamin~Clarenc\inst{\ref{inst5}}
\and Herve~A.~Dole\inst{\ref{inst5}}
\and Nicole~P.H.~Nesvadba\inst{\ref{inst5}}
\and Douglas~Scott\inst{\ref{inst7}}
\and Matthieu~B{\'e}thermin\inst{\ref{inst8},\ref{inst9}}
\and Brenda~Frye\inst{\ref{inst10}}
\and Martin~Giard\inst{\ref{inst4}}
\and Guilaine~Lagache\inst{\ref{inst9}}
\and Ludovic~Montier\inst{\ref{inst4}}}

\institute{European Southern Observatory, ESO Vitacura, Alonso de Cordova 3107,
 Vitacura, Casilla, 19001, Santiago, Chile\label{inst1}
\and
Atacama Large Millimetre/submillimetre Array, ALMA Santiago Central Offices,
 Alonso de Cordova 3107, Vitacura, Casilla, 763 0355, Santiago, Chile
 \email{ruediger.kneissl@alma.cl}\label{inst2}
\and
INAF -- Istituto di Astrofisica Spaziale e Fisica Cosmica Milano,
 via A. Corti 12, 20133 Milano, Italy\label{inst3}
\and
IRAP, Universit\'{e} de Toulouse, CNRS, CNES, UPS, Toulouse, France\label{inst4}
\and
Institut d'Astrophysique Spatiale, CNRS, Univ. Paris-Sud, Universit\'e Paris-Saclay, B\^at. 121, 91405 Orsay, France\label{inst5}
\and 
Departamento de Astronom\'{i}a, Universidad de Concepci\'{o}n, Avenida Esteban
 Iturra s/n, Casilla 160-C, Concepci\'{o}n, Chile\label{inst6}
\and
Department of Physics and Astronomy, University of British Columbia, 6224
 Agricultural Road, Vancouver BC V6T 1Z1, Canada\label{inst7}
\and
European Southern Observatory, Karl-Schwarzschild-Stra{\ss}e 2, D-85748
 Garching, Germany\label{inst8}
\and
Aix Marseille Univ, CNRS, LAM, Laboratoire d'Astrophysique de Marseille,
 Marseille, France\label{inst9}
\and
Department of Astronomy/Steward Observatory, 933 North Cherry Avenue, University of Arizona, Tucson, AZ 85721, USA\label{inst10}
}

\authorrunning{Kneissl et al.}

\date{Received / Accepted}

\abstract {Galaxy clusters at high redshift are key targets for
understanding matter assembly in the early Universe, yet they are
challenging to locate.  A sample of more than 2000 high-$z$ candidate
structures has been found using {\it Planck\/}'s all-sky submillimetre maps,
and a sub-set of 234 have been followed up with {\it Herschel\/}-SPIRE, which
showed that the emission can be attributed to large overdensities of dusty
star-forming galaxies. As a next step, we need to resolve and
characterise the individual galaxies giving rise to the emission seen by
{\it Planck\/} and {\it Herschel\/}, and to find out whether they
constitute the progenitors of present-day, massive galaxy clusters.  Thus,
we targeted the eight brightest {\it Herschel\/}-SPIRE sources in the centre
of the {\it Planck\/} peak PLCK\,G073.4$-$57.5 using ALMA at 1.3\,mm, and
complemented these observations with multi-wavelength data from {\it
Spitzer\/}-IRAC, CFHT-WIRCam in the $J$ and $K_{\rm s}$ bands, and JCMT's
SCUBA-2 instrument.  We detected a total of 18 millimetre galaxies brighter
than 0.3\,mJy within the 2.4\,arcmin$^2$ ALMA pointings, corresponding to an
ALMA source density 8--30 times higher than average background estimates and
larger than seen in typical \lq{}proto-cluster\rq{} fields.  We were able to match
all but one of the ALMA sources to a near infrared (NIR) counterpart.  The four most
significant SCUBA-2 sources are not included in the ALMA pointings, but we
find an $8\,\sigma$ stacking detection of the ALMA sources in the SCUBA-2
map at 850\,$\mu$m.  We derive photometric redshifts, infrared (IR) luminosities,
star-formation rates (SFRs), stellar masses ($\mathcal{M}$), dust temperatures, 
and dust masses for
all of the ALMA galaxies. Photometric redshifts identify two groups each
of five sources, concentrated around $z\,{\simeq}\,1.5$ and 2.4.  The two
groups show two \lq{}red sequences\rq{}, that is similar near-IR [3.6]$\,{-}\,$[4.5] 
colours and different $J\,{-}\,K_{\rm s}$ colours. The majority of
the ALMA-detected galaxies are on the SFR versus $\mathcal{M}$ 
main sequence (MS), and half of the sample is
more massive than the characteristic $\mathcal{M}_\ast$ at the corresponding
redshift. We find that the $z\,{\simeq}\,1.5$ group has total ${\rm
SFR}=840^{+120}_{-100}\,{\rm M}_{\odot}\,{\rm yr}^{-1}$ and
$\mathcal{M}=5.8^{+1.7}_{-2.4}\,{\times}\,10^{11}\,{\rm M}_{\odot}$, and
that the $z\,{\simeq}\,2.4$ group has ${\rm SFR}=1020^{+310}_{-170}\,{\rm
M}_{\odot}\,{\rm yr}^{-1}$ and
$\mathcal{M}=4.2^{+1.5}_{-2.1}\,{\times}\,10^{11}\,{\rm M}_{\odot}$, but the
latter group is more scattered in stellar mass and around the MS.
Serendipitous CO line detections in two of the galaxies appear to match their
photometric redshifts at $z\,{=}\,1.54$. We performed an 
analysis of star-formation efficiencies (SFEs) and 
CO- and mm-continuum-derived gas fractions 
of our ALMA sources, combined with a sample of $1\,{<}\,z\,{<}\,3$ cluster 
and proto-cluster members, and observed trends in both quantities 
with respect to stellar masses and in comparison to field galaxies.} 


\maketitle

\section{Introduction}     \label{sec:introduction}

Hierarchical clustering models of large-scale structure and galaxy formation predict that the progenitors of the most massive galaxies in today's clusters are dusty star-forming galaxies (SFGs) at high redshift \citep[$z\,{\simeq}\,$2--3, e.g.][]{Lilly+99,Swinbank+08}. Observationally, this picture is supported by the clustering measurements \citep{Blain+04} of submillimetre galaxies (SMGs), and by their relative abundance and distribution in known proto-clusters \citep[e.g.][]{Capak+11,Hayashi+12,Casey+15,Hatch+16,Overzier2016}. High-redshift structure-formation studies at millimetre (mm) and submillimetre (submm) wavelength ranges have the advantage of providing access to high redshifts by utilising the steep rise in the warm dust spectrum of infrared galaxies (the \lq{}negative k-correction\rq{}, \citealt{BlainLongair93}; also \citealt{Guiderdoni+97}) and can build on an observed correlation between the total matter density and the cosmic infrared-background fluctuations \citep{PlanckXVIII14}. 

Substantial progress has been made in probing the early formation of massive structures and galaxy clusters through mm/submm observations \citep[see][for a recent discussion]{Casey16}, with a strong emphasis on main sequence (MS) evolution versus starbursts (SB) and mergers \citep[see also][]{Narayanan+15}. Mechanisms for rapid, episodic bursts, suggested to explain how the member galaxies are assembled and grow during cluster formation, can be tested with measurements of mm-galaxy number densities and gas depletion timescales in cluster-forming environments. Likewise, the processes responsible for triggering star formation that is coherent over large spatial scales may depend on environmental effects, which can only be tested using a variety of high quality data over wide areas. 

As proto-clusters are discovered, follow-up observations need to be made to assess their contents, for example observations to trace their cold gas, which provide constraints on the processes of inflow, outflow, and cold gas consumption. Until recently, the limited number of CO detections in high-redshift ($z\,{\simeq}\,$1.5--2) structures had not provided a consensus on the influence of the environment on the gas contents of galaxies \citep{aravena12,wagg12,casasola13,stach17,noble17,coogan18,tadaki14,lee17,dannerbauer17}. However, a recent study by \citet{wang18} of a cluster at $z\,{=}\,2.51$, known as CL\,J1001+0220, has clearly shown that the molecular gas properties of cluster members are correlated with their location, that is with their distance from the cluster core \citep[see also][for XMMXCS\,J2215.9$-$1738 at $z\,{=}\,1.46$]{hayashi17}. Thus galaxies remain relatively gas-rich when they first enter the cluster, but their gas content is rapidly reduced as they approach the cluster centre.
In other words, the environment must play a role in stopping gas accretion and/or reducing and removing gas content \citep{hayashi17,wang18,foltz18}. High-redshift proto-clusters can also be gas-rich; ALMA observations of the proto-cluster around 4C\,23.56 at $z\,{=}\,$2.49 described by \citet{lee17} show that the gas masses and fractions of its members are comparable to those of field galaxies, implying that the total gas density is much higher inside the proto-cluster than in the field.

The {\it Planck\/} satellite has also contributed to the search for proto-clusters; {\it Planck\/} mapped out the whole sky between 30 and 857\,GHz with a beam going down to 5\arcmin\ \citep{PlanckI14}, giving it the capability of detecting the brightest mm/submm regions of the extragalactic sky at Mpc scales. A component-separation procedure using a combination of {\it Planck\/} and IRAS data was applied to the maps outside of the Galactic mask to select over 2000 of the most luminous submm peaks in the cosmic infrared background (CIB), with spectral energy distributions peaking between 353 and 857\,GHz \citep[][the \lq{}PHz\rq{} catalogue]{PlanckXXXIX15}. This selection is distinctly different to the {\it Planck\/} catalogue of cluster candidates detected via the Sunyaev-Zeldovich effect \citep[][the \lq{}PSZ2\rq{} catalogue]{2016A&A...594A..27P}. It targets the bright, far-infrared spectral energy distribution of dust heated by star formation, and therefore selects predominantly rapidly growing galaxies. 234 of these submm peaks (chosen to have S/N$\,{>}\,4$ at 545\,GHz, as well as flux-density ratios $S_{857}\,{/}\,S_{545}\,{<}\,1.5$, and $S_{217}\,{<}\,S_{353}$) were subsequently followed up with {\it Herschel\/}-SPIRE observations between 250 and 500\,$\mu$m, and the half-arcminute (or better) resolution was capable of distinguishing between bright gravitational lenses and concentrations of clustered mm/submm galaxies around redshifts of 2--3 \citep{PlanckXXVII15}. Here, we present the first detailed mm analysis of one of these highly clustered regions, PLCK\,G073.4$-$57.5 (hereafter G073.4$-$57.5), which was observed with ALMA in Cycle 2. We combine near infrared (NIR) and far infrared (FIR) multi-wavelength data with the resolving power of ALMA to identify the individual galaxies responsible for much of the {\it Planck\/} submm flux and to constrain their physical properties. 

This paper on G073.4$-$57.5 is structured as follows. In Sect.~\ref{Sect2} we re-capitulate the features of the {\it Planck\/}/{\it Herschel\/} sample, followed by Sect.~\ref{Sect3}, where we present details of the ALMA observations, data reduction, and results. In Sect.~\ref{sec:observations} we describe the set of multi-wavelength data on G073.4$-$57.5, comprising {\it Herschel\/}-SPIRE, SCUBA-2, {\it Spitzer\/}-IRAC, and CFHT-WIRCam observations. In Sect.~\ref{sec:analysis} we present the analysis of these data, where we estimate the mm galaxy number density of G073.4$-$57.5 and derive the photometric redshifts and IR properties of each galaxy (such as their dust temperatures, dust masses, IR luminosities, star-formation rates, and stellar masses), and in Sect.~\ref{line_detection} we interpret serendipitous line detections. In Sect.~\ref{sec:discussion} we discuss our findings and interpretation. The paper is then concluded in Sect.~\ref{sec:conclusions}. 

In this paper we denote the stellar mass with $\mathcal{M}$ and the characteristic stellar mass with $\mathcal{M}_\ast$. Throughout this paper we use the parameters of the best-fit {\it Planck\/} flat $\Lambda$CDM cosmology \citep{2018arXiv180706209P}, specifically $\Omega_{\rm M}\,{=}\,0.315$, $h\,{=}\,0.674$. In this model 1\arcsec\ at $z\,{=}\,$1.5 (2.4) corresponds to a physical scale of 8.7 (8.3)\,kpc. 

\section{The \textit{Planck}/\textit{Herschel} high-$z$ sample}\label{Sect2}

\begin{figure}[hbtp!]
  \centering
  \includegraphics[width=85mm]{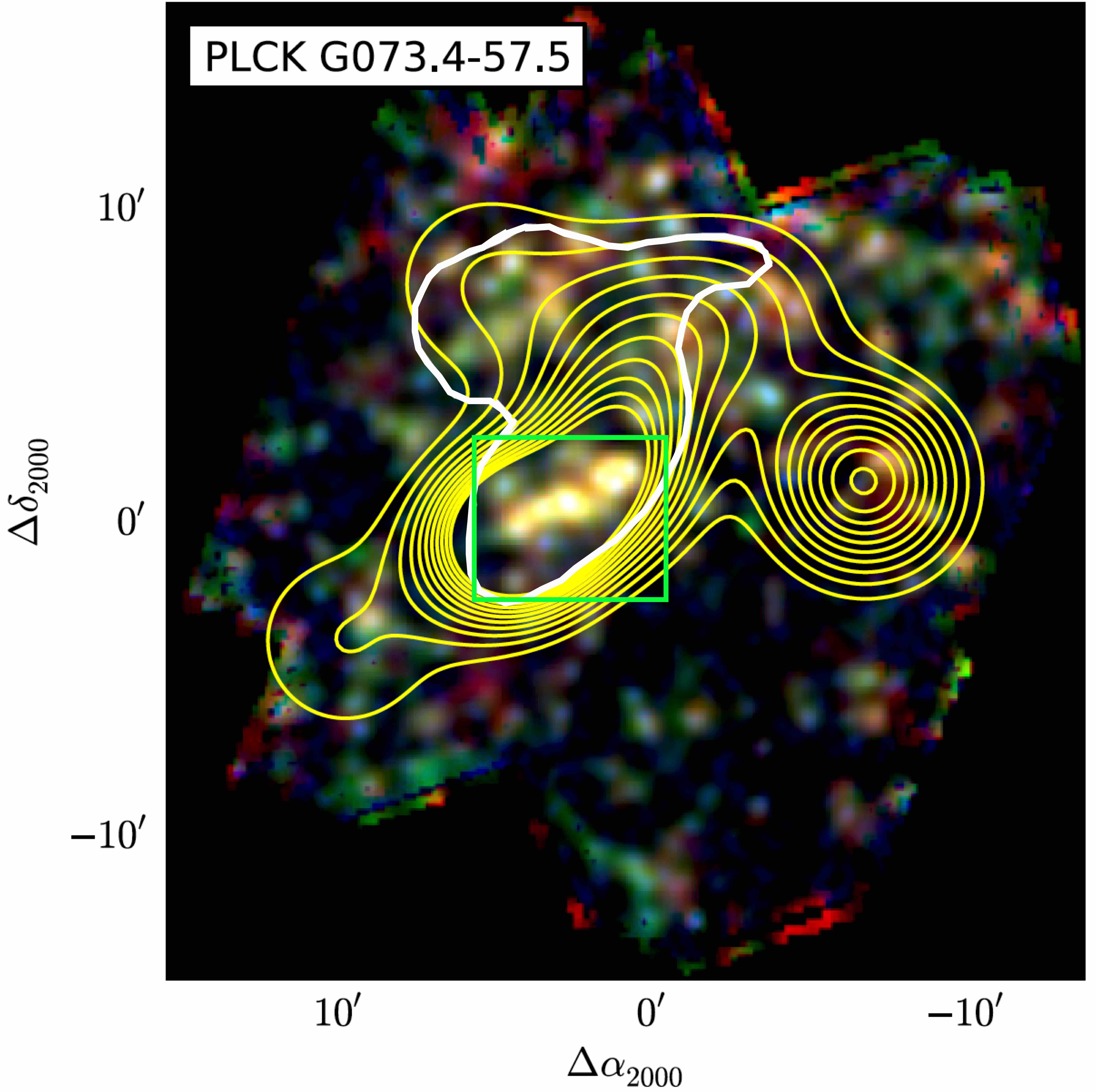}
  \caption[]{\label{Fig0} Three-colour SPIRE image for G073.4$-$57.5 \citep[reproduced from][]{PlanckXXVII15}: blue, 250\,$\mu$m; green, 350\,$\mu$m; and red, 500\,$\mu$m. The white contour shows the region encompassing 50\,\% of the {\it Planck\/} flux density, while the yellow contours are the significance of the overdensity of red (350\,$\mu$m) sources, plotted starting at 2$\,\sigma$ with 1$\,\sigma$ incremental steps. The rectangular area covering the ALMA pointings shown in Fig.~\ref{Fig1} is highlighted in green.}
\end{figure}

\begin{figure*}[htbp!]
  \includegraphics[width=185mm]{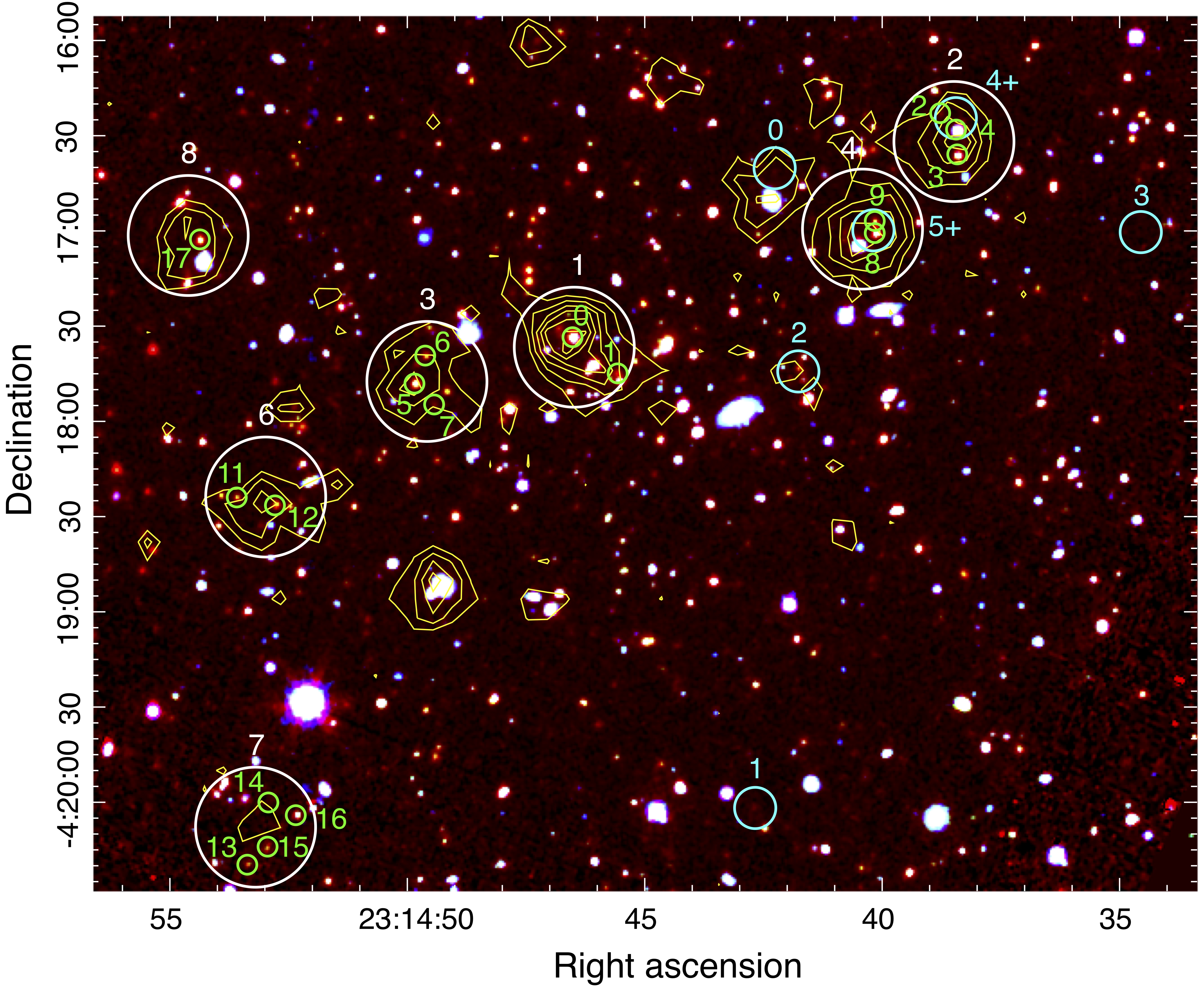}
  \caption{Central region (5.8\arcmin$\,{\times}\,$4.6\arcmin) of G073.4$-$57.5 in a 3-colour image of {\it Spitzer\/}-IRAC 3.6\,$\mu$m (red), combined CFHT-WIRCam/VLT-HAWKI $K$-band (green) and $J$-band (blue), with {\it Herschel\/}-SPIRE 250-$\mu$m contours in yellow (from 0.02\,Jy\,beam$^{-1}$ in 0.0125\,Jy\,beam$^{-1}$ steps) and ALMA galaxy positions shown with green circles of radius 3\arcsec\ (enlarged by a factor of 12 from ALMA's synthesised beam for clarity), labelled according to their source IDs given in Table~\ref{table:2}. 
The ALMA areas that were used for the analysis (0.2 times the primary-beam peak response) are indicated with white circles (37\arcsec\ diameter), labelled according to their field IDs given in Table~\ref{table:2}. Four SCUBA-2 sources centred in the cyan circles (13\arcsec\ diameter, matching the beam size) are labelled according to \citet{MacKenzie+16}; the two SCUBA-2 sources labelled \lq{}4+/5+\rq{} are additionally selected as ${>}\,3\,\sigma$ peaks in the SCUBA-2 maps coincident with ALMA-detected sources. ALMA field 5 (with one detected source, see Fig.~\ref{Fig2}) is located above and to the right of the central region and is not shown in this image.}
  \label{Fig1}
\end{figure*}

A dedicated {\it Herschel\/} \citep{Pilbratt+10} follow-up programme with the SPIRE instrument for 234 {\it Planck\/} targets \citep{PlanckXXVII15} found a significant excess of \lq{}red\rq{} sources (where red means $S_{350}$\,/\,$S_{250}\,{>}\,0.7$ and $S_{500}$\,/\,$S_{350}\,{>}\,0.6$, which is consistent with $z\,{\ga}\,$2 SFGs), in comparison to reference SPIRE fields. Assuming a single common dust temperature for the sources of $T_{\rm d}\,{=}\,35$\,K, IR luminosities of typically $4\,{\times}\,10^{12}$\,L$_\odot$ were derived for each SPIRE source, yielding star-formation rates (SFRs) of around 400\,M$_\odot$\,yr$^{-1}$. If these observed {\it Herschel\/} overdensities are coherent structures, their total IR luminosity would peak at $4\,{\times}\,10^{13}$\,L$_\odot$, or in terms of an SFR, at $4\,{\times}\,10^{3}$\,M$_\odot$\,yr$^{-1}$, that is the equivalent of ten typical sources making up the overdensity. We note that a parallel study of {\it Planck\/} compact sources overlapping within already existing {\it Herschel\/} fields also finds 27 proto-cluster candidates \citep{Greenslade+18}; for earlier such samples see also \citet{Herranz+13}, \citet{Clements+14}, and \citet{Clements+16}.

From the 234 {\it Planck\/}/{\it Herschel\/} high-$z$ sample a small sub-set of 11 {\it Herschel\/} sources are now known to be gravitationally lensed single galaxies \citep{Canameras+15,2018A&A...620A..61C}, including the extremely bright G244.8$+$54.9, greater than 1\,Jy at 350\,$\mu$m. ALMA data for such sources, also aided by HST-based lensing models, have enabled extremely detailed studies of high-$z$ SFGs \citep[e.g.][]{Nesvadba+16,Canameras+17a,Canameras+17b,2018A&A...620A..60C,2018arXiv181204653N}; however, the remaining sources are overdensities of SFGs.

In a recent paper, \citet{MacKenzie+16} have presented SCUBA-2 follow-up of 61 {\it Planck\/}/{\it Herschel\/} targets at 850\,$\mu$m, each observation covering essentially the full emission of the {\it Planck\/} peak. 172 sources are detected in the maps with high confidence (S/N$\,{>}\,4$), and by fitting modified black-body dust spectral energy distributions (SEDs) it is shown that the distribution of photometric redshifts peaks between $z\,{=}\,2$ and $z\,{=}\,3$.

Further studies based on NIR and optical data with the aim of characterising the {\it Planck\/}/{\it Herschel\/} targets have been carried out by \citet{FloresCacho+16} on G95.5$-$61.6 and by \citet{Martinache+18} on a {\it Spitzer\/}-IRAC sample of 82 PHz sources. \citet{FloresCacho+16} are able to conclude that G95.5$-$61.6 consists of two significantly clustered regions at $z\,{\simeq}\,1.7$ and at $z\,{\simeq}\,2.0$, while \citet{Martinache+18} can verify the overdensites seen by {\it Herschel\/} and derive mass estimates suggesting that the PHz sources will become some of the most massive clusters at $z\,{=}\,0$, which further motivates their utility for studying high-redshift clustering.

In the current paper we focus on directly detecting the galaxies responsible for the {\it Planck\/} peak using the high-resolution (sub-)mm imaging capabilities of ALMA. Our target G073.4$-$57.5 was included in the {\it Planck\/}/{\it Herschel\/} sample from the selection of the first public release of the {\it Planck\/} Catalogue of Compact Sources\footnote{We note that for the latest {\it Planck\/} release~\citep{PlanckXXXIX15}, G073.4$-$57.5 lies just inside the more conservatively applied Galactic mask.} with a 545-GHz flux density of $730\pm80\,$mJy. It was included in an ALMA proposal based on the high overdensity of {\it Herschel\/} sources within the {\it Planck\/} contour (Fig.~\ref{Fig0}) and the availability of additional NIR and submm data. 


\section{ALMA observation of G073.4$-$57.5}\label{Sect3}

We received 0.4\,hours of on-source observing time on G073.4$-$57.5 with ALMA in Cycle 2 (PID 2013.1.01173.S, PI R.~Kneissl). We targeted the eight sources found in the SPIRE field that were consistent with a red colour, within the uncertainties, as defined above. A standard Band 6 continuum set-up around 233\,GHz (1.3\,mm) was used, with four 1.78-GHz spectral windows divided into the two receiver sidebands, separated by 16\,GHz (i.e. central frequencies of 224, 226, 240, and 242\,GHz). 34 antennas were available in the array configuration during the time of the observation, and the resulting synthesised beam achieved an angular resolution of $0.56\arcsec\,{\times}\,0.44\arcsec$ (FWHM) with a position angle of $-82.7^\circ$ (turning from north to east for a positive angle). The central sensitivity was approximately $0.06$\,mJy\,beam$^{-1}$ in all eight fields (Fig.~\ref{Fig1} for an overview, and we note that the {\it Herschel\/}-SPIRE IDs, as given in Table~\ref{table:2}); with this sensitivity ALMA can detect all SPIRE sources at any redshift, assuming a dust temperature ${>}\,25$\,K and that all the SPIRE flux comes from a single source, since the detection significance increases at higher redshifts. The observatory standard calibration was used. J2232+1143, a grid-monitoring source, was the bandpass calibrator and Ceres was observed as an additional flux calibrator. All pointings in this data set shared the same phase calibrator, J2306$-$0459. The single pointings were convolved with the primary-antenna-beam pattern (roughly Gaussian with a FWHM$\,{\simeq}\,$25.3\arcsec, assuming $1.13\,\lambda\,{/}\,D$). 

The data were reduced with standard {\tt CASA} tasks \citep{CASA+2007}, including deconvolution, to yield calibrated continuum images with flat noise characteristics for source detection. A S/N$\,{>}\,5$ mask was applied to the primary beam-uncorrected maps with a 2$\,\sigma$ {\tt CLEAN} threshold, yielding 13 sources in six fields, where the detection was based on the peak pixel surface brightness. In addition, the single brightest sources from each of the remaining two fields were included in the sample, since they were both well centred, with S/N$\,{>}\,4.5$. During cross-matching with {\it Spitzer\/} maps, three additional sources were identified with S/N$\,{>}\,4.5$. The final sample, containing 18 ALMA sources with flux densities ${>}\,$0.3\,mJy and S/N$\,{>}\,4.5$, is presented in Table~\ref{table:1}. 

The flux-density results were derived from applying {\tt ImageFitter} to the {\tt CLEAN}ed maps and integrating over each source. They are presented in Table~\ref{table:1}, along with the angular sizes for nine sources that were best fit with an extended profile (and four of which had a major axis  determined with S/N$\,{>}\,3$). In the nine remaining cases the fit for source size did not converge well and these are listed as point sources. In addition, for each source we give the peak flux density at 233\,GHz derived from the beam deconvolved map (which is more accurate for the nine point sources) and the coordinate for the position of the peak surface brightness. We note that ALMA source ID~16, which is on the edge of pointing field 7, has a recovered peak flux density of $0.59\pm0.17\,$mJy\,beam$^{-1}$, that is 3.5$\,\sigma$, and should thus be considered tentative, in spite of the match with a {\it Spitzer\/} source (cf. next section and Fig.~\ref{Fig2}).

\begin{table*}[htbp!]
\caption{Basic properties of the ALMA galaxies detected at 1.3\,mm in G073.4$-$57.5.}              
\label{table:1}      
\centering                                      
\begin{threeparttable}[b]
\begin{tabular}{llllll}          
\hline\hline                        
\noalign{\vskip 3pt}
ALMA& ALMA& Name/Position\tnote{a}& S/N\tnote{b}& $S_\nu$\tnote{c}& Size\tnote{d}\\    
source& field& [ICRS]& & [mJy]& [arcsec]\\
\hline                                   
\noalign{\vskip 3pt}
 0& 1& ALMAU J231446.53$-$041733.5&  7.4& 1.22$\pm$0.20& 0.65/(0.40)\\
 1&  & ALMAU J231445.60$-$041744.4&  7.5& 2.39$\pm$0.48& (0.50/0.30)\\
\hline
\noalign{\vskip 3pt}
 2& 2& ALMAU J231438.78$-$041622.7& 10.3& 1.64$\pm$0.24& 0.61/(0.20)\\
 3&  & ALMAU J231438.42$-$041636.2& 14.8& 1.65$\pm$0.13& 0.44/(0.26)\\
 4&  & ALMAU J231438.36$-$041628.4&  5.6& 0.42$\pm$0.06& p\\
\hline
\noalign{\vskip 3pt}
 5& 3& ALMAU J231449.85$-$041748.1& 22.5& 1.44$\pm$0.17& (0.28/0.20)\\
 6&  & ALMAU J231449.63$-$041739.3&  4.9& 0.33$\pm$0.06& p\\
 7&  & ALMAU J231449.45$-$041754.7&  5.0& 0.34$\pm$0.06& p\\
\hline
\noalign{\vskip 3pt}
 8& 4& ALMAU J231440.15$-$041700.7& 14.7& 1.28$\pm$0.13& 0.47/(0.15)\\
 9&  & ALMAU J231440.14$-$041657.2&  5.0& 0.61$\pm$0.20& (0.97/0.27)\\
\hline
\noalign{\vskip 3pt}
10& 5& ALMAU J231437.03$-$041451.7&  4.6& 0.55$\pm$0.17& (0.53/0.22)\\
\hline
\noalign{\vskip 3pt}
11& 6& ALMAU J231453.61$-$041823.9&  6.9& 0.93$\pm$0.09& p\\
12&  & ALMAU J231452.78$-$041826.1&  6.1& 0.67$\pm$0.06& p\\
\hline
\noalign{\vskip 3pt}
13& 7& ALMAU J231453.37$-$042019.5&  6.0& 1.74$\pm$0.32& (0.49/0.38)\\
14&  & ALMAU J231452.86$-$041959.3&  5.5& 0.75$\pm$0.09& p\\ 
15&  & ALMAU J231452.94$-$042012.7&  4.9& 0.59$\pm$0.10& p\\
16&  & ALMAU J231452.34$-$042004.2&  4.8& 0.59$\pm$0.17& p\\
\hline
\noalign{\vskip 3pt}
17& 8& ALMAU J231454.38$-$041702.9&  4.6& 0.53$\pm$0.07& p\\
\hline                                             
\end{tabular}
     \begin{tablenotes}{\small
       \item[a] Coordinate errors on the ALMA positions are approximately 0.5\arcsec\,/\,(S/N), i.e. 0.1\arcsec\ or better. 
       \item[b] Signal-to-noise ratio in the primary-beam-convolved, {\tt CLEAN}ed detection maps. 
       \item[c] If the source has a \lq{}p\rq{} in the \lq{}Size\rq{} column the flux density comes from the peak pixel, otherwise it is the integrated flux density. In both cases flux densities were derived from primary-beam-deconvolved, {\tt CLEAN}ed maps using the results of the {\tt ImageFitter} routine. 
       \item[d] For extended sources, estimates of the semi-major/semi-minor axes (S/N$\,{<}\,3$ are in brackets); for point sources, a \lq{}p\rq{} is given.} 
     \end{tablenotes}
\end{threeparttable}
\end{table*}

\section{Multi-waveband data}\label{sec:observations}

\subsection{Dust Spectral Energy Distributions}

For the analysis of the SEDs of the far-infrared part of our multi-waveband data we used a modified black-body spectrum given by $L_\nu\,{=}\,N \pi a^2 Q_\nu 4 \pi B_\nu(T)$, where $Q_\nu\,{\propto}\,\nu^{\,\beta}$, $B_\nu(T)$ is the Planck spectrum, $N$ the number of grains, and $a$ the grain size half-diameter \citep{1983QJRAS..24..267H}.\footnote{While we show here that a physically motivated approach exists, we stress that we use the resulting equation in a phenomenological sense, that is with a single normalisation factor per source.} A submm dust opacity spectral index of $\beta\,{\simeq}\,2.0$ is widely used, and lies within the range of theoretical models \citep{Draine11}, as well as empirical fits to nearby galaxies \citep[e.g.][]{Clements+10}, and is close to the local interstellar medium (ISM) value \citep{PlanckE11XXI}. In terms of the observed flux density\footnote{$S_\nu\,{=}\,(1+z) (L_{(1+z) \nu}\,{/}\, L_\nu) L_\nu\,{/}\, (4 \pi D^2_{\rm L}) $.} this gives

\begin{equation}
S_{\nu} \propto \frac{N \nu^{3+\beta} (1+z)^{4+\beta} D^{-2}_{\rm L}}{{\rm exp}[h \nu (1+z)/(k T_{\rm d})] - 1},
\label{equ1}
\end{equation}
\noindent
where $D_{\rm L}$ is the luminosity distance. Following \citet{Scoville+2014,Scoville+2016} we can adopt a direct proportionality between the flux in the Rayleigh-Jeans regime and the ISM (i.e. \ion{H}{i}, H$_{2}$, and He) mass, with $\kappa_\nu ({\rm ISM})\,{/}\,\kappa_\nu ({\rm dust})\,{=}\,M_{\rm ISM}\,{/}\,M_{\rm dust}$ (${\simeq}\,100$).  Then

\begin{equation}
S_{\nu} = \frac{1.17 x}{e^{x} - 1} \left(\frac{M_{\rm ISM}}{10^{10} {\rm M}_{\odot}}\right)  \left( \frac{\nu}{353\,{\rm GHz}} \right)^{2+\beta} (1+z)^{3+\beta} \left( \frac{\rm Gpc}{D_{\rm L}} \right)^2\,{\rm mJy}, 
\label{equ2}
\end{equation}
\noindent 
where $x\,{=}\,0.484 (35\,{\rm K}\,{/}\,T_{\rm d})\, (\nu\,{/}\,353\,{\rm GHz})\, (1+z)$.

\subsection{\textit{Herschel}-SPIRE}

G073.4$-$57.5 was observed with {\it Herschel\/}-SPIRE at 250, 350, and 500\,$\mu$m (where the corresponding angular resolutions are 18\arcsec, 25\arcsec, and 36\arcsec, respectively) as part of the dedicated follow-up programme of 234 {\it Planck\/} sources \citep{PlanckXXVII15}. The images reached 1$\,\sigma$ (instrument + confusion) noise levels of 9.9\,mJy at 250\,$\mu$m, 9.3\,mJy at 350\,$\mu$m, and 10.7\,mJy at 500\,$\mu$m. 

\begin{table*}[htbp!]
\caption{{\it Herschel\/}-SPIRE sources observed with ALMA in the G073.4$-$57.5 field.}              
\label{table:2}      
\centering                                      
\begin{threeparttable}[b]
\begin{tabular}{llllll}          
\hline\hline                        
\noalign{\vskip 3pt}
ALMA& SPIRE& $S_{350}$\tnote{a}& $S_{350}$\,/\,$S_{250}$\tnote{b}& $S_{500}$\,/\,$S_{350}$\tnote{c}& $S_{1300}$\tnote{d}\\
field& source& [mJy]& & & [mJy]\\
\hline                                   
\noalign{\vskip 3pt}
1&  1& 83$\pm$10& 0.90$\pm$0.21& 0.78$\pm$ 0.22& 3.6$\pm$0.5\\
2&  3& 64$\pm$10& 0.96$\pm$0.31& 1.07$\pm$ 0.23\tnote{f}& 3.6$\pm$0.2\\
3&  7& 56$\pm$10& 1.35$\pm$0.58& 0.74$\pm$ 0.33& 2.1$\pm$0.1\\
4& 11& 50$\pm$10& 0.67$\pm$0.23$^*$& 1.36$\pm$0.26\tnote{f}& 1.9$\pm$0.2\\
5& 13& 50$\pm$10& 0.87$\pm$0.34& 0.73$\pm$ 0.37& 0.6$\pm$0.2\\
6& 15& 49$\pm$10& 1.12$\pm$0.50& 0.97$\pm$ 0.32& 1.6$\pm$0.1\\
7& 19& 44$\pm$10& 1.32$\pm$0.71& 0.25$\pm$ 1.04$^*$& 3.6$\pm$0.4\\
8& 25\tnote{e}& 41$\pm$10& 0.68$\pm$0.29$^*$& 0.31$\pm$0.92$^*$& 0.5$\pm$0.1\\
\hline                                             
\end{tabular}
     \begin{tablenotes}{\small
      \item[a] 350-$\mu$m flux density of SPIRE sources in mJy. SPIRE source IDs~19 and 25 have uncertain detections at 500\,$\mu$m. 
       \item[b/c] Colours of SPIRE sources. An asterisk indicates sources not red enough to formally pass the criteria of~\citet{PlanckXXVII15}, although they would within 1$\,\sigma$. 
       \item[d] Integrated 233-GHz flux density of ALMA galaxies (i.e. summed over the individual integrated flux density estimates) in each {\it Herschel\/} source. 
       \item[e] This {\it Herschel\/} source lies outside the iso-surface brightness contour encompassing 50\,\% of the {\it Planck\/} peak flux density at 545\,GHz.
       \item[f] The same 500-$\mu$m flux density was assigned to both SPIRE sources 3 and 11, instead of deblending the flux in the 500-$\mu$m image. If we split the 500-$\mu$m flux density among the two sources proportionally to their 350-$\mu$m flux density, we obtain $S_{500}\,{/}\,S_{350}\,{=}\,$0.60 for both sources, instead of 1.07 and 1.36 for 3 and 11, respectively.} 
     \end{tablenotes}
\end{threeparttable}
\end{table*}

As discussed in the previous section, the SPIRE analysis revealed the presence of several red sources, compatible with a $z\,{\simeq}\,2$ structure, centred approximately on SPIRE source ID~7 (i.e. ALMA field 3) and highly elongated in the NW-SE direction. 
A modified black-body fit of only the {\it Herschel\/} data for SPIRE sources 3, 7, and 15 (ALMA fields 2, 3, and 6) was consistent with $z\,{\simeq}\,2$, assuming a dust temperature of $T_{\rm d}\,{=}\,35$\,K. Table~\ref{table:2} lists the SPIRE sources targeted with ALMA, along with their measured flux densities at 350\,$\mu$m, the colours relative to 250 and 500\,$\mu$m, and the sum per field of the 1.3-mm flux density resolved into the individual galaxies seen with ALMA.

\subsection{JCMT SCUBA-2}

As part of a SCUBA-2 follow-up of 61 {\it Planck\/} high-$z$ candidates \citep{MacKenzie+16}, G073.4$-$57.5 was observed at 850\,$\mu$m with approximately 10\arcmin\ diameter \lq{}daisy\rq{}-pattern scans, thus covering the whole {\it Planck\/} region. The imaging (with a matched filter applied) reached a minimum rms depth of 1.6\,mJy\,beam$^{-1}$. Table~\ref{table:3} lists the sources identified in \citet{MacKenzie+16}, as well as their peak flux densities. These include all sources with S/N$\,{>}\,4$ and within the Planck beam (i.e. the area in the Planck 353\,GHz map, where the flux density was greater than half the peak flux density). We also include two additional sources that we have identified as having pronounced flux-density peaks coincident with the detected ALMA sources in fields 2 and 4, but with $3\,{<}\,$S/N$\,{<}\,4$ in the SCUBA-2 data. These two additional, lower significance SCUBA-2 sources (labelled \lq{}4+\rq{} and \lq{}5+\rq{} in Table~\ref{table:3} and Fig.~\ref{Fig1}) are well matched to ALMA sources (although blended in the SCUBA-2 map). The apparent clustering of SCUBA-2 sources in Fig.~\ref{Fig1} may indicate a physical concentration of bright submm sources around ALMA field 4. The ratios of the integrated flux densities, ALMA/SCUBA-2, for ALMA fields 2 and 4 are consistent with a modified black-body spectrum for $z\,{=}\,2.0$, $\beta\,{=}\,2.0$, and $T_{\rm d}\,{=}\,30$\,K. Conversely, for the other ALMA sources we would not necessarily expect strong individual detections in the SCUBA-2 data, given the sensitivity and confusion levels. Because of this we performed a stacking analysis by summing the flux densities in the matched-filtered SCUBA-2 maps at the positions of all the ALMA-detected mm sources, obtaining a significant signal of (56$\pm$11)\,mJy, or (4.0$\pm$0.5)\,mJy per source from a weighted average. 

\begin{table}[htbp!]
\caption{SCUBA-2 sources in G073.4$-$57.5.}              
\label{table:3}      
\centering                                      
\begin{threeparttable}[b]
\begin{tabular}{cccc}          
\hline\hline                        
\noalign{\vskip 3pt}
SCUBA-2& RA& Dec& $S_{850}$\\
ID & [h:m:s]& [d:m:s]& [mJy]\\
\hline                                   
\noalign{\vskip 3pt}
 0& 23:14:42.3& $-$04:16:40& 10.4$\pm$1.8\\
 1& 23:14:42.6& $-$04:20:00& 13.6$\pm$2.5\\
 2& 23:14:41.8& $-$04:17:44&  8.3$\pm$2.0\\
 3& 23:14:34.6& $-$04:17:00&  7.2$\pm$1.8\\
4+& 23:14:38.4& $-$04:16:25&  7.1$\pm$1.6\\
5+& 23:14:40.2& $-$04:17:00&  5.5$\pm$1.7\\
\hline                                             
\end{tabular}
\end{threeparttable}
\end{table}

\subsection{Spitzer IRAC}\label{subsecspitzer}

G073.4$-$57.5 was observed with {\it Spitzer\/}-IRAC in GO11 (PID 80238, PI H.~Dole), along with 19 other promising (i.e. high S/N and \lq{}red\rq{}) {\it Planck\/} sources with complimentary {\it Herschel\/} data. The observations involved a net integration time of 1200\,s per (central) sky pixel at 3.6\,$\mu$m (hereafter \lq{}channel~1\rq{}) and 4.5\,$\mu$m (hereafter \lq{}channel~2\rq{}) over an area of about $5\arcmin\,{\times}\,5\arcmin$, and two additional side fields of the same area covered only in channel~1 or in channel~2. The area mapped in both channels with 2\arcsec\ angular resolution is well matched to the angular size of one {\it Planck\/} beam and covers the full area of interest. 

Source extraction in the IRAC mosaics was performed using {\tt SExtractor} \citep{BertinArnouts96}, with the IRAC-optimised parameters of \citet{Lacy+05}. The detection threshold was set to 2$\,\sigma$. A choice was made not to filter the image due to the high density of sources. Photometry was performed using the {\tt SExtractor} dual mode with the channel-2 mosaic as the detection image. Given the relative depth of channel~1 compared to channel~2, a detection at the longer wavelength can be sufficient to confirm that the source is red \citep[i.e. selecting galaxies at $z\,{>}\,1.3$, see][]{Papovich08}, where \lq{}red\rq{} in this context is defined as [3.6]$\,{-}\,$[4.5]$\,{>}\,-0.1$ (in AB magnitudes). Aperture photometry was performed in a 2\arcsec\ radius circular aperture, and aperture corrections were applied. The catalogues were then cut to 50\,\% completeness in channel~2 (at 2.5\,$\mu$Jy). The surface density of IRAC red sources was computed in a circle of radius 1\arcmin\ around SPIRE source ID~1 (which is the brightest red source in the {\it Herschel\/}-SPIRE field and central to the structure of bright {\it Herschel\/} sources selected for the ALMA pointings). The resulting surface density estimate is 14.6\,arcmin$^{-2}$. When compared to the field value derived from the {\it Spitzer\/} ultra deep survey (SpUDS) data at the same depth, which has a mean source density of 9.2\,arcmin$^{-2}$ (and a standard deviation of 2.2\,arcmin$^{-2}$), this corresponds to an overdensity of approximately $2.5\,\sigma$ \citep{Martinache+18}.

\begin{figure*}[htbp!]
\centering                                      
 	\includegraphics[width=5.5cm]{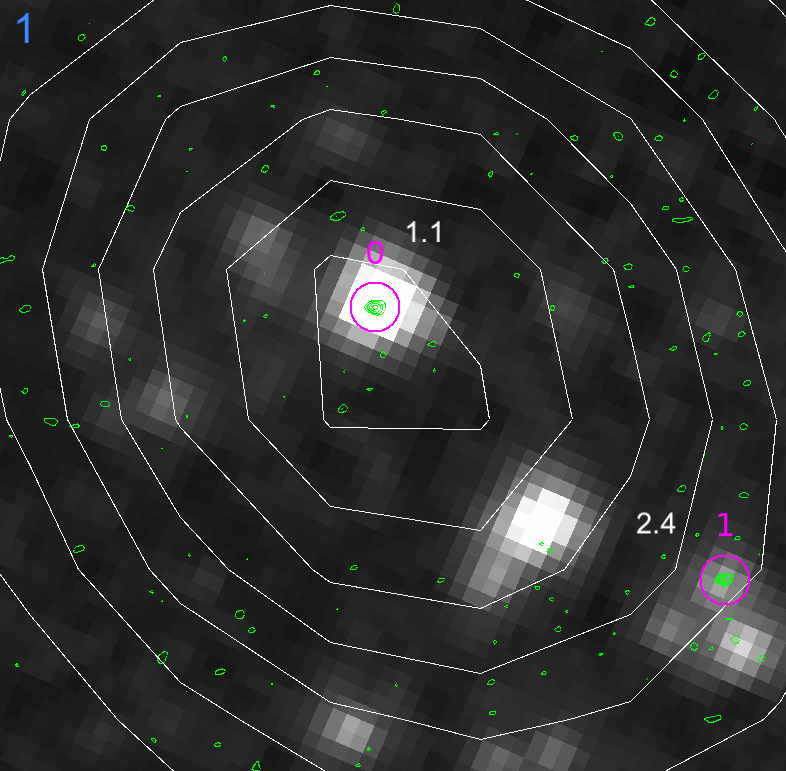}
	\includegraphics[width=5.5cm]{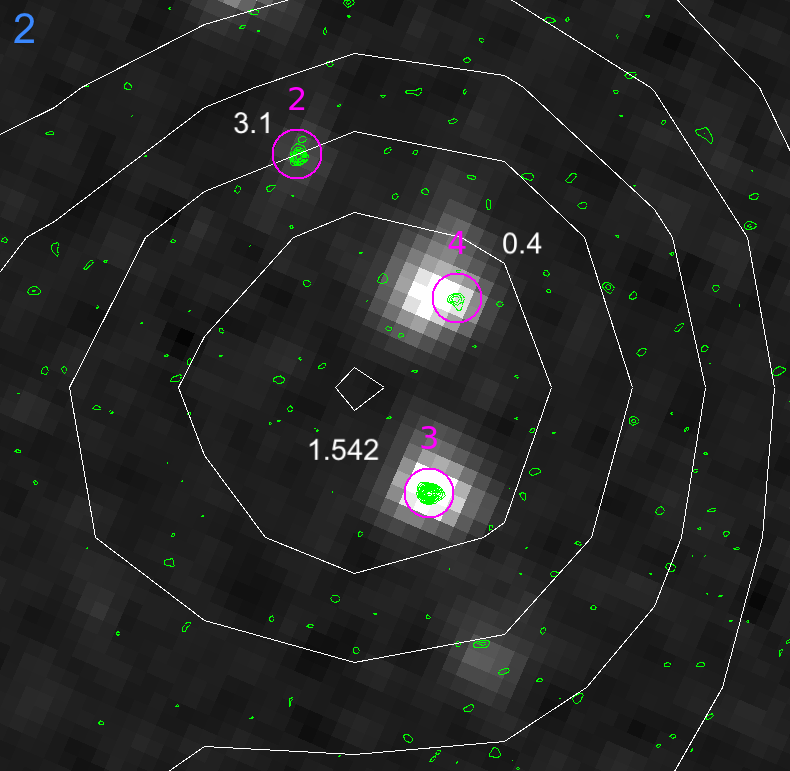}
 	\includegraphics[width=5.5cm]{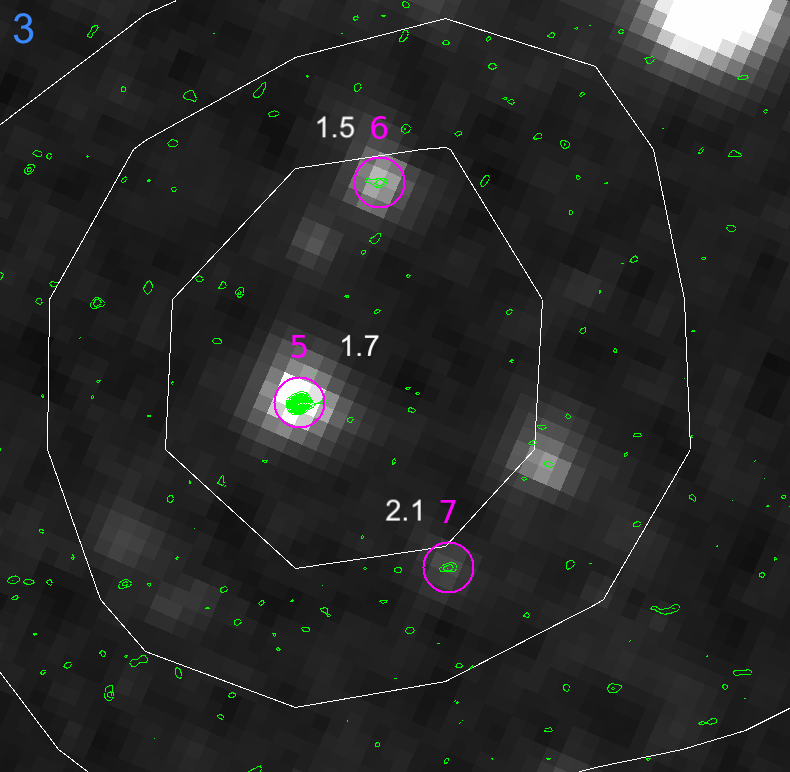}
	\vspace{0.06cm}\\
 	\includegraphics[width=5.5cm]{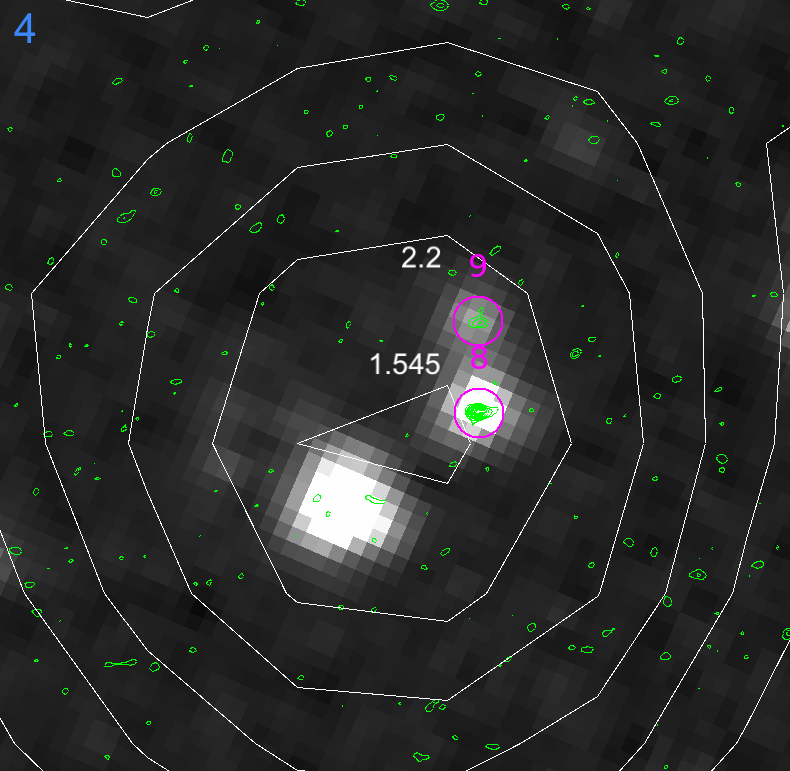}
	\includegraphics[width=5.5cm]{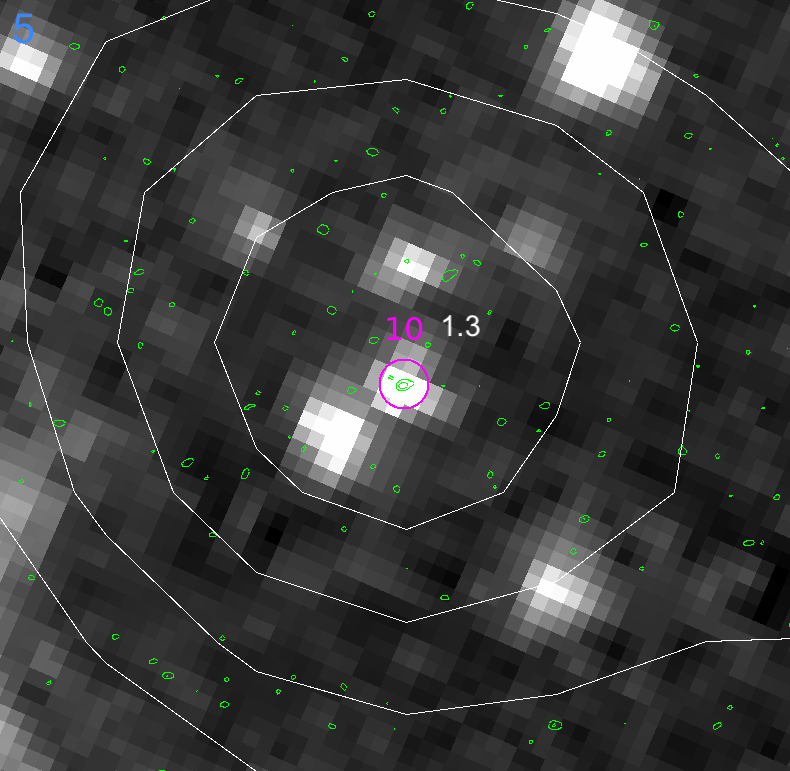}
 	\includegraphics[width=5.5cm]{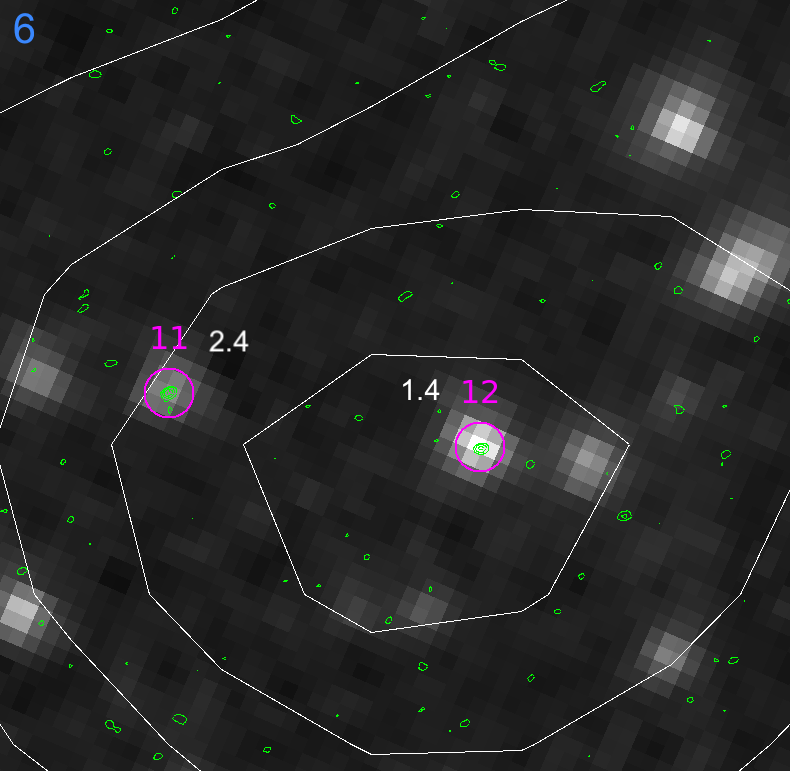}
	\vspace{0.06cm}\\
 	\includegraphics[width=5.5cm]{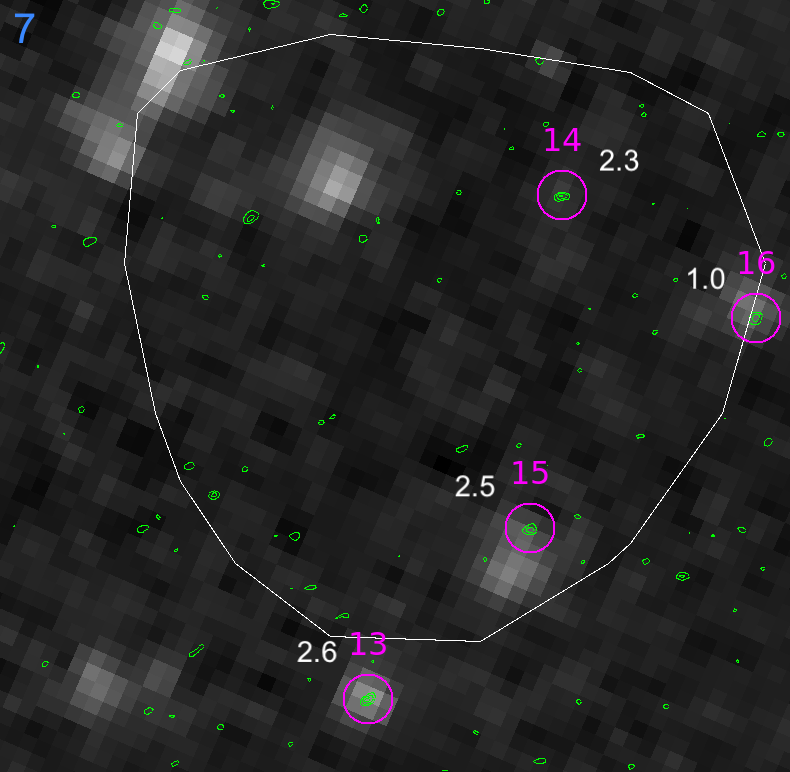}
 	\includegraphics[width=5.5cm]{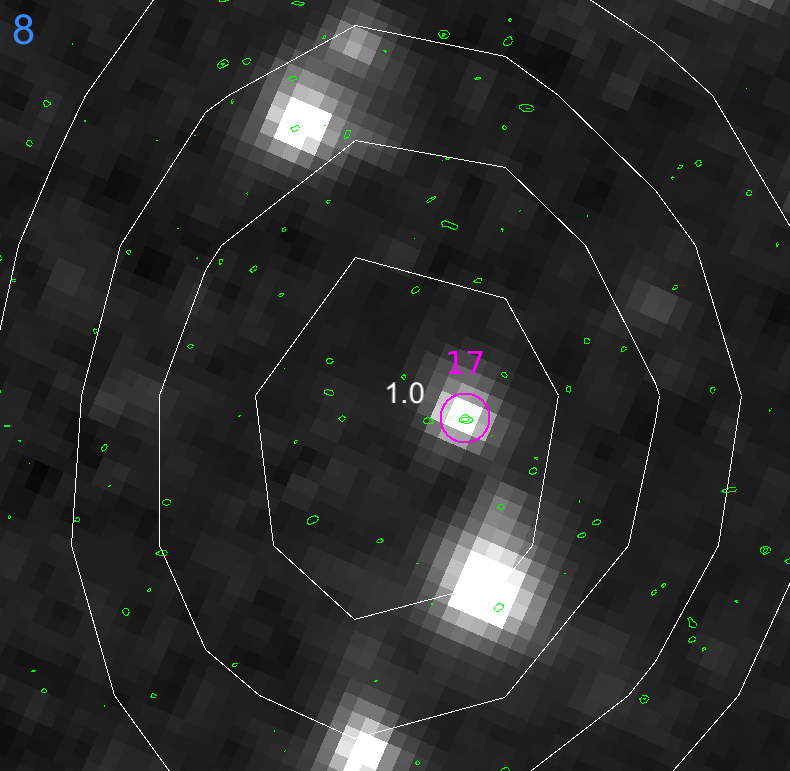}
	\caption{{\it Spitzer\/} channel-2 postage stamps ($30\arcsec\,{\times}\,30\arcsec$, with 2\arcsec\ angular resolution) in grey scale (from 0.35 to 0.6\,MJy\,sr$^{-1}$) of the eight ALMA fields. White contours show the {\it Herschel\/} 250-$\mu$m surface brightness (from 0.01\,Jy\,beam$^{-1}$ in 0.005\,Jy\,beam$^{-1}$ or 1$\,\sigma$ steps). Green contours represent the ALMA surface brightness (from 0.12\,mJy\,beam$^{-1}$ in 0.06\,mJy\,beam$^{-1}$ or 1$\,\sigma$ steps). The detected ALMA galaxies are labelled with 1\arcsec\ radius magenta circles, their photometric redshifts derived in Sect.~\ref{sec:analysis} (or spectroscopic redshifts for ALMA IDs~3 and 8, Sect.~\ref{line_detection}) are given in white, and the ALMA fields are numbered in blue.}
     \label{Fig2}
\end{figure*}

The ALMA detections have a match in at least the channel-2 image (Fig.~\ref{Fig2}), apart from galaxy ID~14, where there is emission in the {\it Spitzer\/} channel-2 map, but not significant enough to claim a detection. Most of the counterparts have a positional difference of $d\,{<}\,0.4\arcsec$, except for three ALMA galaxies: ID~4 (0.6\arcsec); ID~16 (0.7\arcsec); and ID~15 (1.1\arcsec). In these cases the IRAC emission is seen to be extended (likely composites of two sources), with the ALMA source position still matching the detectable surface brightness of the IRAC source. It is also worth pointing out that these three galaxies (IDs~4, 15, and 16) match to blue IRAC sources. We note that the significant counterparts for ALMA IDs~2 and 7 appear weak in contrast to Fig.~\ref{Fig2}. 

\begin{figure*}[htbp!]
	\resizebox{\hsize}{!}{\includegraphics{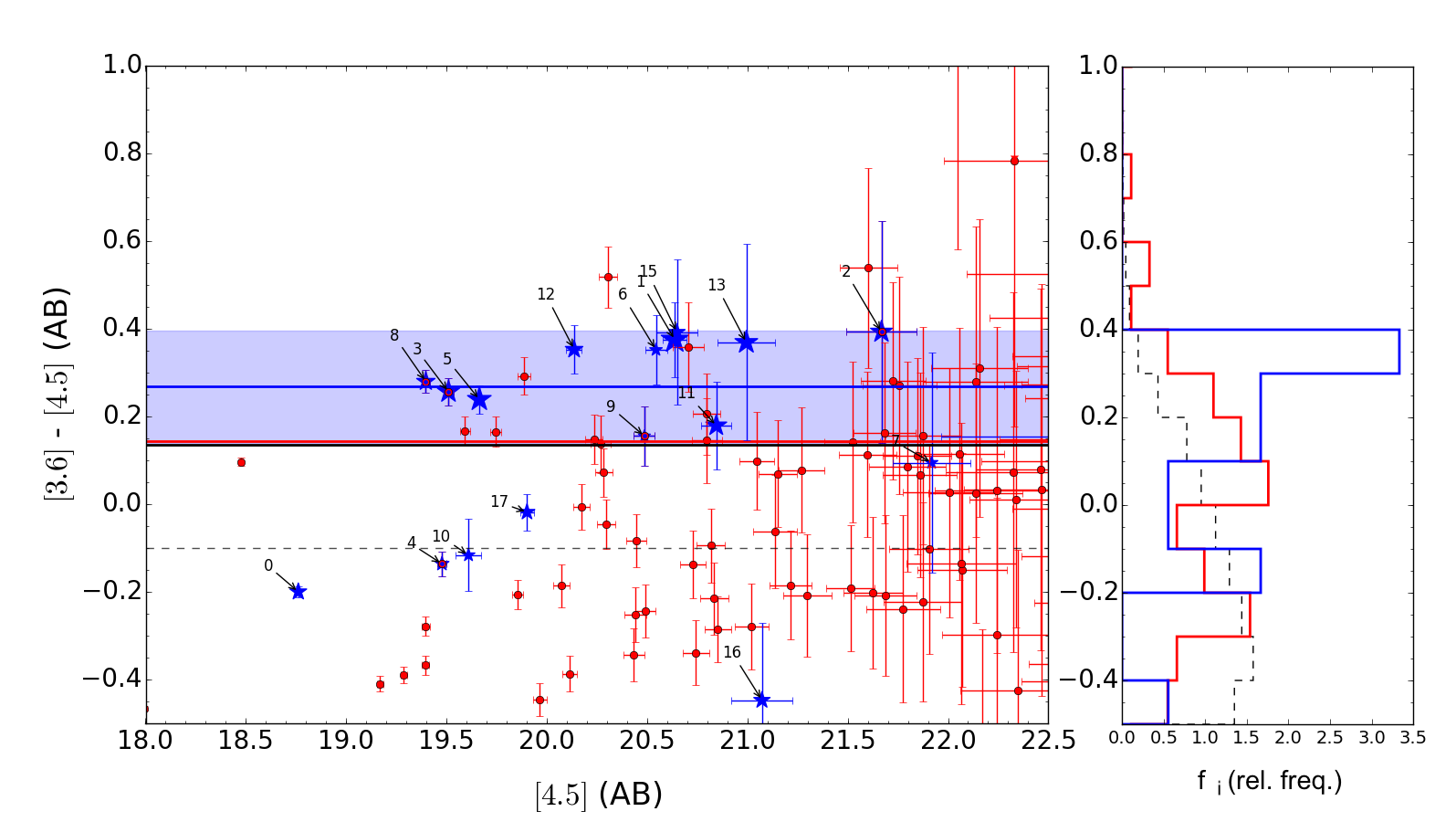}}
	\caption{{\it Left}: Colour ([3.6]$\,{-}\,$[4.5]) versus magnitude ([4.5]) diagram for IRAC sources (red points) located within 1\arcmin\ of SPIRE source 11, and for all the ALMA sources (blue stars); \lq{}[3.6]\rq{} means \lq{}channel 1\rq{} and \lq{}[4.5]\rq{} is \lq{}channel 2\rq{}. Numbers indicate the ALMA source IDs, as in Table~\ref{table:1}. The black dashed line indicates a colour of $-0.1$ (red sources are defined to have colours above this value), the red line indicates the median colour of IRAC red sources within 1\arcmin\ of SPIRE source 11 (0.14\,mag), and the blue line indicates the median colour for the ALMA sources matched to IRAC red sources (0.27\,mag). The dispersions around these median values are 0.15 and 0.13\,mag, respectively (the latter is indicated by the blue region). The solid black line indicates the colour of a single stellar population formed at $z_{\rm f}\,{=}\,5$, passively evolved to redshift $z\,{=}\,1.5$ \citep[from][]{bc03}, but extinction and possible metallicity effects have not been considered. Most ALMA sources lie on a sequence in this colour-magnitude plane, a characteristic feature of high-$z$ structures \citep[e.g.][]{muzzin13,Rettura+14}. We note that ALMA ID~14 is not plotted here, because it was not detected in the {\it Spitzer\/}-IRAC data. {\it Right}: Normalised distribution of the colour of IRAC sources: the red line corresponds to sources within 1\arcmin\ of SPIRE source 11; the blue line shows the ALMA sources; and the black dashed line shows the distribution of the colours of general sources in the COSMOS field for comparison. There is a significant excess of red galaxies around SPIRE source 11, in particular for the ALMA detections (further details in the text).}
     \label{Fig3}
\end{figure*}

Searching for a stellar bump sequence \citep{muzzin13} in the colour-magnitude diagram (Fig.~\ref{Fig3}) of sources lying in a circle of radius 1\arcmin\ (balancing increasing numbers versus avoiding confusion) around SPIRE source 11, we found a median colour of IRAC red sources of 0.14\,mag, and a dispersion of 0.15\,mag. Most ALMA matches exhibit distinctly redder colours, with a median of 0.27\,mag ([3.6]$\,{-}\,$[4.5]), and a dispersion of 0.13\,mag. Such colours are compatible with a $z\,{\simeq}\,1.7$ structure \citep{Papovich08}, but the scatter is high. SPIRE source 11 was chosen because it lies at the centre of an overdensity of IRAC sources (and indeed the majority of the SCUBA-2 detections are clustered around there, see Fig.~\ref{Fig1}).

Comparing the colour distributions shown in the right panel of Fig.~\ref{Fig3} between ALMA sources, IRAC sources around SPIRE source 11, and all IRAC sources from the COSMOS field for reference, the Kolmogorov-Smirnov statistic gives large deviations (the highest for the ALMA sources versus the COSMOS sources) with probabilities less than 0.02 for the ALMA sources versus the IRAC sources around SPIRE source 11 and less than $2\,{\times}\,10^{-5}$ relative to the IRAC sources from COSMOS; thus there is strong evidence that the sources are not drawn from the same distributions. 

\subsection{CFHT WIRCam J and K}

G073.4$-$57.5 was observed by CFHT-WIRCam at 1.3\,$\mu$m ($J$ band) and 2.1\,$\mu$m ($K_{\rm s}$ band) in projects PID 13BF12 and PID 14BF08 (PI H.~Dole). The total integration times were 9854\,s and 4475\,s for the $J$ and $K_{\rm s}$ bands, respectively. The area covered was $25\arcmin\,{\times}\,25\arcmin$, and the central $18\arcmin\,{\times}\,19\arcmin$ was selected in order to exclude the edges with high noise. For this analysis we extracted sources using {\tt SExtractor} in dual mode with detection in the $K_{\rm s}$ band, reaching $K_{\rm lim} = 22.94\pm0.01$ (AB, statistical error only) and $J_{\rm lim} = 24.01\pm0.01$, at a threshold level of 2.5$\,\sigma$ (50\,\% completeness). The completeness level was determined by placing 1000 simulated point sources at random positions, then using {\tt SExtractor} to detect them and measure the percentage of recovered objects. By applying this procedure 10 times per filter, we derived the statistical error. The photometry was performed in a 2\arcsec\ radius circular aperture and we applied the aperture correction in the same way as with the IRAC data. All sources flagged by {\tt SExtractor} in the $K_{\rm s}$ band (except for blended ones), representing 11\,\% of the catalogue, were removed from the analysis. We then matched the resulting catalogue with the 18 {\it Spitzer\/}-IRAC+ALMA sources and found 13 matches within 0.6\arcsec\ (consistent with the seeing of the CFHT data). The five unmatched sources are IDs~1 and 10 (best match separation ${>}\,$2.5\arcsec), and 2, 7, and 14 (not detected in $K_{\rm s}$). 

\begin{figure*}[htbp!]
	\resizebox{\hsize}{!}{\includegraphics{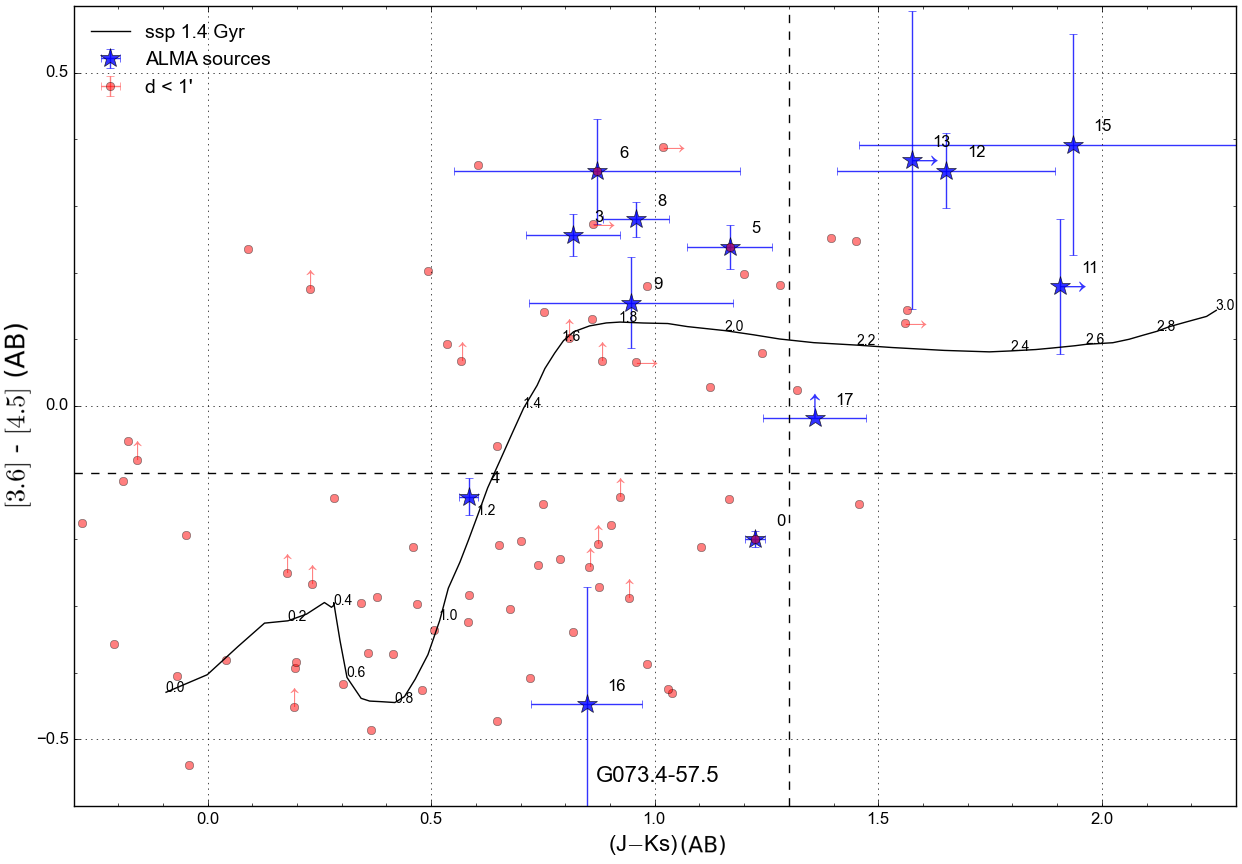}}
	\caption{{\it Spitzer\/}-IRAC versus CFHT WIRCam colour-colour diagram, with a track drawn from \citet{bc03} for a 1.4-Gyr simple stellar population, numbered by redshift. The horizontal and vertical dashed lines indicate the \citet{Papovich08} and \citet{Franx+03} criteria, respectively, representing colours of $z\,{\simeq}\,1.3$ and $z\,{\simeq}\,2$ galaxies (cf.~the labels on the model curve). A redshift around $z\,{=}\,1.6$--2.6 is indicated for the majority of the ALMA galaxies. Arrows represent 2$\,\sigma$ limits for the sources not detected in any channel or band.} 
     \label{Fig5}
\end{figure*}

In Fig.~\ref{Fig5} we summarise the evidence that the majority of ALMA sources lie at redshifts $z\,{\simeq}\,2$ following the colour-redshift criteria of \citet{Papovich08} and \citet{Franx+03}, and the evolutionary state predictions for a 1.4-Gyr simple stellar population (corresponding to a formation redshift $z_{\rm f}\,{=}\,3.5$ for an observed redshift of $z\,{=}\,2$, approximately applicable for the majority of the ALMA-detected galaxies). The galaxies with ALMA IDs~3, 5, 6, 8, and 9 appear to be more consistent with a redshift below 2, whereas the colours of IDs~11, 12, 13, 15, and possibly 17 seem to indicate redshifts above 2 (while having larger uncertainties). ALMA IDs~0, 4, and 16 may be interlopers at lower redshift ($z\,{\leq}\,1$).

\subsection{WISE}

Additional mid-IR data were obtained from the AllWISE catalogue \citep{wright10,mainzer11} using a search radius of 3\arcsec. Six galaxies are detected in the W1 (3.4\,$\mu$m) or W2 (4.5\,$\mu$m) bands, one in the W3 (12\,$\mu$m) band, and none in the W4 (22\,$\mu$m) band (for details see Table~\ref{phot_data}).

\subsection{Pan-STARRS}

We also searched the Pan-STARRS ($grizy$) DR1 data\footnote{\url{https://panstarrs.stsci.edu}} \citep{chambers16}, since even upper limits can provide additional constraints to the fits. The upper limits in AB mags are $g=23.3$, $r=23.2$, $i=23.1$, $z=22.3$, and $y=21.4$. Only ALMA ID~4 is detected in the $r$, $i$, $z$, and $y$ bands. The full set of available photometric data is reported in Table~\ref{phot_data}.

\subsection{VLA FIRST}

A potential contribution from a radio-loud active galactic nucleus (AGN) can 
be investigated using the radio maps at 1.4\,GHz of the VLA FIRST
Survey~\citep{Becker+1995}.  The 5$\,\sigma$ threshold of the VLA FIRST
survey, above which a point source is considered detected, is 0.75\,mJy. 
Such a limit corresponds to an FIR luminosity $\geq$10$^{13}$\,L$_{\odot}$ at
redshift $\geq$1.5 for a radio-quiet AGN or a star-forming galaxy assuming a
logarithmic FIR-to-radio luminosity ratio $q_{\rm IR}$\,=\,2.4 and a radio slope
$\alpha_{\rm radio}$\,=\,$-0.8$~\citep{ivison+10}.  We find no strong evidence
of flux at the positions of the ALMA galaxies.  This is consistent with the
estimated FIR luminosities (Table~\ref{bestfit_params}), which are all
below such a value.  We can thus affirm that none of the detected ALMA
sources is a radio-loud AGN. The highest peak brightness of
0.62\,mJy\,beam$^{-1}$ is seen within the 5.4\arcsec\ synthesised beam from
the position of ALMA ID~12, but this is still below the detection threshold
and consistent (log(L$_{\rm IR}$/L$_{\odot}$)$\la$13.2 at $z$\,=\,1.4
versus an estimated log(L$_{\rm IR}$/L$_{\odot}$)=12.12$\pm$0.06) with the
FIR luminosity estimated for this source (Table~\ref{bestfit_params}). 
However, since radio-quiet AGNs and SFGs both lie on the
FIR-to-radio relation, we cannot claim that the tentative radio detection of
ALMA ID~12 is due to AGN or star-forming activity. Furthermore, the two 
ALMA galaxies, whose fit is consistent with an obscured AGN template are not 
ID 12.

\section{Analysis}\label{sec:analysis}

\subsection{Source counts} 

Since we have targeted only the brightest {\it Herschel\/} sources found within this {\it Planck\/} peak, we can only make a qualitative comparison to known average ALMA mm source counts in order to discuss the approximate overdensity in sources of these regions. 

Each ALMA field has been searched for sources within a 37\arcsec\ diameter circle, over which the noise increases from the centre outwards by up to a factor of 5. The area of each field is 0.30\,arcmin$^{2}$, and adding together the eight fields we obtain a total survey area of 2.4\,arcmin$^{2}$, or $6.6\,{\times}\,10^{-4}$\,deg$^{2}$. For the count estimate we take the eight sources in our sample with flux densities above 0.9\,mJy. The effective area over which these sources can be detected is 75\,\% of the total (i.e. where rms is ${<}\,S_\nu$\,/\,(S/N)$\,{=}\,$0.9\,mJy\,/\,4.5$\,{=}\,$0.2\,mJy). Thus, the surface density is 8\,/\,(0.75$\,{\times}\,6.6\,{\times}\,10^{-4}$\,deg$^{2}$)$\,{=}\,$1.6$\,{\times}\,10^{4}$\,deg$^{-2}$. 

Comparing to recent blank-field counts of ALMA sources \citep[e.g.][]{Hatsukade+2016,Dunlop+16}, serendipitous counts derived from various archival data \citep[e.g.][]{Hatsukade+13, Ono+2014, Carniani+15, Oteo+16, Fujimoto+2016}, or source numbers found in lensing cluster fields \citep[e.g.][]{Gonzalez+2016,2018A&A...620A.125M}, we estimate an expected 1.2-mm source density of 0.6--$2\,{\times}\,10^3$\,deg$^{-2}$, where the lower estimate \citep[from][]{Oteo+16} is derived from a relatively large area of different fields used for the serendipitous searches, which might be expected to reach beyond the effects of cosmic variance. Thus, the number of sources we find in the ALMA pointings of G073.4$-$57.5 is a factor of 8--30 higher than estimates of the average number of mm sources in the sky. 

In terms of the total numbers of mm/submm sources in the G073.4$-$57.5 field, 18 are identified with ALMA (even without a complete mosaic of the total emission region of the {\it Planck\/}/{\it Herschel\/} peak), and an additional four from SCUBA-2, for a total of 22 mm/submm sources in the area of the {\it Planck\/} peak. In comparison, typical \lq{}proto-cluster\rq{} overdensities, not selected by their high integrated submm flux, do not show the same abundance. Examples include the COSMOS $z\,{=}\,2.47$ structure \citep{Casey+13,Casey+15} and the SSA22 $z\,{=}\,3.09$ structure \citep{Chapman+01,Umehata+15}, each of which contains 12 sources \citep[][in particular their table~1, for a comprehensive summary of SFGs in several overdense regions]{Casey16} at a comparable depth, although the SCUBA-2 850-$\mu$m data for the COSMOS structure are not as deep, at 0.8\,mJy rms \citep[][]{Casey+13}. In a more recent study of the SSA22 structure \citet{Umehata+2016} find 18 ALMA sources (${>}\,5\,\sigma$) at 1.1\,mm, but over an area of $2\arcmin\,{\times}\,3\arcmin$ and with a depth of 0.06--0.1\,mJy, much wider and overall somewhat deeper (given their shorter wavelength and homogeneous coverage) than our data, requiring approximately 16 times our on-source time with comparable numbers of antennas and conditions. At a similar depth and area to our selected eight pointings, this would correspond to about four detections. A comparison with the $z\,{=}\,1.46$ cluster XCS J2215.9$-$1738 studied by \citet{stach17} with the same ALMA on-source time in a 1\,arcmin$^2$ central mosaic shows a similar number of sources (14, with 12 likely members), but they are all weaker (${<}\,$1\,mJy). They find a total SFR of 850\,M$_{\odot}$\,yr$^{-1}$, which is lower than the ${\ga}\,$2700\,M$_{\odot}$\,yr$^{-1}$ in our sample (cf. Sect.~\ref{sec:analysis}). The SCUBA-2 sources in XCS J2215.9$-$1738 break up mostly into groups of two to three ALMA sources, similar to the SCUBA-2 and {\it Herschel\/} sources in G073.4$-$57.5. The early ($z{=}\,4.00$) proto-cluster found by \citet{Oteo+18} with 10 galaxies, on the other hand, has a higher SFR of ${\ga}\,$6500\,M$_{\odot}$\,yr$^{-1}$.

We can conclude that for our data, contamination by the average background source is expected to be small, amounting to about one to three galaxies not related to the {structure(s) causing the {\it Planck\/} peak. We discuss later whether gravitational lensing could affect the number counts. 

\subsection{Spectral energy distributions and photometric redshifts}

The detection, in most cases, of several ALMA galaxies (with sub-arcsecond accuracy) per single {\it Herschel\/} target allows us to employ a deblending technique to estimate the {\it Herschel\/}-SPIRE fluxes of these galaxies, which can then be used to fit SEDs and derive several physical properties. To accomplish this we use a combination of a recently developed algorithm called {\tt SEDeblend} \citep{MacKenzie+16}, specifically designed for confused FIR imaging, and the {\tt EAZY} code \citep{Brammer+2008}, to estimate source photometric redshifts and find the best fits to their multi-wavelength SEDs.

We first applied {\tt EAZY} to all available flux measurements for each source (excluding those from {\it Herschel\/}-SPIRE, which are initially too confused to be useful) to obtain posterior probability distributions for photometric redshifts.  For the 850-$\mu$m data from SCUBA-2, the flux density from SCUBA-2 ID~4+ was assigned proportionally to the ALMA flux of IDs~2 and 4 (ID~3 also falls into field 2, but is 11\arcsec\ from the SCUBA-2 position, see Fig.~\ref{Fig1}), and the flux density from SCUBA-2 ID~5+ was assigned proportionally to ALMA IDs~8 and 9.  Similarly, the {\it Herschel\/}-SPIRE flux densities were initially assigned according to the ALMA flux-density ratios of the constituent galaxies. The library of SED templates employed by {\tt EAZY} covers the full optical--mm spectral range and a wide variety of galaxy types, including early-type galaxies, SFGs, SBs, AGNs, and SMGs, for a total of 37 templates \citep[23 from the SWIRE library, and 14 from the zLESS compilation,][]{polletta07,danielson17}.

We then used the resulting photometric redshift posterior probability distributions as inputs to {\tt SEDeblend}. To summarise briefly, {\tt SEDeblend} reconstructs the {\it Herschel\/}-SPIRE 250-$\mu$m, 350-$\mu$m, and 500-$\mu$m images and the SCUBA-2 850-$\mu$m image by placing a point source multiplied by the appropriate instrumental point-spread function at each location of a detected ALMA galaxy and adding a constant background offset, then uses a Markov chain technique to simultaneously fit for galaxy SED parameters. The ALMA images were not reconstructed, since the much greater angular resolution there, after {\tt CLEAN}ing, leads to essentially no source blending. The model takes into account each {\it Herschel\/}-SPIRE instrumental transmission function (typically amounting to a 10\,\% flux correction), and considers calibration uncertainties by multiplying the flux in each band by a nuisance parameter, whose prior is a Gaussian function with a mean of 1.0 and a standard deviation given by each instrument's quoted calibration uncertainty. The SEDs are modelled as modified black-bodies (Eq.~\ref{equ1}) at a redshift $z$ with a temperature $T_{\rm d}$, an overall normalisation constant, and the dust emissivity-index is fixed at $\beta\,{=}\,2.0$.  For more details on {\tt SEDeblend} we refer to \citet{MacKenzie+16}.

For the fitting, a Markov chain Monte Carlo (MCMC) algorithm with Gibbs sampling and adaptive step-sizing was used to maximise a Gaussian likelihood function calculated pixel-by-pixel for the SPIRE and SCUBA-2 images, and source-by-source for the ALMA flux measurements. The chain was run for 120,000 iterations and the first 20,000 iterations were removed as the \lq{}burn-in\rq{} sequence. We set a sufficiently wide uniform prior on the amplitudes of the modified black-body SEDs and the background levels to leave them effectively unconstrained, and a uniform prior between 10 and 100\,K on the dust temperatures \citep[since no galaxies have been observed to lie outside this range, e.g.][]{Dale+2012, Swinbank+2014}. To remove the degeneracy between temperature and redshift in the modified black-body model, the photometric redshift posterior probability distributions from {\tt EAZY} in the previous step were input as the prior for the new photometric redshifts.

From the resulting Markov chain we derived the {\it Herschel\/}-SPIRE flux densities in each band by evaluating $S_\nu (\nu_{\it b})$, with {\it b} labelling the band, from each iteration within the MCMC algorithm (thus obtaining the marginal likelihood) and calculate the maximum likelihood and 68\,\% confidence interval; these are reported in Table~\ref{table:4}. In a few cases the 68\,\% confidence interval extends to 0\,mJy (IDs~11, 15, and 16). For these cases, and when the measured flux is ${<}\,2\,\sigma$, we assigned an upper limit equal to the maximum likelihood flux plus 3$\,\sigma$. In the SED fitting procedure and in Fig.~\ref{bestfit_seds} we also considered the SPIRE confusion limits \citep[i.e. 5.8, 6.3, and 6.8\,mJy at 250, 350, and 500\,$\mu$m, respectively; ][]{nguyen10} when dealing with upper limits. For consistency, we check the mm/submm properties (such as dust temperature, FIR luminosity, and SFR, discussed in the following section) derived from {\tt SEDeblend} with those derived in Sect.~\ref{FIRparam} and find them to be in generally good agreement. 

\begin{table}[htbp!]
\caption{Deblended SCUBA-2 and {\it Herschel\/}-SPIRE flux densities for the individual ALMA-detected galaxies. In cases where the measured flux density is ${<}\,2\,\sigma$, we report an upper limit corresponding to the maximum likelihood flux density plus 3$\,\sigma$.}          
\label{table:4}      
\renewcommand{\arraystretch}{1.2}
\centering                                      
\begin{tabular}{lrrrr}          
\hline\hline                        
ID & $S_{250}$& $S_{350}$& $S_{500}$& $S_{850}$\\
 & [mJy]& [mJy]& [mJy]& [mJy]\\
\hline                                   
\noalign{\vskip 3pt}
\,\,0 & 71$^{+ 3}_{- 3}$& 48$^{+ 2}_{- 2}$& 22$^{+ 1}_{- 1}$& 4.7$^{+0.4}_{-0.4}$\\ 
\,\,1 & 19$^{+ 3}_{- 3}$& 29$^{+ 1}_{- 1}$& 23$^{+ 1}_{- 1}$& 7.8$^{+0.6}_{-0.6}$\\ 
\,\,2 & $<$12           & 10$^{+ 3}_{- 3}$& 10$^{+ 1}_{- 1}$& 4.3$^{+0.4}_{-0.4}$\\ 
\,\,3 & 25$^{+ 3}_{- 3}$& 25$^{+ 3}_{- 3}$& 16$^{+ 1}_{- 1}$& 4.6$^{+0.2}_{-0.2}$\\ 
\,\,4 & 31$^{+ 5}_{- 5}$& 16$^{+ 2}_{- 2}$&  6$^{+ 1}_{- 1}$& 1.0$^{+0.3}_{-0.3}$\\ 
\,\,5 & 21$^{+ 4}_{- 4}$& 25$^{+ 2}_{- 2}$& 17$^{+ 1}_{- 1}$& 5.0$^{+0.3}_{-0.3}$\\ 
\,\,6 & 14$^{+ 3}_{- 3}$& 11$^{+ 2}_{- 2}$&  6$^{+ 1}_{- 1}$& 1.4$^{+0.2}_{-0.2}$\\ 
\,\,7 &  9$^{+ 4}_{- 4}$&  9$^{+ 2}_{- 2}$&  5$^{+ 1}_{- 1}$& 1.3$^{+0.2}_{-0.2}$\\ 
\,\,8 & 64$^{+ 4}_{- 4}$& 43$^{+ 3}_{- 3}$& 20$^{+ 2}_{- 2}$& 4.2$^{+0.3}_{-0.3}$\\ 
\,\,9 & $<$8            &  6$^{+ 2}_{- 2}$&  7$^{+ 1}_{- 1}$& 2.8$^{+0.4}_{-0.4}$\\ 
10 & 46$^{+ 4}_{- 4}$& 35$^{+ 3}_{- 3}$& 17$^{+ 1}_{- 1}$& 3.9$^{+0.4}_{-0.4}$\\ 
11 & $<$32           & $<$41           & $<$13           & 2.7$^{+0.4}_{-0.4}$\\ 
12 & 28$^{+ 4}_{- 4}$& 22$^{+ 2}_{- 2}$& 12$^{+ 1}_{- 1}$& 2.8$^{+0.3}_{-0.3}$\\ 
13 &  9$^{+ 4}_{- 4}$& 16$^{+ 3}_{- 3}$& 14$^{+ 1}_{- 1}$& 5.1$^{+0.8}_{-0.8}$\\ 
14 & 18$^{+ 3}_{- 3}$& 18$^{+ 2}_{- 2}$& 10$^{+ 1}_{- 1}$& 2.7$^{+0.3}_{-0.3}$\\ 
15 & $<$1.4          & $<$5            & $<$7            & 1.3$^{+0.5}_{-0.5}$\\ 
16 & $<$1.3          & $<$4            & $<$6            & $<$2.4\\
17 & 40$^{+ 3}_{- 3}$& 24$^{+ 2}_{- 2}$& 10$^{+ 1}_{- 1}$& 1.9$^{+0.2}_{-0.2}$\\ 
\noalign{\vskip 3pt}
\hline                                             
\end{tabular}\\
\smallskip
\end{table}

We lastly re-ran {\tt EAZY} with the now deblended {\it Herschel\/}-SPIRE and SCUBA-2 flux densities included. The best-fit templates and SEDs are shown in Fig.~\ref{bestfit_seds} in the Appendix, and photometric redshifts and associated uncertainties are listed in Table~\ref{bestfit_params}. The redshift uncertainties correspond to the range of redshifts with a probability higher than half the maximum value. 

Reasonably good fits (median $\chi^2_{\rm red}=0.35$) are obtained for
most of the galaxies, with the exception of IDs~14 and 16. For these sources, which are found in ALMA region 7, it seems plausible that the source blending is simply too substantial to be overcome; four galaxies are sharing a combined flux half that of most of the other regions, where there are three or fewer galaxies. We thus suggest caution in the interpretation of galaxies 13 through 16.

We also note that the two galaxies within region 4, IDs~8 and 9, are the most closely-spaced pair in the data, separated by less than a pixel in the SPIRE maps. This leads to strong degeneracies between the best-fit SED parameters found by {\tt SEDeblend}, which may result in unreliable flux estimates. However, the redshift derived by our procedure for galaxy 8, $z\,{\simeq}\,1.3$, is similar to the ALMA CO spectroscopic redshift (details in Sect.~\ref{line_detection}).

The best-fit templates selected by {\tt EAZY} for all of our sources come from the zLESS library, with only three exceptions, where a template is from the SWIRE library. In one of these three cases (ID~14), the best-fit template corresponds to the prototypical ULIRG Arp\,220, and is thus similar to those in the zLESS library. The presence of a significant FIR peak and red optical colours probably favour this type of template compared to those of passive, spiral, and starburst galaxies. In another case (ID~6), the best-fit template corresponds to Mrk\,231, another well-known ULIRG that contains an obscured AGN, and in the final case (ID~15) the best-fit template is that of an obscured Seyfert galaxy. IDs~6 and 15 are among the reddest sources in the [3.6]$\,{-}\,$[4.5] colour, implying that the peak of the stellar component \citep[at 1.6\,$\mu$m in the rest-frame,][]{sawicki02} must be redshifted to $\lambda\,{>}\,$4.5\,$\mu$m, unless an AGN-heated hot dust component contributes to their mid-IR emission. In the case of ID~6 the FIR SED implies a redshift $z\,{\la}\,1.5$; in order to fit both the red IRAC colour and the FIR SED, an AGN template is favoured, since it can reproduce the red IRAC colour through the contribution of a hot dust component. The FIR SED of ID~15 is not very well constrained, but its relatively low emission compared to the NIR emission, as well as its extremely red colours, favour a hybrid template where both the stellar and AGN components are visible; however, a starburst template at approximately the same redshift yields a similarly good fit, so the presence of an AGN in this source is uncertain.

We validate our photometric redshifts by first checking the SCUBA-2 flux densities predicted for all ALMA galaxies based on the fits. We find that our best-fit SEDs give a total SCUBA-2 flux of 58\,mJy, in good agreement with the stacking result of (56$\pm$11)\,mJy. 

The photometric redshift distribution of all ALMA sources is shown in Fig.~\ref{redshift_dist}, along with the combined probability distribution function (PDF), that is the sum of the likelihoods output from {\tt EAZY}. Both the single photo-$z$ distribution and the combined PDF of the ALMA sources show two redshift concentrations, one at $z\,{\simeq}\,$1.5 and the other at $z\,{\simeq}\,$2.4 (vertical dashed lines in Fig.~\ref{redshift_dist}). The redshift distribution suggests that G073.4$-$57.5 might contain two structures overlapping along the line of sight.

\begin{figure}[htbp!]
\includegraphics[width=9.cm]{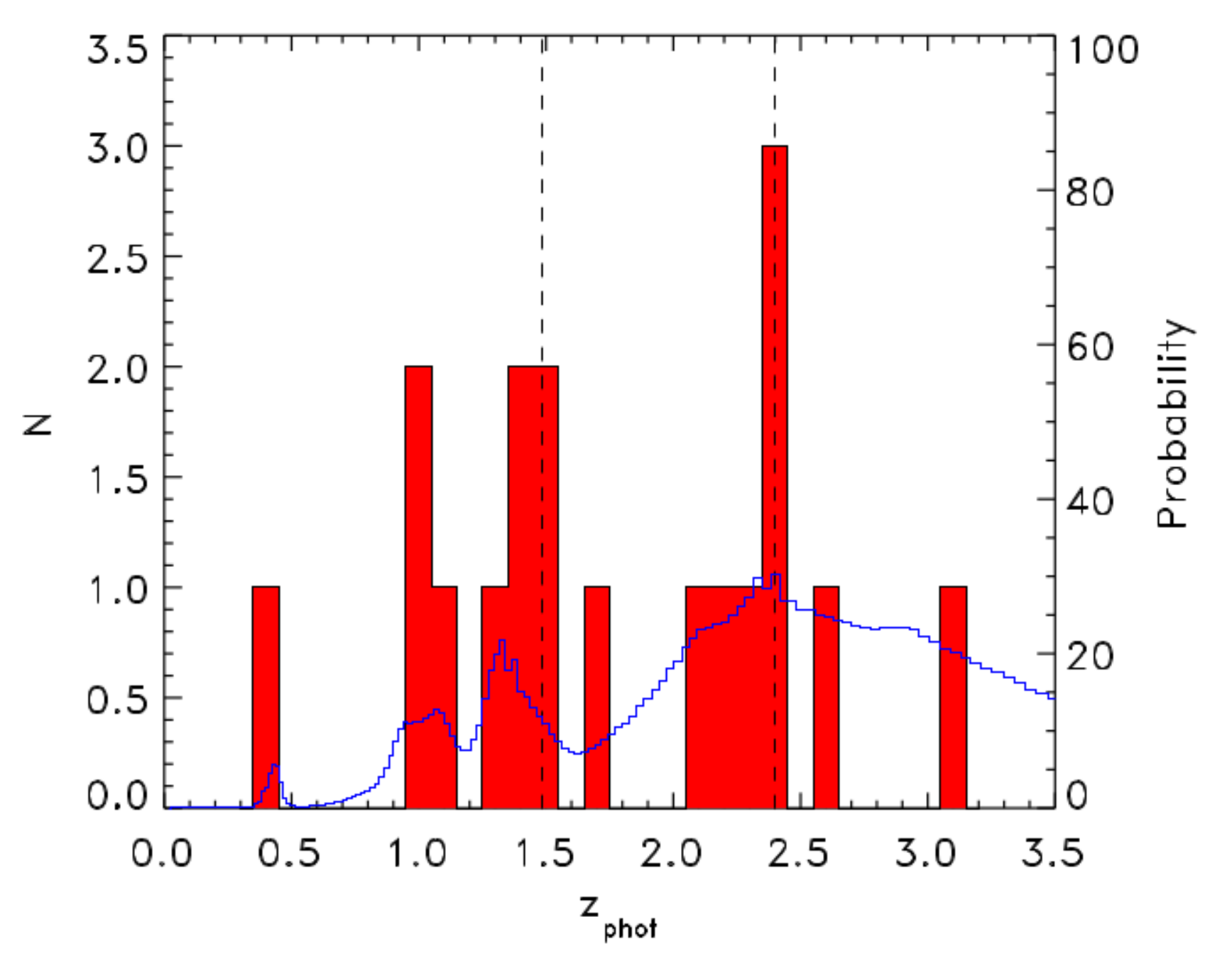}
\caption{Photometric redshift distribution of the maximum-likelihood solutions (filled red histogram) and combined probability density function (PDF, i.e. sum of the individual source likelihoods; solid blue curve) obtained with {\tt EAZY} for all ALMA galaxies in the G073.4$-$57.5 field.  Both the single photo-$z$ distribution and the combined PDF of the ALMA sources show two clear redshift concentrations, one at $z\,{\simeq}\,$1.5 and the other at $z\,{\simeq}\,$2.4 (indicated by the vertical dashed lines).}
\label{redshift_dist}
\end{figure}

\subsection{FIR-derived parameters}\label{FIRparam}

The total (8--1000\,$\mu$m) IR luminosities ($L_{\rm IR}$), dust masses ($M_{\rm d}$), dust temperatures ($T_{\rm d}$), and SFRs of our ALMA sources were estimated by fitting their FIR-mm SEDs with single-temperature modified black-body models. Fits were performed using the {\tt cmcirsed} package \citep{Casey12} and assuming the photometric redshifts derived above (or, when available, the CO spectroscopic redshifts, see Sect.~\ref{line_detection}) and a dust emissivity-index $\beta$ equal to 2.0 \citep[][see the purple curve in Fig.~\ref{bestfit_seds}]{pokhrel16}. Uncertainties on $L_{\rm IR}$ and $T_{\rm d}$ were derived by fitting the SPIRE data and assuming the SPIRE flux plus (minus) 1$\,\sigma$ at 250\,$\mu$m and the ALMA flux plus (minus) 1$\,\sigma$ at 233\,GHz to obtain the best-fit with the highest (lowest) temperature. These two best fits are shown as red (warmest) and cyan (coldest) dashed curves in Fig.~\ref{bestfit_seds}, respectively. From the IR luminosities, SFR estimates were derived assuming the relationship in \citet{kennicutt98}, modified for a Chabrier initial mass function (IMF) \citep{chabrier03}, that is SFR[M$_\odot$\,yr$^{-1}$]$\,{=}\,9.5\,{\times}\,10^{-11}\,L_{\rm IR}$[L$_\odot$]. 

The FIR-derived parameters $L_{\rm IR}$, $M_{\rm d}$, $T_{\rm d}$, and SFR are listed in Table~\ref{bestfit_params}. The majority (${\ga}\,$70\,\%) of the ALMA galaxies are classified as ULIRGs \citep[$L_{\rm IR}\,{\geq}\,10^{12}$\,L$_\odot$; ][]{sanders88a}, with consequently large (${\ga}\,$100\,M$_\odot$\,yr$^{-1}$) SFRs. The highest SFRs (${>}\,$300\,M$_\odot$\,yr$^{-1}$) are measured in IDs~8 ($z_{\rm CO}\,{=}\,$1.5449, see Sect.~\ref{line_detection}), and 1 ($z_{\rm phot}\,{=}\,2.42^{+0.15}_{-0.14}$). In Sect.~\ref{ms_analysis} we find that, in spite of the large SFRs, most ALMA galaxies lie on the SFR--$\mathcal{M}$ MS \citep[][]{speagle14}.  The dust temperatures, with an average of $\left\langle T_{\rm d}\right\rangle\,{=}\,(27\pm5)$\,K, are within the expected range for normal SFGs at $z\,{\ga}\,1$ \citep{magnelli14}, and the dust masses are within the expected range of $10^8$--$10^9$\,M$_\odot$ \citep{popping17,remy_ruyer14}.

In order to estimate stellar masses we fit only the Pan-STARRS-WIRCam-IRAC SED, fixing the redshift to the photo-$z$ or to the spec-$z$, when available, using the {\tt Hyper-$z$} code \citep{bolzonella00} and the composite stellar population models from \citet{bc03}, then assuming a Chabrier IMF. 
The estimated stellar masses are also listed in Table~\ref{bestfit_params}.
The reported uncertainties are likely to underestimated because they do not take into account the redshift uncertainty or the choice of IMF, synthetic models, and fitting method~\citep[e.g.][]{kannappan07,barro11}.  In addition, the 
reported uncertainties are obtained from the best-fit template and do not
consider the likelihoods associated with the full set of models (i.e. full
PDF).  We thus caution using these stellar masses, 
especially when the uncertainties are less than 0.1\,dex.  In the following
analysis we use these estimates only to compare 
our sources with well known relations from the literature.
The scatter associated with these relations is likely to be larger
than the neglected additional uncertainties, and moreover the systematic
uncertainties are less important when considering relative measurements,
so that our interpretations should still be valid.
In the next section we compare the estimated stellar masses and SFRs of our ALMA galaxies to those of typical SFGs.

\begin{table*}[htbp!]
\caption{Best-fit SED parameters and 1$\,\sigma$ uncertainties.}
\label{bestfit_params}.
\renewcommand{\arraystretch}{1.5}
\centering
\begin{threeparttable}[b]
\begin{tabular}{rccccccc} 
\hline\hline
ID& $z_{\rm phot}$& log($L_{\rm IR}$)\tnote{a}& SFR\tnote{b}& $T_{\rm d}$& log($M_{\rm d}$)& log($\mathcal{M}$)\tnote{c}& log($M_{\rm ISM}$)\tnote{d}\\ 
& &  [L$_{\odot}$]& [M$_{\odot}$\,yr$^{-1}$]& [K]& [M$_{\odot}$]& [M$_{\odot}$]& [M$_{\odot}$]\\ 
\hline
  0& 1.06$^{+0.06}_{-0.06}$& 12.27$^{+0.04}_{-0.04}$& 179$^{+ 18}_{-17}$& 26.6$^{+0.8}_{-0.8}$& 8.73$^{+0.03}_{-0.03}$& 11.53$^{+0.01}_{-0.01}$& 11.30$^{+0.07}_{-0.08}$\\
  1& 2.42$^{+0.15}_{-0.14}$& 12.53$^{+0.03}_{-0.03}$& 320$^{+ 20}_{-18}$& 27.0$^{+0.7}_{-0.7}$& 9.16$^{+0.04}_{-0.03}$& 10.56$^{+0.10}_{-0.02}$& 11.58$^{+0.08}_{-0.10}$\\
  2& 3.05$^{+0.30}_{-0.28}$& 12.30$^{+0.08}_{-0.04}$& 191$^{+ 38}_{-17}$& 27.6$^{+2.0}_{-1.2}$& 8.95$^{+0.07}_{-0.10}$& 10.86$^{+0.14}_{-0.58}$& 11.38$^{+0.05}_{-0.06}$\\
  3& 1.543\tnote{e}& 12.06$^{+0.06}_{-0.05}$& 109$^{+ 16}_{-12}$& 22.8$^{+1.1}_{-1.0}$& 8.98$^{+0.05}_{-0.05}$& 10.59$^{+0.39}_{-0.01}$& 11.47$^{+0.03}_{-0.03}$\\
  4& 0.43$^{+0.04}_{-0.04}$& 11.01$^{+0.20}_{-0.17}$&   9$^{+  5}_{- 3}$& 20.9$^{+3.0}_{-2.3}$& 7.90$^{+0.11}_{-0.12}$& 10.58$^{+0.02}_{-0.01}$& 10.61$^{+0.06}_{-0.07}$\\
  5& 1.74$^{+0.22}_{-0.22}$& 12.20$^{+0.05}_{-0.04}$& 151$^{+ 18}_{-13}$& 25.0$^{+0.9}_{-0.9}$& 8.93$^{+0.04}_{-0.04}$& 11.32$^{+0.11}_{-0.52}$& 11.38$^{+0.03}_{-0.03}$\\
  6& 1.45$^{+0.15}_{-0.15}$& 11.88$^{+0.13}_{-0.11}$&  71$^{+ 24}_{-15}$& 28.7$^{+2.6}_{-2.1}$& 8.22$^{+0.08}_{-0.08}$& 10.95$^{+0.08}_{-0.19}$& 10.77$^{+0.07}_{-0.09}$\\
  7& 2.14$^{+0.38}_{-0.39}$& 12.07$^{+0.13}_{-0.07}$& 110$^{+ 39}_{-17}$& 32.7$^{+3.2}_{-2.0}$& 8.19$^{+0.08}_{-0.10}$& 10.61$^{+0.14}_{-0.73}$& 10.75$^{+0.07}_{-0.08}$\\
  8& 1.545\tnote{e}& 12.60$^{+0.05}_{-0.05}$& 381$^{+ 48}_{-41}$& 32.0$^{+1.3}_{-1.2}$& 8.69$^{+0.04}_{-0.04}$& 11.06$^{+0.07}_{-0.10}$& 11.35$^{+0.03}_{-0.03}$\\
  9& 2.21$^{+0.21}_{-0.22}$& 11.81$^{+0.12}_{-0.05}$&  61$^{+ 18}_{- 6}$& 23.4$^{+2.1}_{-1.1}$& 8.76$^{+0.07}_{-0.09}$& 10.60$^{+0.38}_{-0.02}$& 11.03$^{+0.12}_{-0.17}$\\
 10& 1.27$^{+0.11}_{-0.11}$& 12.24$^{+0.04}_{-0.04}$& 164$^{+ 15}_{-14}$& 27.2$^{+0.6}_{-0.7}$& 8.68$^{+0.03}_{-0.02}$& 11.03$^{+0.06}_{-0.02}$& 10.98$^{+0.13}_{-0.20}$\\
 11& 2.43$^{+0.29}_{-0.29}$& 12.31$^{+0.30}_{-0.26}$& 195$^{+193}_{-87}$& 30.7$^{+6.2}_{-4.7}$& 8.63$^{+0.15}_{-0.16}$& 11.26$^{+0.08}_{-0.18}$& 11.17$^{+0.04}_{-0.04}$\\
 12& 1.40$^{+0.10}_{-0.10}$& 12.12$^{+0.06}_{-0.05}$& 126$^{+ 18}_{-14}$& 28.0$^{+1.2}_{-1.1}$& 8.52$^{+0.04}_{-0.05}$& 11.09$^{+0.06}_{-0.18}$& 11.06$^{+0.04}_{-0.04}$\\
 13& 2.63$^{+0.25}_{-0.25}$& 12.34$^{+0.05}_{-0.03}$& 207$^{+ 27}_{-14}$& 27.0$^{+1.5}_{-1.2}$& 9.00$^{+0.07}_{-0.08}$& 10.46$^{+0.34}_{-0.05}$& 11.42$^{+0.07}_{-0.08}$\\
 14& 2.30$^{+0.15}_{-0.17}$& 12.43$^{+0.07}_{-0.06}$& 255$^{+ 43}_{-34}$& 33.5$^{+1.8}_{-1.7}$& 8.51$^{+0.06}_{-0.06}$& 10.23$^{+0.17}_{-0.84}$& 11.09$^{+0.05}_{-0.06}$\\
 15& 2.45$^{+0.37}_{-0.35}$& 11.68$^{+0.19}_{-0.20}$&  44$^{+ 24}_{-16}$& 24.0$^{+4.0}_{-3.7}$& 8.60$^{+0.20}_{-0.19}$& 11.21$^{+0.14}_{-0.63}$& 10.99$^{+0.07}_{-0.08}$\\
 16& 1.02$^{+0.18}_{-0.18}$& 10.71$^{+0.22}_{-0.21}$&   4$^{+  3}_{- 1}$& 14.4$^{+4.0}_{-2.9}$& 8.62$^{+0.30}_{-0.36}$& 10.67$^{+0.06}_{-0.25}$& 11.16$^{+0.13}_{-0.18}$\\
 17& 0.98$^{+0.08}_{-0.08}$& 12.02$^{+0.09}_{-0.08}$&  99$^{+ 22}_{-17}$& 28.6$^{+1.8}_{-1.6}$& 8.28$^{+0.06}_{-0.06}$& 10.89$^{+0.05}_{-0.05}$& 10.94$^{+0.05}_{-0.06}$\\
\hline
\end{tabular}
\begin{tablenotes}
{\small
\item[a] 8--1000\,$\mu$m (rest-frame) luminosity derived by fitting the FIR SED with a single-temperature modified black-body model.
\item[b] SFR derived from $L_{\rm IR}$, assuming the relationship in \citet{kennicutt98} modified for a Chabrier IMF, i.e. SFR[M$_\odot$\,yr$^{-1}$]$\,{=}\,9.5\,{\times}\,10^{-11}\,L_{\rm IR}$[L$_\odot$].
\item[c] Stellar mass derived from fitting the Pan-STARRS-WIRCam-IRAC SED with the models of \citet{bc03}.
\item[d] ISM mass derived from the ALMA 233-GHz flux density (Table~\ref{table:1}) and using Eq.~(\ref{equ2}) \citep{Scoville+2016}.
\item[e] The photometric redshifts for IDs~3 and 8 are 1.70$^{+0.23}_{-0.20}$ and 1.33$^{+0.07}_{-0.07}$, respectively; however, these best-fit parameters have been derived assuming the reported CO redshifts.
}
\end{tablenotes}
\end{threeparttable}
\end{table*}

\subsection{Relationship to main-sequence galaxies}\label{ms_analysis}

In Fig.~\ref{Mstar_z}, we compare the derived stellar masses with the expected values of the characteristic mass $\mathcal{M}_\ast$ obtained by fitting the Schechter mass function of SFGs in multiple redshift intervals between 0.2 and 4.0 \citep{Davidzon+17}. Nine ALMA galaxies have stellar masses comparable with the expected $\mathcal{M}_\ast$ values, the other nine galaxies are instead more massive than the expected $\mathcal{M}_\ast$ values, with stellar masses above $3\,{\times}\,10^{11}$\,M$_{\odot}$, implying that they have become quite massive early on. 

In  Fig.~\ref{starburstiness}, we show the location of our sources with respect to the MS relation. To accentuate the offset from the MS (i.e. the \lq{}starburstiness\rq{}) for our ALMA galaxies, we plot the IR-derived SFR normalised by the expected SFR based on the MS at each source redshift, as parameterised by \citet{speagle14}, as a function of redshift. The grey region corresponds to the scatter around the MS, which is about a factor of 3.  The majority (13 out of 18, or 72\,\%) of our ALMA galaxies lie within this factor of 3 of the main sequence, while two ALMA galaxies lie below this region (IDs~15 and 16) and three lie above it (IDs~1, 8, and 14). These latter three sources (where it must be noted that the SFRs derived for ID~14 may not be reliable) are thus experiencing enhanced star-forming activity, consistent with being SB galaxies.  It is indeed typically assumed that SB galaxies are offset by a factor of ${\simeq}$3--4 or more from the MS \citep{elbaz11,rodighiero11}.

From this analysis, we notice the following. Firstly, the sources in the redshift concentration around $z\,{\simeq}\,$1.5 (IDs~3, 5, 6, 8, and 12, shown as large stars in Fig.~\ref{starburstiness}) are mostly on the MS and more massive than the expected $\mathcal{M}_\ast$. Conversely, there is a group of three galaxies (IDs~1, 13, and 14) around $z\,{\simeq}\,$2.3--2.6, corresponding to the second most prominent redshift concentration (Fig.~\ref{redshift_dist}), with large starburstiness values (SFR$\,{\geq}\,3\,{\times}\,$SFR$_{\rm MS}$), but stellar masses consistent with or below the expected $\mathcal{M}_\ast$. These two redshift concentrations might be associated with two structures, one in the background, at $z\,{\simeq}\,$2.4, where galaxies are actively forming stars and are still building their stellar masses, and one in the foreground, at $z\,{\simeq}\,$1.5, where most galaxies have reached the end of their stellar mass build-up and their activity level is relatively low. Finally, we note that the most massive galaxies (IDs~0, 5, 11, and 15) are on the MS, or below it, as expected for objects close to the end of their active phase. Two of these galaxies are at $z\,{\simeq}\,$2.4 and might thus be in the same structure as IDs~1, 13, and 14. If true, then we would have two types of member of the $z\,{\simeq}\,$2.4 structure: one starbursting, with less-than-expected stellar mass (IDs~1, 13, and 14); and the other lying on or below the MS and with greater-than-expected stellar mass (IDs~11 and 15).

The total SFRs of these two structures are 840$^{+120}_{-100}$\,M$_{\odot}$\,yr$^{-1}$ and 1020$^{+310}_{-170}$\,M$_{\odot}$\,yr$^{-1}$ for the $z\,{\simeq}\,1.5$ and $z\,{\simeq}\,2.4$ structures, respectively, and the associated total stellar masses are (5.8$^{+1.7}_{-2.4})\,{\times}\,10^{11}$\,M$_{\odot}$ and (4.2$^{+1.5}_{-2.1})\,{\times}\,10^{11}$\,M$_{\odot}$. These numbers yield starburstiness values of 2.4, and 1.9, respectively, thus consistent with the MS at their redshifts.

\begin{figure} 
\centering
\includegraphics[width=\linewidth]{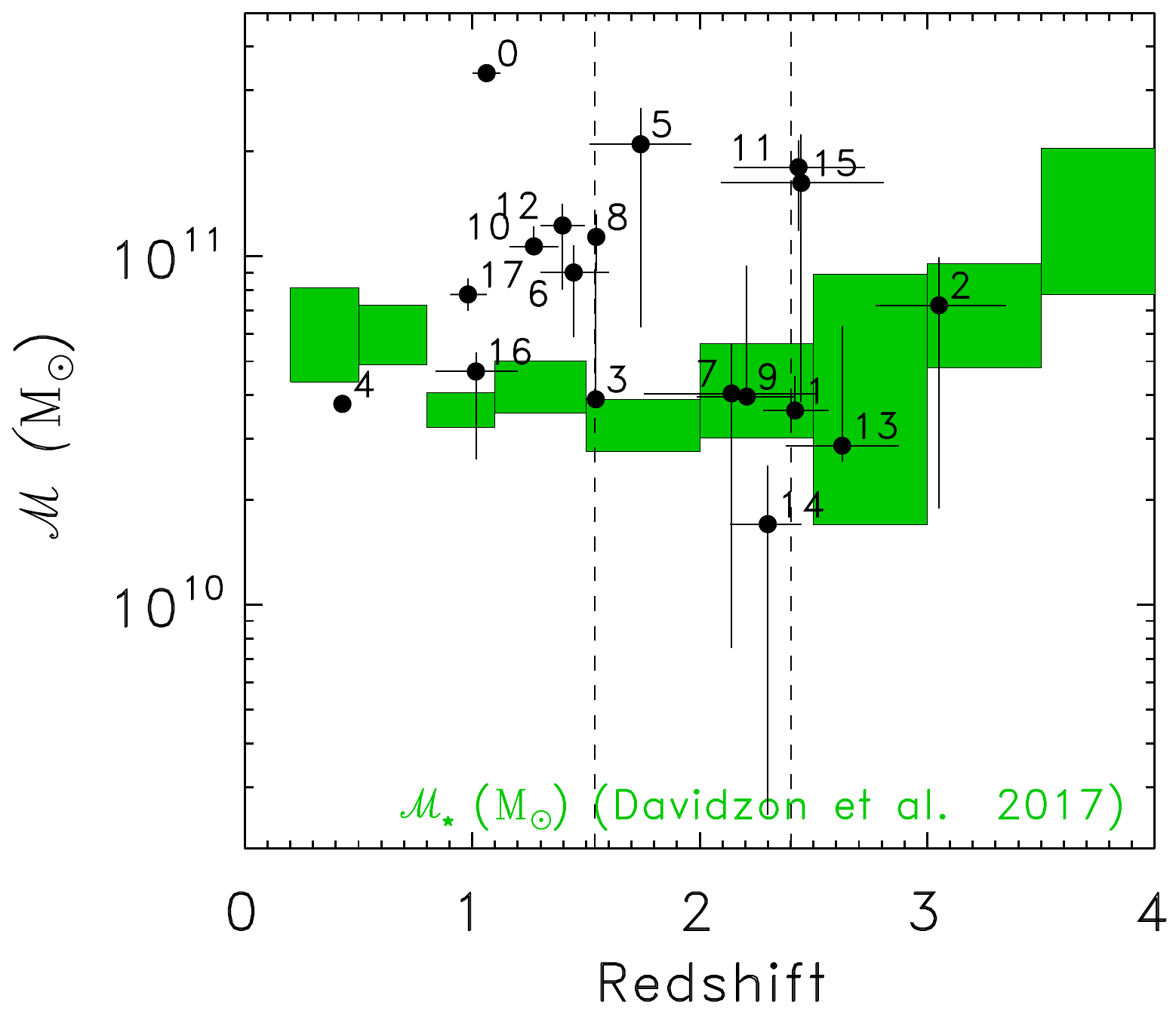}
\caption{Stellar mass as a function of photometric redshift 
for the 18 ALMA galaxies. The green rectangles represent the expected values of the characteristic mass $\mathcal{M}_\ast$ and their uncertainties obtained by fitting the mass function of SFGs with a Schechter function in the redshift ranges 0.2--0.5, 0.5--0.8, 0.8--1.1, 1.1--1.5, 1.5--2.0, 2.0--2.5, 2.5--3.0, 3.0--3.5, and 3.5--4.0 \citep{Davidzon+17}. The source IDs are labelled next to the corresponding symbols.}
\label{Mstar_z} 
\end{figure}

\begin{figure} \centering 
\includegraphics[width=\linewidth]{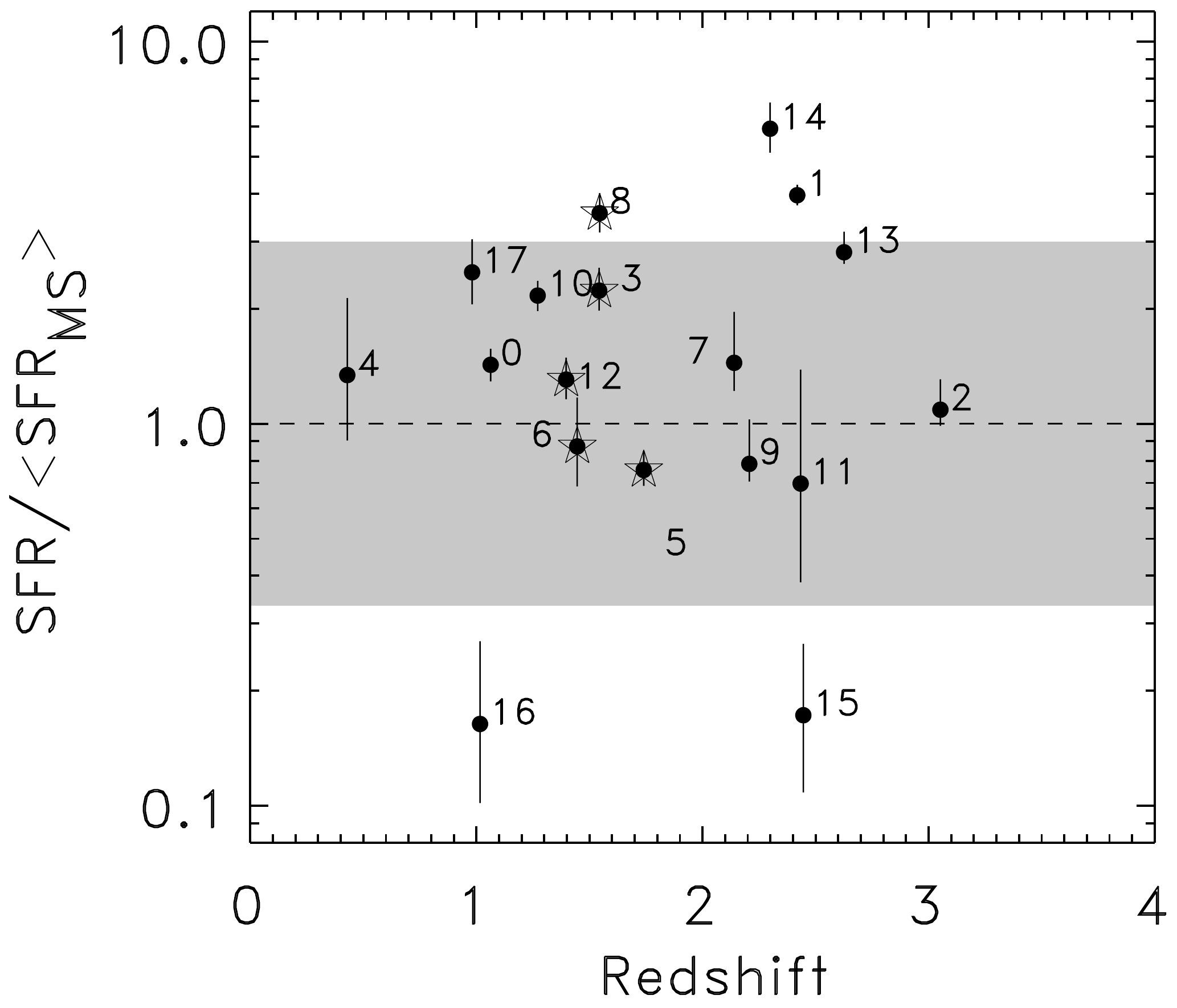}
\caption{\lq{}Starburstiness\rq{}, which is the ratio of the star-formation rate to the SFR expected for a source on the MS (using the relation at the respective redshift), plotted against redshift. A factor of 3 around the MS is indicated by the grey region. Stars highlight galaxies assumed to be at $z\,{\simeq}\,1.54$.}
\label{starburstiness} 
\end{figure}

\section{Serendipitous line detections}\label{line_detection}

\subsection{ALMA galaxies ID~3 and 8}

Spectral cubes of the ALMA primary-beam-convolved continuum (128 channels for each of the four 2-GHz wide spectral windows) were made for the eight fields, with a spectral binning of width 0.08\,GHz, giving 25 frames, for the line search. The spectra were analysed in the local standard of rest (kinematic, i.e. LSRK) with 64 frames per spectral window. Fluxes quoted in the text are beam corrected. 

ALMA galaxy ID~3, the brightest mm galaxy located in ALMA field 2, shows the detection of a strong line at (226.656\,${\pm}$\,0.009)\,GHz (line peak in Fig.~\ref{Fig6a}, top panel). We find an integrated flux density of $(2.5\,{\pm}\,0.2)\,{\rm Jy}\,{\rm km}\,{\rm s}^{-1}\,{\rm beam}^{-1}$ at the spatial peak, and $(2.9\,{\pm}\,0.2)\,{\rm Jy}\,{\rm km}\,{\rm s}^{-1}$ in an extended aperture (Fig.~\ref{Fig6b}), with a line width of $(417\,{\pm}\,31)\,{\rm km}\,{\rm s}^{-1}$ for the FWHM in the Gaussian fit (Table~\ref{table:5}). Using the physical size of this source as 0.44$\arcsec$, the semi-major axis from Table~\ref{table:1}, the dynamical mass can be estimated as $M_{\rm dyn}\,{=}\,(417\,{\rm km}\,{\rm s}^{-1})^{2}\,{\times}\,3.8\,{\rm kpc}\,/\,G$ ${=}\,1.5\,{\times}\,10^{11}\,{\rm M}_\odot$, as compared to a stellar mass of $\mathcal{M}\,=\,3.9^{+5.7}_{-0.1}\,{\times}\,10^{10}$\,M$_\odot$ (from Table~\ref{bestfit_params}).

\begin{figure}[htbp!]
	\includegraphics[width=8.92cm]{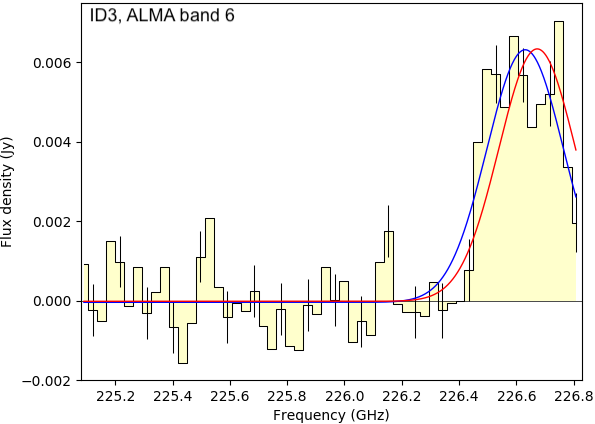}\\
	\includegraphics[width=9.0cm]{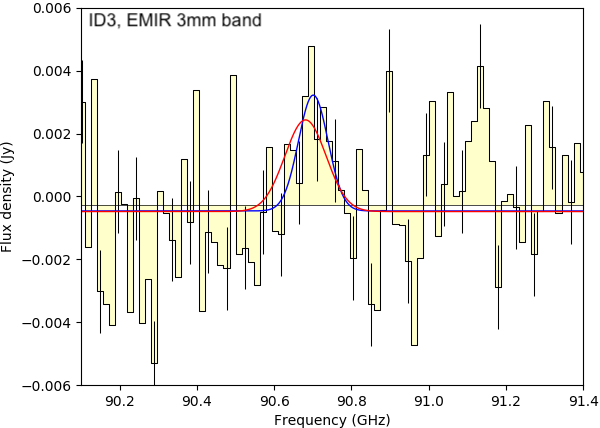}\\
	\includegraphics[width=8.85cm]{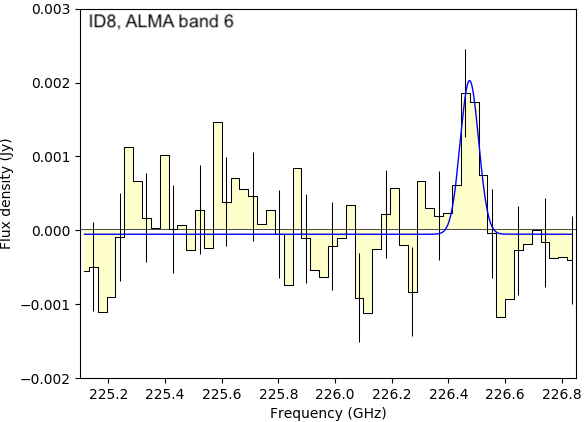}\\
	\caption{Spectra of the two ALMA galaxies ID~3 (top and middle) and ID~8 (bottom), showing the serendipitous line detections, consistent with a CO(5--4) transition at $z\,{=}\,1.54$. The blue Gaussian profiles show the best fits to each individual line. The red Gaussian profiles for ID~3 show the best combined fit to the CO(5--4) line in the ALMA spectrum (top) and the CO(2--1) line in the IRAM/EMIR spectrum (middle). The offset between the fitted line centres seen in the EMIR data is small, 66\,km\,s$^{-1}$, and could be due to the low S/N, the edge of the ALMA spectral window, or a physical difference between the transitions. Representative error bars per bin are shown for every third bin. We note that we have applied the standard flagging of edge channels in the ALMA spectral window for ID~3, which otherwise could introduce systematic uncertainties.} 
     \label{Fig6a}
\end{figure}

\begin{figure*}[htbp!]
	\includegraphics[width=5.7cm]{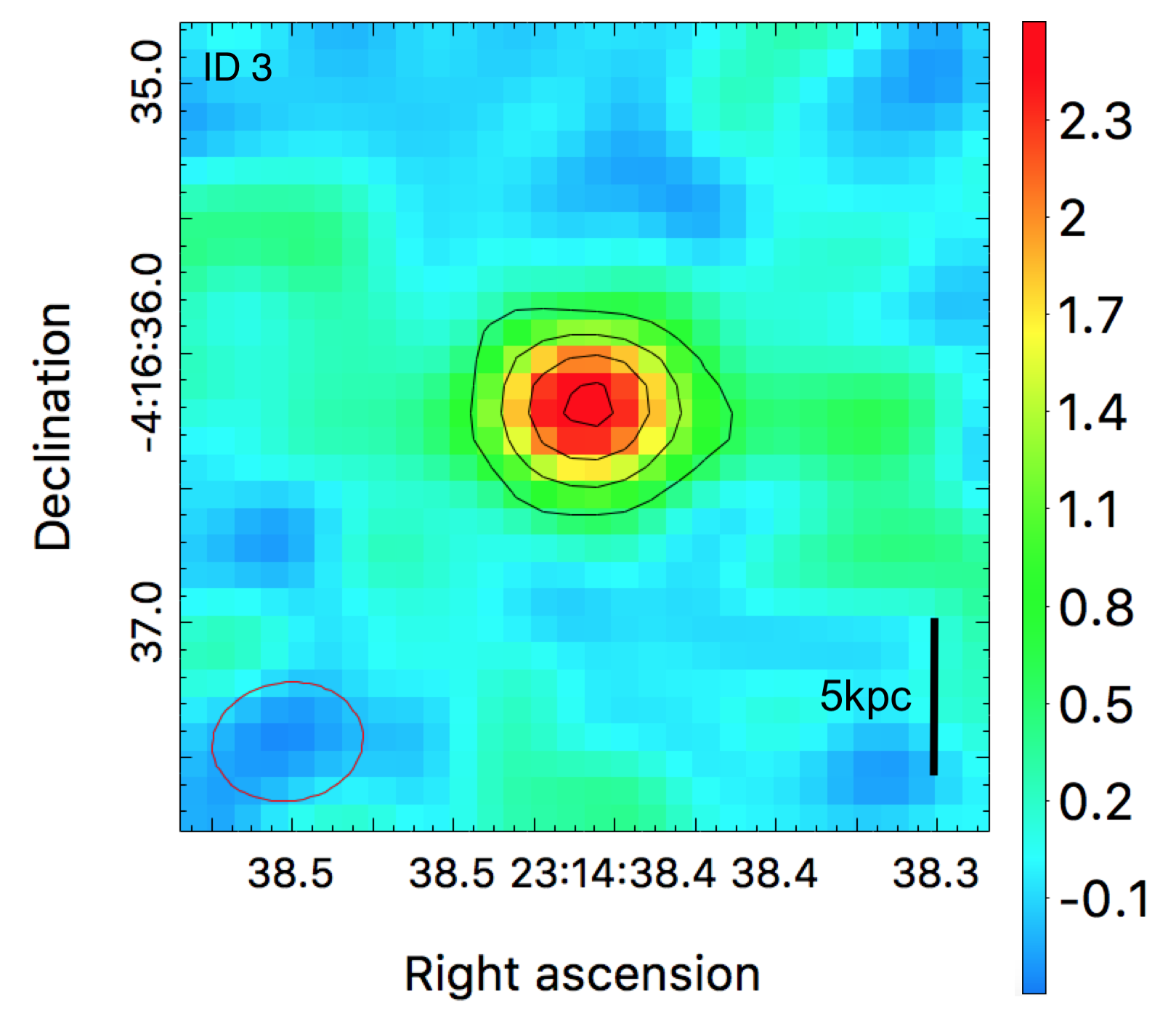}
	\hspace{0.1cm}
	\includegraphics[width=5.7cm]{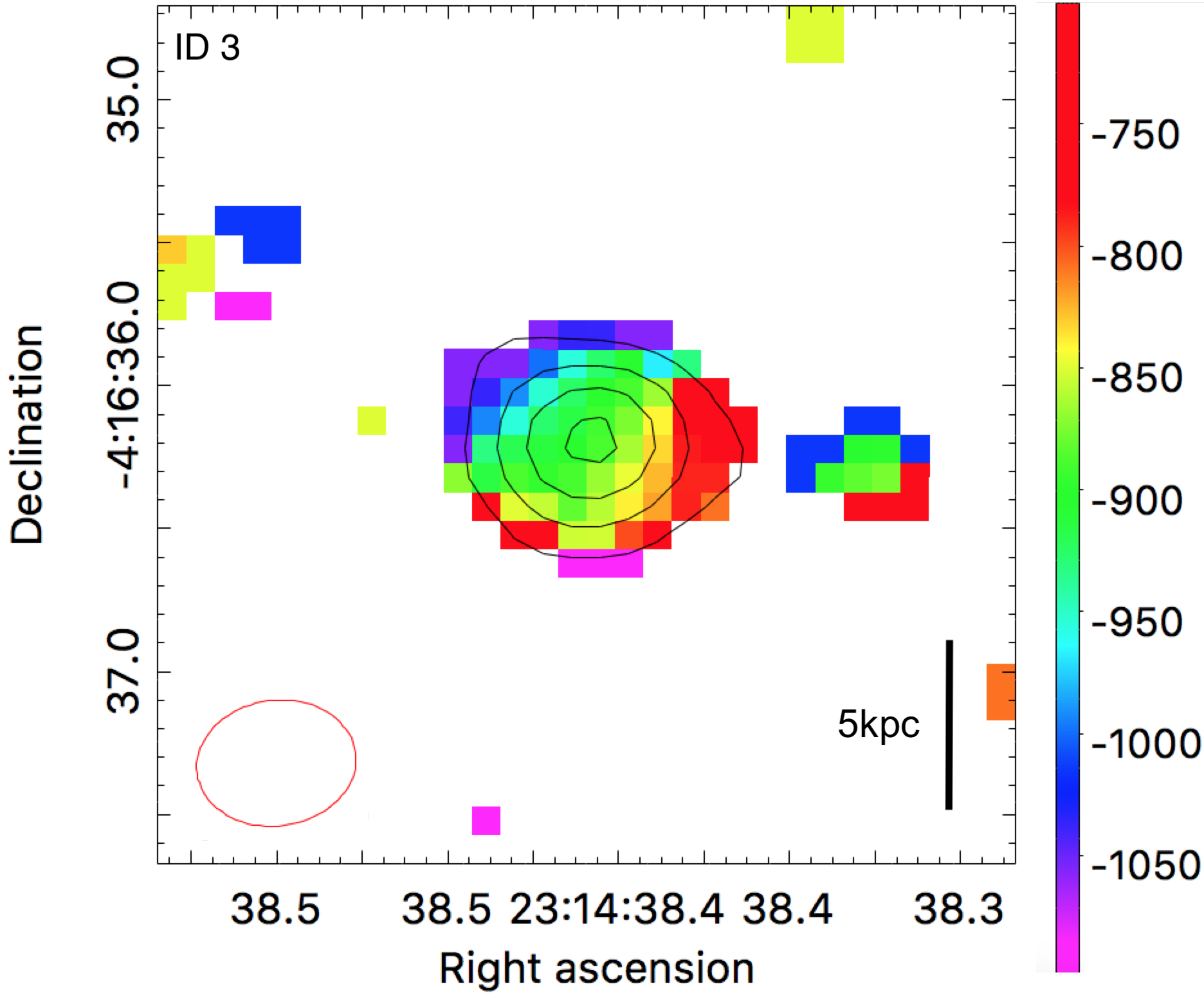}
	\hspace{0.1cm}
	\includegraphics[width=6.4cm]{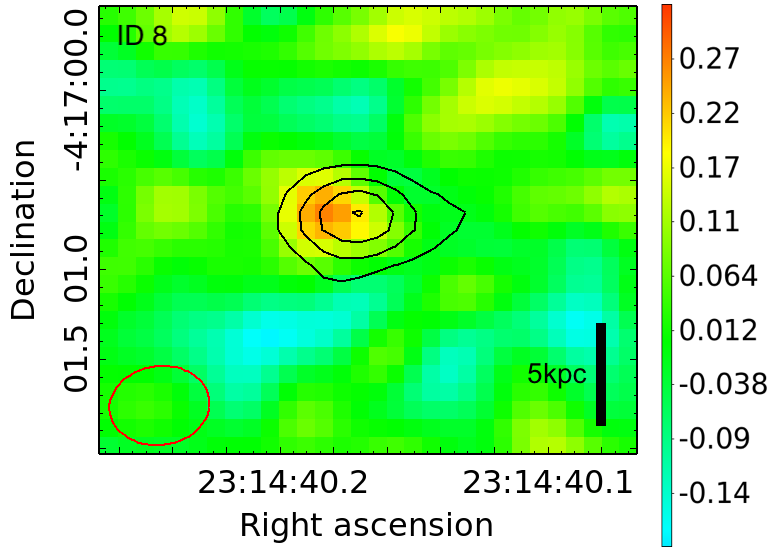}
	\caption{Images for ID~3 (on the left) and ID~8 (on the right) of the integrated line emission in Jy\,km\,s$^{-1}$\,beam$^{-1}$ (where the continuum has been subtracted). In both cases line and continuum emission (i.e. the black contours from 3$\,\sigma\,{=}\,$0.18\,mJy in 3$\,\sigma$ steps) coincide. The middle panel shows (for the stronger line of ID~3 only) the first-moment image in km\,s$^{-1}$, along with continuum contours for reference. The FWHM of the synthesised beam ($0.56\arcsec\,{\times}\,0.44\arcsec$) is shown with red ellipses and a 5-kpc bar is shown in black for reference.} 
     \label{Fig6b}
\end{figure*}

The galaxy shows a smooth velocity gradient from north-east to south-west (Fig.~\ref{Fig6b}, middle panel), but it is only barely resolved spatially. CO transitions are known to be bright for mm and submm galaxies \citep[e.g.][]{CarilliWalter2013,Vieira+2013}, and would correspond to the redshifts $z\,{=}\,$1.034 CO(4--3), 1.542 for CO(5--4), 2.051 for CO(6--5), and 2.559 for CO(7--6), if we keep with the most plausible range of $z\,{\simeq}\,$1--3. 

Associating the observed line with the CO(5--4) transition appears the most plausible conclusion, since it provides the closest match for the photometric redshift of ALMA galaxy ID~3. However, we now briefly discuss other interpretations. The higher redshift transitions ($J\,{>}\,6$, corresponding to $z\,{>}\,2.5$) would yield poorer agreement with the photometric redshifts, and in addition may be expected to be much weaker. The \ion{C}{I}(2--1) line would provide a direct identification, as its rest frequency of 809.34\,GHz is very close to the rest frequency of the CO(7--6) line, which has a rest frequency of 806.65\,GHz; however, this is not possible with our observation, since the expected 227.4-GHz (sky frequency) line would lie inside the sideband separation. Moreover, a redshift around $z\,{\simeq}\,1$ does not seem consistent with the colour and photo-$z$ results of most of our galaxies (apart from those identified as interlopers). On the other hand, CO(6--5) appears possible, though not favoured by the photometric redshifts within their errors. 

\subsection{IRAM-30m/EMIR CO redshift}

Observations of G073.4$-$57.5 were carried out using the heterodyne receiver EMIR \citep{Carter+12} on the IRAM 30-m telescope between 13 and 16 September 2016 and (PID 077-16, PI C.~Martinache). We used the 3-mm band (E090) to search for CO transitions. The frequencies covered were 74--82\,GHz and 90--98\,GHz. For the backends, we simultaneously used the wideband line multiple autocorrelator (WILMA, 2-MHz spectral resolution) and the fast Fourier transform spectrometer (FTS200, 200-kHz resolution). Given that the observed object, SPIRE source 3 (ALMA field 2, see Table~\ref{table:2}), is a point source, observations were performed in wobbler-switching mode with a throw of 30\arcsec. The FWHM of IRAM 30-m/EMIR is 27\arcsec\ at 91\,GHz, comparable to the {\it Herschel\/}-SPIRE beam at 350\,$\mu$m (25\arcsec). The total integration time was 300\,min. For calibration, pointing, and focusing we used Jupiter, Mars, and bright quasars. Data reduction was performed with the help of the {\tt CLASS} package in {\tt GILDAS} \citep{gildas13}. Baseline-removed spectra were co-added using the inverses of the squares of the individual noise levels as weights. We then fit the co-added spectra with a Gaussian profile and derived the line position, the peak flux, and the line width (FHWM).  The results are presented in Table~\ref{table:5}. 

In Fig.~\ref{Fig6a}, middle panel, we show the EMIR spectrum, together with the best-fit Gaussian curves for the EMIR data and the combined EMIR and ALMA data. We note a significant (4.7$\,\sigma$) detection very close (90\,km\,s$^{-1}$ separation) to the expected frequency of 90.674\,GHz for the CO(2--1) transition. We take the EMIR spectrum and the joint fit result as a strong indication for a CO(5--4) line in ALMA and a redshift of $z\,{=}\,1.5423\pm0.0001$ (Table~\ref{table:5}); this is dominated by the high S/N ratio in the ALMA data (fitted Gaussian curves in Fig.~\ref{Fig6a}, top panel). We assume, of course, that the EMIR line comes from ALMA ID~3 and not from another galaxy within the larger beam, and also not from another molecular species, since either of these options would be a rather unlikely coincidence. 

\begin{table*}[htbp!]
\caption{Spectral fitting results for ALMA galaxies IDs~3 and 8}              
\label{table:5}      
\centering                                      
\begin{tabular}{llccccc}          
\hline\hline                        
\noalign{\vskip 3pt}
ALMA& Data& $S_{\rm peak}$& Line width& Redshift& Offset& Assumed\\
ID& & [mJy]& FWHM [${\rm km}\,{\rm s}^{-1}$]& & [mJy]& transition\\
\hline                                   
\noalign{\vskip 3pt}
3& ALMA& $6.4\pm0.4$& $417\pm31$& $1.54248\pm0.00010$& $-0.04\pm0.10$& CO(5--4)\\
3& EMIR& $3.7\pm0.8$& $296\pm74$& $1.54173\pm0.00010$& $-0.46\pm0.12$& CO(2--1)\\
3& ALMA joint& $6.4\pm0.3$& $416\pm28$& $1.54229\pm0.00009$& $-0.08\pm0.08$& CO(5--4)\\
3& EMIR joint& $2.9\pm0.3$& $416\pm28$& $1.54229\pm0.00009$& $-0.47\pm0.08$& CO(2--1)\\
8& ALMA& $2.0\pm0.5$& $101\pm31$& $1.54452\pm0.00004$& $-0.05\pm0.08$& CO(5--4)\\
\hline                                             
\end{tabular}
\end{table*}

\subsection{CO line properties}\label{CO_analysis}

Under the assumption that the detected line in ALMA is indeed CO(5--4), the CO luminosity can be calculated as \citep{Solomon+1997}
\begin{equation}
L'_{\rm CO} = \frac{c^2}{2 k} S_{\rm CO} \left( \Delta V\right) \nu^{-2}_{\rm sky} D^2_{\rm L} (1+z)^{-3}.
\end{equation}
\noindent
Using the linewidth estimate and peak intensity from the joint fit (Table~\ref{table:5}), we find $L'_{\rm CO}\,{=}\,(1.5\,{\pm}\,0.1)\,{\times}\,10^{10}$\,K\,km\,s$^{-1}$\,pc$^{2}$, consistent with the integrated line flux density of Fig.~\ref{Fig6b}. The CO(2--1) luminosity for the EMIR line, also using the joint fit results, is $L'_{\rm CO}\,{=}\,(4.2\,{\pm}\,0.4)\,{\times}\,10^{10}$\,K\,km\,s$^{-1}$\,pc$^{2}$, giving a ratio of $r_{54/21}\,{=}\,0.36$ relative to the CO(5--4) transition luminosity, consistent with the values measured for typical SMGs \citep[e.g.][]{CarilliWalter2013}.
We also find tentative evidence for a faint line (S/N$\,{\simeq}\,4.4$ over four channels with two-channel Hanning smoothing) in ALMA galaxy ID~8 (the brightest detection in ALMA field 4), which has very similar NIR properties to those of ALMA galaxy ID~3 (Figs.~\ref{Fig3} and~\ref{Fig5}). There is a (spatially unresolved) peak of intensity $(0.274\,{\pm}\,0.062)\,{\rm Jy}\,{\rm km}\,{\rm s}^{-1}$ at (226.474\,${\pm}$\,0.004)\,GHz in Fig.~\ref{Fig6b}. In the Gaussian fit the linewidth is (101\,${\pm}\,31)\,{\rm km}\,{\rm s}^{-1}$ (Table~\ref{table:5}), and the redshift is $z\,{=}\,1.54452\,{\pm}\,0.00004$ for the same CO(5--4) transition. Using the parameters of the fit the line luminosity is $L'_{\rm CO}\,{=}\,(1.1\,{\pm}\,0.4)\,{\times}\,10^{9}\,{\rm K}\,{\rm km}\,{\rm s}^{-1}\,{\rm pc}^2$. The dynamical mass estimate is $M_{\rm dyn}\,{=}\,9.2\,{\times}\,10^{9}$\,M$_\odot$, that is much smaller than the expected stellar mass of $\mathcal{M}\,{=}\,1.1\,{\times}\,10^{11}$\,M$_\odot$ (from Table~\ref{bestfit_params}). The near coincidence of the frequency with that of ID~3 argues for the reality of this weaker line. 

For these two galaxies, with the simple assumptions that $L'_{\rm
CO(1{-}0)}\,{=}\,L'_{\rm CO(5{-}4)}\,{/}\,r_{54/10}$, with
$r_{54/10}\,{=}\,0.32\,{\pm}\,0.05$ \citep[the median brightness temperature
ratio derived for SMGs by][]{bothwell13}, we derive gas masses of
$(4.7\,{\pm}\,0.8)\,\alpha_{\rm CO}\,{\times}\,10^{10}\,{\rm M}_\odot$
(ID~3) and $(3.5\,{\pm}\,1.5)\,\alpha_{\rm CO}\,{\times}\,10^{9}\,{\rm
M}_\odot$ (ID~8).  Assuming $\alpha_{\rm CO}\,{=}\,4.36\,{\rm
M}_\odot\,/\,({\rm K}\,{\rm km}\,{\rm s}^{-1}\,{\rm pc}^2$), more typical of an
MS galaxy~\citep{Bolatto+2013}, $M_{\rm
gas}\,{=}\,(2.0\,{\pm}\,0.4)\,{\times}\,10^{11}\,{\rm M}_\odot$ (ID~3), and
$(1.5\,{\pm}\,0.6)\,{\times}\,10^{10}\,{\rm M}_\odot$ (ID~8); on the other
hand, $\alpha_{\rm CO}\,{=}\,0.8\,{\rm M}_\odot\,/\,({\rm K}\,{\rm km}\,{\rm
s}^{-1}\,{\rm pc}^2$), more typical for SB galaxies \citep{Solomon+1997}, yields
$(3.7\,{\pm}\,0.7)\,{\times}\,10^{10}\,{\rm M}_\odot$ (ID~3), and
$(2.8\,{\pm}\,1.2)\,{\times}\,10^{9}\,{\rm M}_\odot$ (ID~8). However, the large 
difference in the assumed conversion factors also indicates that there is a range 
of uncertainty.

The gas content and star-formation efficiency (SFE) of galaxies
in high-redshift overdensities are of great interest, since they allow us to
constrain the mechanisms that trigger, regulate, and quench their
star-formation activity.  In addition, a comparison with galaxies in the
field, in clusters, and in other proto-clusters can provide insights into
the role played by the environment.  To this end, we collected CO data from
the literature for galaxies in clusters \citep[62 galaxies in 11
clusters,][]{wang18,rudnick17,stach17,aravena12,casasola13,castignani18,coogan18,hayashi17,webb17}
and in proto-clusters \citep[16 galaxies in four 
proto-clusters,][]{dannerbauer17,ivison13,tadaki14,lee17}, and of SMGs and
AGN in the $z\,{=}\,$1--3 redshift range \citep[31 SMGs, 15 AGN, and 38
obscured AGN,][]{perna18}.  In order to represent normal star-forming
galaxies and SB galaxies, we used the SFR--$\mathcal{M}$
relation\footnote{The relationship assumed here for MS galaxies is the
following:
\vspace{-0.3cm}
\begin{align}
\log({\rm SFR}_{\rm MS})&= \left[\,(0.84\pm 0.02) - (0.026\pm 0.003) \times t(z)\,\right] \times \log(\mathcal{M})
\nonumber\\
&-\left[\,(6.51\pm0.24) - (0.11\pm0.03) \times t(z)\,\right],
\label{eq_ms}
\end{align}\\
\noindent
\vspace{-1.0cm}\\
where $t(z)$ is the age of the Universe in Gyr at redshift $z$,
$\mathcal{M}$ is the stellar mass in M$_{\odot}$, and SFR$_{\rm MS}$ is the
star-formation rate in M$_{\odot}$\,yr$^{-1}$.} 
from~\citet{speagle14}, and the $L'_{\rm CO}\,{-}\,L_{\rm IR}$
relation from MS and SB galaxies as derived by~\citet{sargent14}
\footnote{The predicted CO(1--0) line luminosity scales with to the
total IR luminosity as:
\vspace{-0.3cm}
\noindent
\begin{equation}
\log(L'_{{\rm CO(1-0)}}) = (0.54\pm0.02) + (0.81\pm0.03) \times \log(L_{\rm IR}),
\label{eq_co_lir_ms}
\end{equation}\\ 
\vspace{-0.95cm}\\
\noindent 
for MS galaxies, while for SB galaxies\\
\vspace{-1cm}\\
\begin{equation} 
\log(L'_{\rm CO(1-0)}) = (0.08^{+0.15}_{-0.08}) + (0.81\pm0.03) \times \log(L_{\rm IR}),
\label{eq_co_lir_sb}
\end{equation}\\
\noindent
\vspace{-0.95cm}\\
with $L_{\rm IR}$ in L$_{\odot}$, and $L'_{\rm CO(1-0)}$ in
K\,km\,s$^{-1}$\,pc$^{2}$.}

For estimating molecular gas masses, we converted all high transition
CO luminosities to $L'_{\rm CO(1-0)}$, when not available, using the
median brightness temperature ratios derived by \citet{bothwell13}.  In
order to investigate the gas fractions, we considered only the cluster and
proto-cluster galaxies for which a stellar-mass estimate was available (14
clusters galaxies and 14 proto-cluster galaxies).  Extending the molecular
gas analysis to all of our ALMA-detected galaxies, we also considered ISM
masses estimated from the mm continuum at 233\,GHz by applying
Eq.~(\ref{equ2}) \citep[][see Table~\ref{bestfit_params}]{Scoville+2016}. 
This method, based on the continuum level of the Rayleigh-Jeans tail
associated with the ISM thermal emission, is limited to galaxies within a
certain redshift and mass range and for limited dust temperatures, but is
less affected by the CO kinematics, clumpiness, and metallicity that affects
the CO excitation level \citep{Scoville+2014,Scoville+2016}.  In the
following analysis, we consider gas masses derived from the CO line for
ALMA IDs~3 and 8, and ISM masses derived from the ALMA continuum for all
of the ALMA sources.

In Fig.~\ref{sfe_lir}, we show the SFE, defined as the ratio between
$L_{\rm IR}$ (which is proportional to the SFR) and $L'_{\rm CO(1-0)}$
(which is proportional to $M_{\rm gas}$), as a function of $L_{\rm IR}$ for
all of the examples from the literature, as well as ALMA IDs~3 and 8.  We
also show the expected SFE for main sequence and starburst galaxies
as derived from Eqs.~(\ref{eq_co_lir_ms}) and (\ref{eq_co_lir_sb}).
The SFEs of field SMGs, AGN, obscured AGN, and proto-cluster galaxies are
consistent with either the MS or the SB relation.  Cluster galaxies exhibit,
on average, smaller IR luminosities than the other sub-samples, and their
SFEs cover a wider range, from 0.8\,dex lower to 0.9\,dex higher than the
expected MS values.  The two galaxies detected by ALMA in CO show different
behaviour: the SFE of ID~3 is 0.36\,dex lower than expected according to the
MS relation; but, on the other hand, the SFE of ID~8 is among the highest
observed, 0.65\,dex higher than the SB-expected value.

SFE is inversely proportional to the gas depletion time, modulo a normalisation factor that depends on the value of $\alpha_{\rm CO}$, that is $\tau_{\rm depl}$\,=\,$M_{\rm gas}\,{/}\,$SFR\,=\,1.05$\,{\times}\,$10$^{10} \alpha_{\rm CO} L'_{\rm CO(1-0)}\,{/}\,L_{\rm IR}$. In this paper, we prefer not to discuss the gas depletion times because this definition does not take into account the variety of processes (gas accretion and removal) that might be relevant in overdense environments, and thus the estimated values might be misleading. We instead examine the SFE as defined here and compare it with the values reported in the literature. As a reference for guidance, we note that SFE values consistent or above the expected SB values (Fig.~\ref{dsfe_sdfr}) would imply fast depletion times, consistent with bursty star-formation activity ($\tau_{\rm depl}\,{<}\,100\,$Myr), while SFE values consistent with the MS are longer and consistent with secular evolution (between 0.5 and 1.5\,Gyr).

\begin{figure} 
\centering
\includegraphics[width=\linewidth]{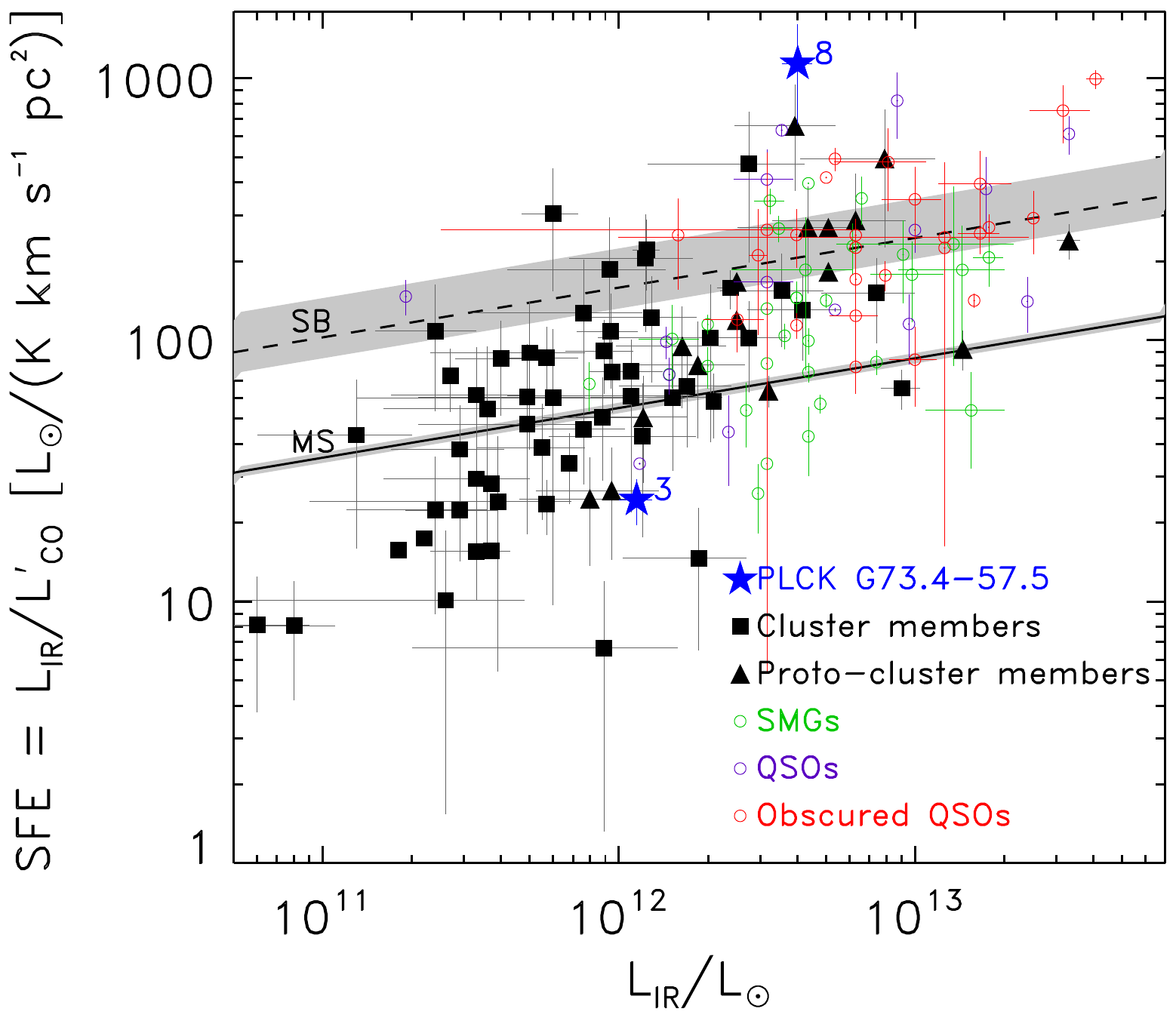}
\caption{SFE ($\equiv$\,$L_{\rm IR}\,{/}\,L'_{\rm CO}$) as a function of $L_{\rm IR}$ for ALMA IDs~3 and 8 (blue filled stars), and for sources from the literature with $1\,{<}\,z\,{<}\,3$ (black filled squares: cluster galaxies, black filled triangles: proto-cluster galaxies, green open circles: SMGs, purple open circles: AGN, and red open circles: obscured AGN; see Sect.~\ref{CO_analysis}). The solid and dashed lines represent, respectively, the average relations for MS and SB galaxies derived by \citet{sargent14} and given in Eqs.~(\ref{eq_co_lir_ms}) and (\ref{eq_co_lir_sb}).}
\label{sfe_lir}
\end{figure} 

\begin{figure} 
\centering
\includegraphics[width=\linewidth]{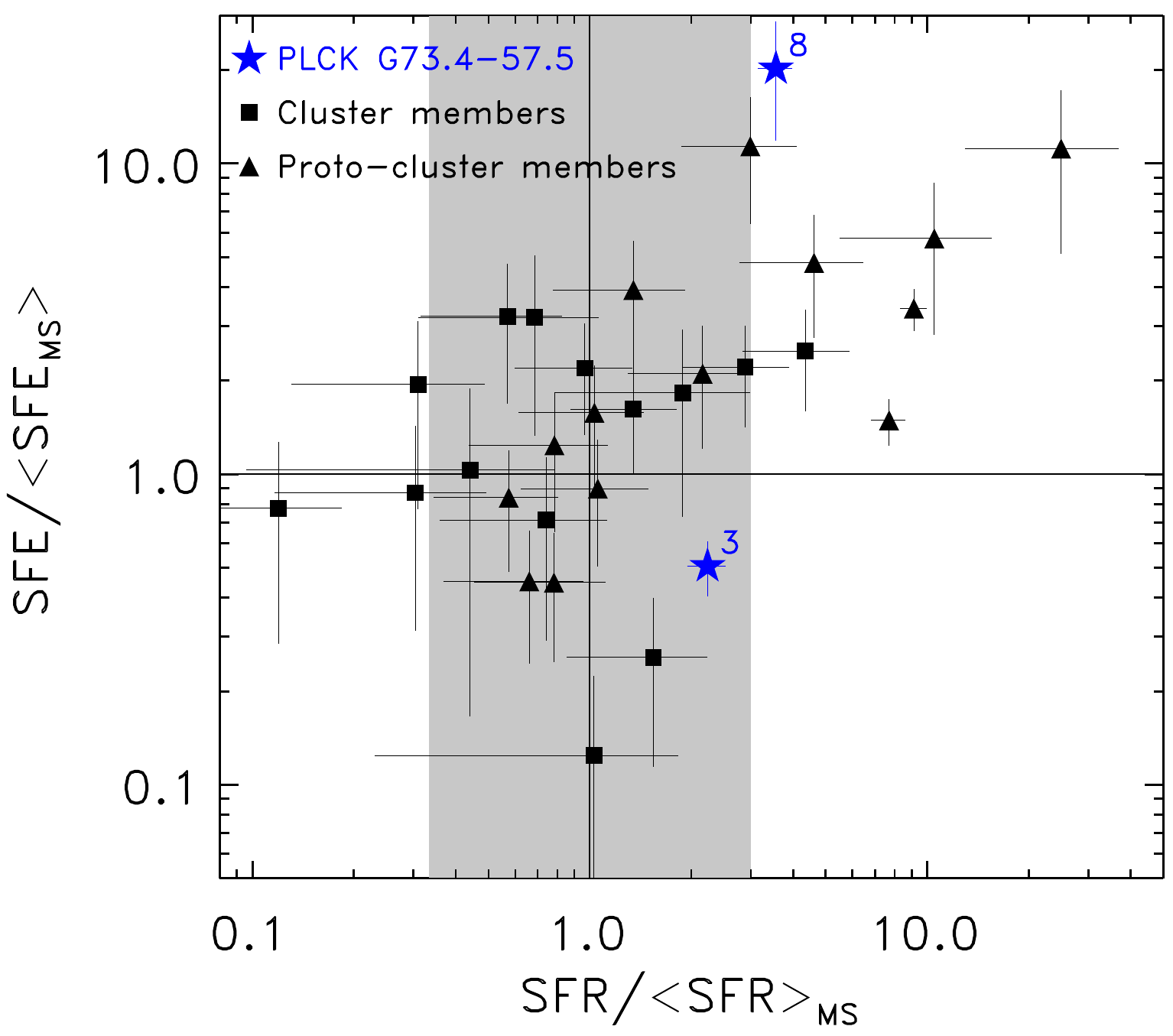}
\caption{SFE normalised to the expected SFE assuming the relation valid for
MS galaxies (Eq.~\ref{eq_co_lir_ms}) as a function of offset from the MS
(starburstiness) for ALMA IDs~3 and 8 (filled stars), and for sources
from the literature with $1\,{<}\,z\,{<}\,3$ (filled squares: cluster
galaxies, filled triangles: proto-cluster galaxies cluster galaxies). 
The horizontal and vertical solid lines represent, respectively, SFEs
and starburstiness values of MS galaxies, and the grey region indicates SFRs
consistent with the MS within a factor of 3.}
\label{dsfe_sdfr} 
\end{figure} 

The wide range of SFEs in cluster galaxies could be due to processes
that favour star formation, such as gas accretion and cooling, or that
hamper it through gas removal or heating.  Molecular gas studies of
high-redshift clusters show a significant suppression of molecular gas for
all the massive cluster galaxies close to the centre (within the core
radius).  This indicates that the environment plays a role in stopping gas
accretion and/or reducing/removing gas content
\citep{hayashi17,wang18,foltz18,socolovsky18,2019A&A...623A..48C}. 
On the other hand, ALMA observations of the proto-cluster 4C23.56 at
$z\,{=}\,$2.49 \citep[see][]{lee17} suggest gas masses and fractions of its
members consistent with those of field galaxies, implying a higher gas
density in the proto-clusters than in the field by an order of magnitude,
due to the overdensity.

In the rest of this section, we investigate whether this is also true for
the ALMA-detected galaxies by comparing their molecular gas content with
the expected values for MS and SB galaxies \citep[see ][]{sargent14}, and
with those observed in other cluster and proto-cluster members for which
both CO and stellar masses are available.

\begin{figure} \centering
\includegraphics[width=\linewidth]{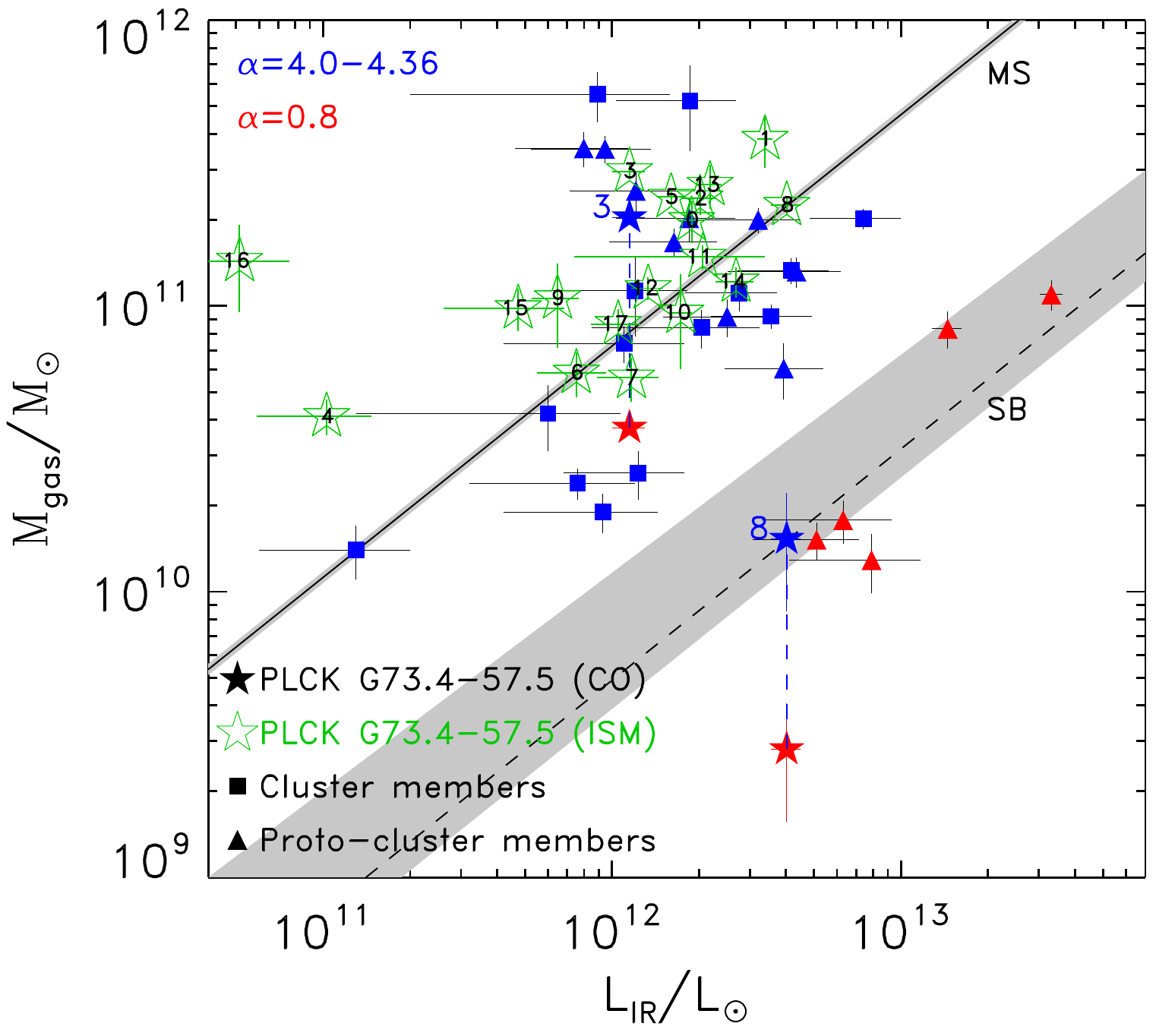}
\caption{Molecular gas mass, derived from the CO luminosity (in red if
$\alpha_{\rm CO}\,{=}\,$0.8, and in blue if $\alpha_{\rm CO}\,{=}\,$4.0--4.36), 
as a function of $L_{\rm IR}$ for ALMA IDs~3 and 8 (filled stars)
and galaxies in clusters (filled squares) or in proto-clusters (filled
triangles).  Gas masses refer to ISM masses derived from the mm continuum
for all ALMA sources (green open stars with annotated IDs).  The solid and
dashed lines represent the expected relations for MS and SB galaxies,
respectively, as derived from Eqs.~(\ref{eq_co_lir_ms}) and
(\ref{eq_co_lir_sb}), and assuming $\alpha_{\rm CO}$\,=\,4.36 (for MS
galaxies) or 0.8 (for SB galaxies, ${\rm SFR}\,{>}\,3\times{\rm SFR}_{\rm MS}$ for our sample)~\citep{sargent14}.  The grey areas represent 1$\sigma$ uncertainties in the
theoretical parameters of the relations.}
\label{mgas_lir} 
\end{figure}

\begin{figure} \centering
\includegraphics[width=\linewidth]{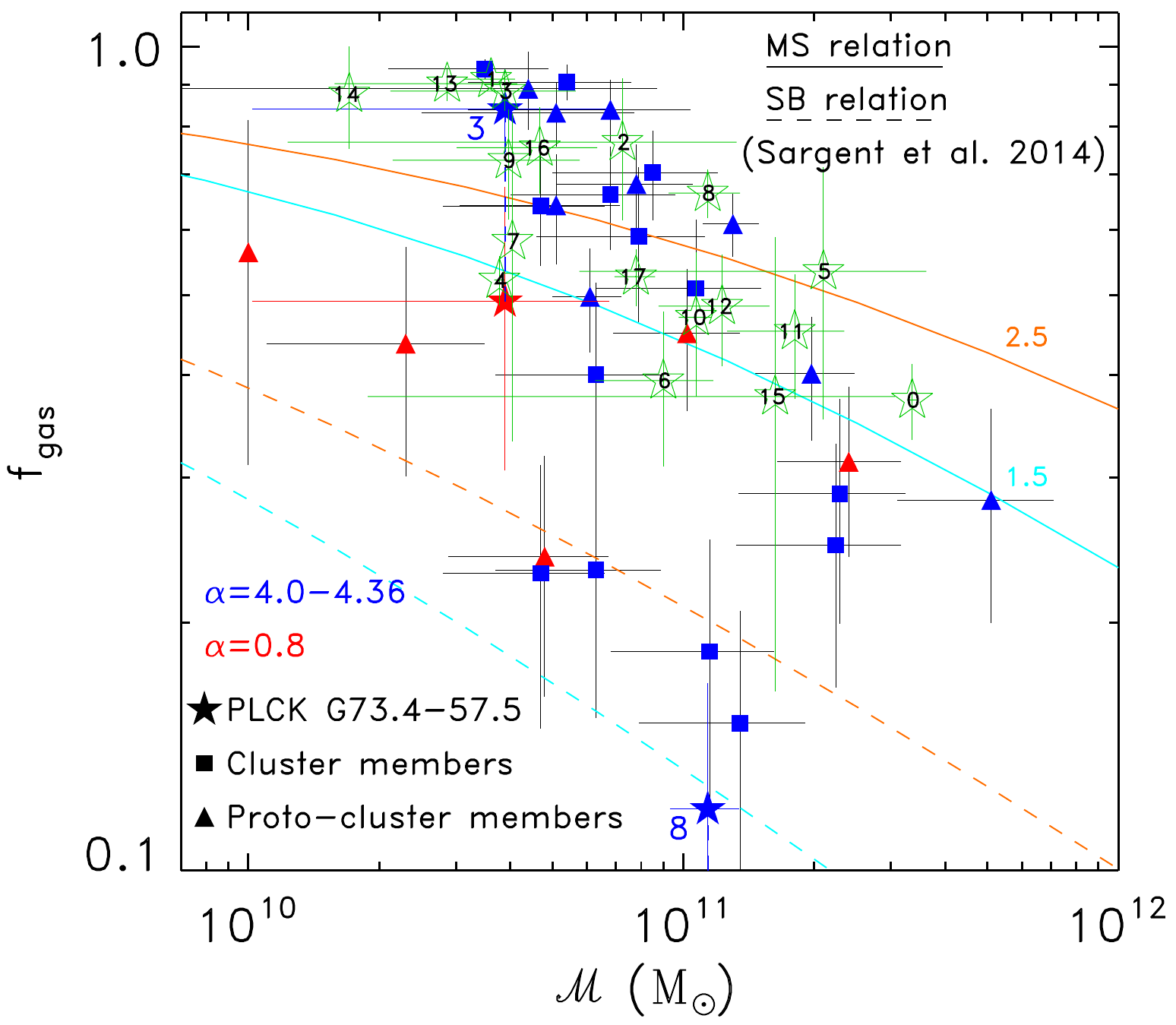}
\caption{Gas fraction $f_{\rm gas}\,{=}\,M_{\rm gas}\,{/}\,(M_{\rm
gas}\,{+}\,\mathcal{M}$) as a function of $\mathcal{M}$.  The symbols are
defined as in Fig.~\ref{mgas_lir}.  The solid and dashed curves represent
the expected relations for MS and SB galaxies, respectively, at $z$\,=\,1.5
(cyan) and at $z$\,=\,2.5 (orange).}
\label{fgas_mstar} 
\end{figure} 

\begin{figure} \centering
\includegraphics[width=\linewidth]{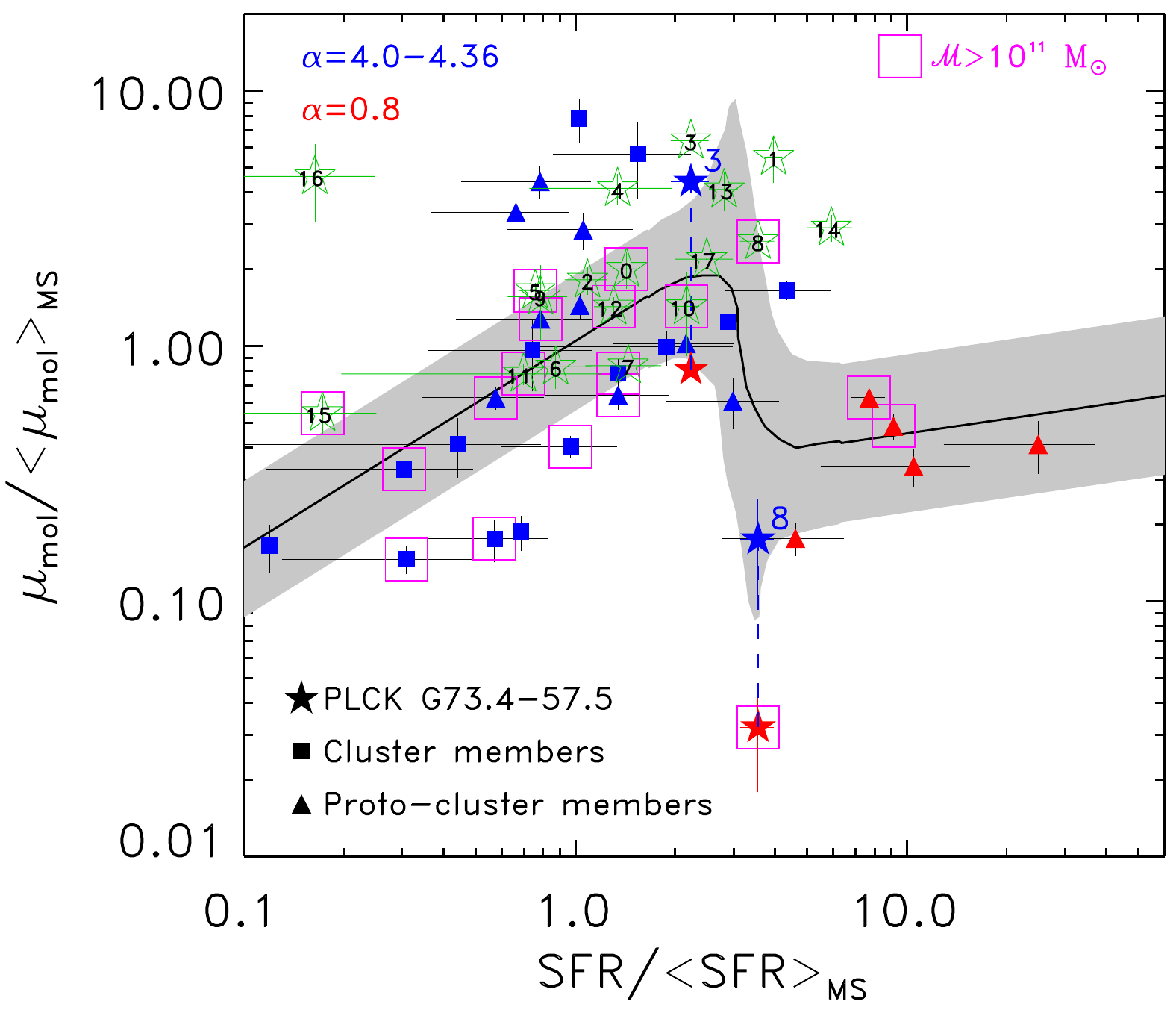}
\caption{Normalised molecular gas mass fraction 
$\mu_{\rm mol}\,{\equiv}\,M_{\rm gas}/\mathcal{M}$
as a function of the normalised SFR.  The symbols are defined as in
Fig.~\ref{mgas_lir}. The black solid curve and grey area represent the
values predicted by the 2-SFM framework described in~\citet{sargent14}.
Sources with $\mathcal{M}\,{>}$\,10$^{11}$\,M$_{\odot}$ are shown with 
magenta large squares. The grey area shows the expected 1$\sigma$ scatter 
around the average molecular gas mass fraction.}
\label{dfgas_dsfr} 
\end{figure} 

In Fig.~\ref{mgas_lir} we show the molecular gas mass derived using CO
luminosities for ALMA IDs~3 and 8, cluster galaxies, and proto-cluster
galaxies, as a function of $L_{\rm IR}$, and the expected values for MS
or SB galaxies~\citep{sargent14}.  Gas masses are derived using $M_{\rm
gas}\,{=}\,\alpha_{\rm CO} L'_{\rm CO(1-0)}$.  The adopted $\alpha_{\rm CO}$
value for each galaxy is that reported in earlier
studies \citep{tadaki14,dannerbauer17,wang18,lee17,ivison13}, and had been 
fixed to either $\alpha_{\rm CO}$\,=\,0.8, typical of SB galaxies, or $\alpha_{\rm
CO}$\,=\,4.0--4.36, typical of normal SFGs.  We remind that we 
consider as an SB any source with
${\rm SFR}\,{>}\,3\times{\rm SFR}_{\rm MS}$,
but this definition might not match what was found in other published samples.
In the case of our CO-detected sources, ALMA IDs 3 and 8, we show gas masses
assuming both $\alpha_{\rm CO}$\,=\,0.8, and 4.36. We also show the full
ALMA sample assuming that $M_{\rm ISM}$ is equivalent to the gas mass
\citep[see Eq.~\ref{equ2},][]{Scoville+2016}. Most of the objects are within
a factor of 3 from the expected relations~\citep{sargent14}, with
a few exceptions that are either richer or poorer in gas. For our selection
of galaxies in the mm range, the ISM mass estimates of the ALMA galaxies are
scattered around the MS relation (mostly above), similar to the galaxies
from the literature.  ALMA ID~3 is richer in gas than expected from the 
relations for either MS or SB galaxies with the respectively assumed 
$\alpha_{\rm CO}$ values. The ISM mass estimate is 
similar (within 2$\sigma$\footnote{For ID\,3, we estimate 
M$_{\rm ISM}$\,=\,(3.0$\pm$0.2)\,10$^{11}$\,M$_{\odot}$, 
and M$_{\rm mol}$\,=\,(2.0$\pm$0.4)\,10$^{11}$\,M$_{\odot}$, 
thus the difference between M$_{\rm ISM}$ and M$_{\rm mol}$ 
for ID\,3 is (1.0$\pm$0.5)\,10$^{11}$\,M$_{\odot}$.}) to the 
molecular gas mass derived from the CO line assuming 
$\alpha_{\rm CO}$\,=\,4.36. Thus a line ratio
CO(5--4)/CO(1--0) of $r_{54/10}\,{\simeq}\,0.3$, as indicated also by the
EMIR data and assumed here \citep[e.g. $r_{54/21}\,{=}\,0.38$
in][]{bothwell13}, and an $\alpha_{\rm CO}$ of 4.36 \citep[as for spiral
galaxies, see][]{Bolatto+2013}, are plausible for ID~3, since they bring the
gas CO and ISM mass estimates into agreement. This suggests that the
molecular gas in ID~3 has properties similar to that of SMGs. The
molecular gas derived from the CO line in ID~8 is instead significantly
lower than expected for both an MS or an SB galaxy, while its ISM mass falls
exactly on the MS relation.  This discrepancy suggests that the assumed CO
excitation might not be adequate for this source.  Indeed it is well known
that there are large uncertainties involved with the conversion factors
\citep[see e.g.][for CO excitations]{Daddi+2015}, up to perhaps a factor of 5.

For ID~8 the mm continuum is extended (cf.\
Table~\ref{table:1}), contrary to the situation for its line emission 
(Fig.~\ref{Fig6b}). For ID~3 the sizes of the continuum and CO emission are
instead in good agreement.  Comparing their ISM masses to those derived from
the CO luminosity, in the case of ID~8 it appears that the gas mass could be
substantially underestimated.

We could not find any evidence for similar line emission from the other sources that have photo-$z$ estimates at $z\,{\simeq}\,1.5$. If their redshifts were to be confirmed and they do not fall outside of the spectral window, the lack of CO detection would imply similarly small gas masses and fractions as for ID~8.

In Fig.~\ref{fgas_mstar}, we examine the gas fraction, defined as
$f_{\rm gas}\,{=}\,M_{\rm gas}\,{/}\,(M_{\rm gas}\,{+}\,\mathcal{M})$, as a
function of $\mathcal{M}$.  We also show the expected relations for MS and
SB galaxies at $z$\,=\,1.5 and 2.5 (since all of the sources from the
literature lie at approximately $z\,{\simeq}\,2.5$, and our CO-detected
sources are at $z$\,=\,1.5), as derived assuming the relations reported
in Eqs.~(\ref{eq_ms})--(\ref{eq_co_lir_sb}). It is interesting to note
that even if a decrease in gas fraction is expected at increasing
stellar masses, this trend is more prominent than expected here, with the
majority of normal SFGs with $\mathcal{M}\,{>}\,10^{11}$\,M$_{\odot}$
having lower than predicted gas fractions (six out of eight $versus$ four out of 15 with
lower $\mathcal{M}$).  For $\mathcal{M}\,{<}\,10^{11}$\,M$_{\odot}$, the gas
fraction is characterised by a wide range, with fractions that can be up to
a factor of 3 lower, and 1.4 higher than expected (seven and four out of 15 normal
SFGs show, respectively, higher and lower than expected gas fractions).  The
two CO-detected ALMA sources follow the same trend; the less massive, ID~3,
has a higher gas fraction, while the more massive, ID~8, is gas deficient. 
Gas fractions derived from the mm-continuum (ISM masses) of the ALMA sources
also decrease with $\mathcal{M}$, faster than expected.  The different
behaviour as a function of stellar mass, is not due to a redshift
difference, since all the examples we selected from the literature lie at
approximately the same redshift and the ALMA sources have redshifts between
1 and 3 in both stellar-mass groups.  The trend is observed for both cluster
and proto-cluster galaxies, and we can thus tentatively state that cluster
and proto-cluster galaxies are, on average, gas poorer than isolated field galaxies
for $\mathcal{M}\,{>}\,10^{11}$\,M$_{\odot}$.

To further investigate this mass dependency and take into account 
each source offset from the MS, we analyse MS-normalised 
quantities~\citep[see][]{sargent14, Genzel+15,Scoville+2016,Tacconi+18}. 
In particular, we show in Fig.~\ref{dfgas_dsfr} the normalised gas fraction,
defined as $\mu_{\rm mol}/\big\langle\mu_{\rm mol}\big\rangle_{\rm MS}$,
where $\mu_{\rm mol}\,{\equiv}\,M_{\rm mol}/\mathcal{M}$,
and the offset from the MS in terms of starburstiness, that is
${\rm SFR}/\big\langle{\rm SFR}_{\rm MS}\big\rangle$, for our ALMA
sources and for cluster and proto-cluster galaxies from the literature, and
compare them with the values predicted by the \lq{}2-Star Formation Mode\rq{} 
(2-SFM) framework formulated by~\citet{sargent14}.  
The 2-SFM predicted values are independent of stellar mass and redshift 
and take into account the type (MS or SB) of source that 
dominates as a function of starburstiness. 
The 2-SFM framework assumes a continuous distribution of $\alpha_{\rm CO}$ 
conversion factors that varies according to the position of a source in the 
SFR-$\mathcal{M}$ diagram, rather than two distinct values for MS and SB galaxies. 
The diagram in Fig.~\ref{dfgas_dsfr} confirms that most 
MS galaxies with $\mathcal{M}\,{>}$\,10$^{11}$\,M$_{\odot}$ (shown as
magenta squares) have gas fractions below the expected values (i.e. only one
out of eight sources with $\mathcal{M}\,{>}\,$10$^{11}$\,M$_{\odot}$ is above
the relation, while five are below).  The same is not
true among SB galaxies.  ISM mass fractions are also systematically lower
for the ALMA sources with $\mathcal{M}\,{>}\,$10$^{11}$\,M$_{\odot}$,
although still consistent with the 2-SFM relation.  However, we caution 
the use of mm-continuum ISM masses as proxies for CO molecular gas 
masses as they might be affected by systematic uncertainties. For example, 
in the two cases of our sample for which CO masses are available, ISM masses
are higher than the CO masses, possibly suggesting that the ISM-derived
masses are overestimates. Keeping this caveat in mind, we conclude that 
gas fractions based on ISM masses tend to be lower for galaxies with
$\mathcal{M}\,{>}\,10^{11}\,{\rm M}_{\odot}$.  Our analysis also indicates 
a notable difference, both in excess (for cluster and proto-cluster
members), and in deficiency (only for cluster galaxies), between measured
and predicted gas fractions.  The observed wide range of gas fractions may 
suggest that in dense environments these change quickly, or at least
faster than their SFRs.

We note that this analysis is not aimed at obtaining a comprehensive picture of
SFEs and gas fractions in field, cluster, and proto-cluster galaxies, but is
instead simply aimed at comparing our ALMA sources to what is available in
the literature (keeping in mind that those samples are often biased towards
the most gas-rich members).  To limit this bias, we also include data from
two clusters for which a large number of members with deep CO data are
available \citep{rudnick17,hayashi17} in the SFE analysis; however, stellar
masses are not available for the latter, so they are missing in our
gas-fraction analysis. A future analysis of the gas fraction in an unbiased
sample of $1\,{<}\,z\,{<}\,3$ galaxies in clusters and proto-clusters could
reveal the processes and the timescales that affect the cold gas as a function
of environment. 

\section{Discussion}\label{sec:discussion}

We have found that the ALMA-detected galaxies comprising this {\it Planck\/} peak are primarily main-sequence galaxies that break-up into redshift groups, with the galaxies associated with individual {\it Herschel\/} flux regions not necessarily falling into the same redshift ranges.

In ALMA fields 5 and 8 only a single ALMA galaxy is observed, and these are thus simple to interpret. In both cases the fluxes are attributed to a low redshift galaxy ($z\,{\la}\,$1.3). The other ALMA fields contain multiple sources and in most of the cases they are at different redshifts. The only exceptions are IDs~5 and 6 in field 3, and IDs~13, 14, and 15 in field 7. The tentative picture that emerges puts ALMA fields 1, 5 and 8 at $z\,{\simeq}\,1$, ALMA fields 2, 3, 4, and 6 at $z\,{\simeq}\,1.5$, and ALMA fields 1, 6, and 7 at $z\,{\simeq}\,2.4$.  But reality may not be quite so simple, as evidenced by the various \lq{}interlopers\rq{} found in several of the fields, which seem to indicate that some regions cannot really be categorised as belonging to one group or the other, but contain mixtures of high and low redshift galaxies. Broadly speaking, however, the nature of G073.4$-$57.5 seems to be at least two line-of-sight structures at $z\,{\simeq}\,1.5$ and $z\,{\simeq}\,2.4$, and probably a few other galaxies at different redshifts. This conclusion seems reasonable, since G073.4$-$57.5 appears similar to the {\it Planck\/} peak G95.5$-$61.6 \citep[][]{FloresCacho+16}, in the sense that it is a superposition of two independent structures on the sky (in the case of G95.5$-$61.6, the two groups are at $z\,{=}\,1.7$ and $z\,{=}\,2.0$). It also ties in with the simulation results of \citet{Negrello+17}, who found that the number of {\it Planck\/} peaks exceeds the number of single massive haloes expected at redshifts $z\,{=}\,1.5$--3, and that the majority of {\it Planck\/} peaks are consistent with being superpositions of proto-clusters. 

We have also investigated our photometric redshift distribution in the context of the sky simulation catalogue of \citet{Bethermin+17}. First, using the whole catalogue, the expected number of detections in 1.8\,arcmin$^2$ is 2.9 sources, broadly consistent with the average source counts we adopt from the literature, and giving a 6-times higher observed source density. It is beyond the scope of this paper to attempt a modelling of all the selection effects that enter into the final detection with ALMA, or a detailed comparison of the clustering we find within the photometric redshift distribution; however, one thing we can easily do is compare the simulated redshift distribution for all sources, renormalised to the observed number of sources (and in coarse redshift bins of $\Delta z\,{=}\,0.85$) with the observed distribution.  We find that we see no significant difference, given the relatively large Poisson errors (the Kolmogorov-Smirnov $p$-value is 0.22); in the $z\,{=}\,0.85$--1.7 redshift bin 3.0 more sources are observed than expected in the renormalised distribution (9.9 times higher than the simulated numbers) and in the $z\,{=}\,1.7$--2.55 redshift bin 1.6 more sources are observed than expected when renormalised (8.0 times higher compared to simulated numbers).

Going back to the {\it Spitzer\/} and CFHT colour-magnitude and colour-colour diagrams (Figs.~\ref{Fig3} and \ref{Fig5}), we see that four out of five member candidates at $z\,{\simeq}\,$1.5 (i.e. ALMA IDs~3, 5, 6, and 8) exhibit similar $J\,{-}\,K_{\rm s}$ and [3.6]$\,{-}\,$[4.5] colours, in good agreement with the photo-$z$ results and our interpretation. The other $z\,{\simeq}\,1.5$ member candidate, ID~12, shows consistency only in the IRAC colour, but not in the NIR colour. Similarly, ALMA IDs~1, 11, 13, and 15, member candidates in the $z\,{\simeq}\,2.4$ structure, show similar IRAC and NIR colours (the latter is not available for ID~1). The other member candidate, ID~14, was not included in the colour analysis because it is undetected in the NIR.

The SED-fitting procedure used in this paper is based on an unusual set of filters that combine four wavelengths in the NIR (1--5\,$\mu$m) and five in the FIR-mm (250--1300\,$\mu$m). Most high-$z$ dusty galaxies (and obscured AGN) are characterised by red NIR colours, thus the redshift is mostly constrained by the peak location of the FIR component. In five cases the stellar peak is clearly visible in the NIR SED (IDs~0, 4, 10, 16, and 17) and the resulting redshift is below 1.3, and in all other cases the stellar peak is redshifted to $\lambda\,{>}\,4.5\,\mu$m, or alternatively an AGN might be present. Because of the lack of mid-IR data, the presence of an AGN cannot be easily determined, and thus it cannot be completely ruled out. Our best fits suggest that two sources are AGN, and each of them is associated with one of the two structures: ID~6 with the structure at $z\,{=}\,1.5$; and ID~15 with that at $z\,{=}\,2.4$. In high-$z$ proto-clusters, accretion onto a supermassive black hole is expected to be favoured with respect to the field because of the presence of cold gas, high galaxy density, and modest relative velocities. Previous work targeted at quantifying the AGN fraction in proto-clusters yields conflicting results \citep[e.g.][]{krishnan17,lehmer13,digby10,macuga19}. Our results, based on the SED fits, indicate that one out of five ($20\,{\pm}\,9$\,\%) members of each structure is an AGN.  Although this fraction is consistent with the findings in other high-$z$ structures \citep[i.e. 17\,\%,][]{lehmer13,krishnan17}, our result is only suggestive. In order to quantify its significance we would firstly need to confirm the membership of each structure, and, secondly quantify the AGN fraction in the field by applying the same technique.

While our work is carried out in the context of searching for \lq{}proto-clusters\rq{}, that is early structures that are not yet virialised, with our findings of two overdensities of mm galaxies we cannot be sure if either is in fact already a \lq{}cluster\rq{}. We have no information about the thermal state of any intra-cluster medium, since the limits available from the X-ray {\it ROSAT} all-sky survey and the {\it Planck\/} Sunyaev-Zeldovich cluster survey are too weak; our overdensities should perhaps be identified simply as \lq{}structures\rq{}.

The two possible structures revealed by our selection in the mm range at $z\,{\simeq}\,$1.5 and 2.4 have members distributed over six ALMA fields. Those at $z\,{\simeq}\,$1.5 are distributed over a narrow region 4\arcmin\ long (5.2 comoving Mpc at $z\,{=}\,$1.5), connecting fields 2, 3, 4, and 6. The member candidates of the $z\,{\simeq}\,$2.4 structure are distributed over fields 1, 6, and 7, corresponding to a region of about 2$\arcmin\,{\times}2\arcmin$ (3.4\,Mpc$\,{\times}\,$3.4\,Mpc comoving at $z$\,=\,2.4). The total SFRs of the two structures at $z$\,=\,1.5, and 2.4 are 840$^{+120}_{-100}$\,M$_{\odot}$\,yr$^{-1}$ and 1020$^{+310}_{-170}$\,M$_{\odot}$\,yr$^{-1}$, respectively; the total stellar masses are 5.8$^{+1.7}_{-2.4}\,{\times}\,10^{11}$\,M$_{\odot}$ and 4.2$^{+1.5}_{-2.1}\,{\times}\,10^{11}$\,M$_{\odot}$, respectively. The members in the lower-$z$ structure are all quite massive and located on the MS. Those in the higher-$z$ structure are instead a mix of massive and more typical sources that are, respectively, on the MS or above. These results are based on photometric redshifts with considerable uncertainty, but if confirmed they would indicate that the members in the low-$z$ structure are mature and close to the end of their active growth, while some of those in the high-$z$ structure are still active (SB-like) and growing. The analysis of SFEs and gas fractions, possible for only two galaxies, IDs~3 and 8, both in the $z$=1.5 structure, yield values consistent with the MS relation for ID~3 and inconsistent with either the MS or SB relations for ID~8. The CO luminosity in the latter source is anomalously low, casting some doubts on its association with the observed continuum. The gas fractions of our ALMA sources, based on the dust continuum, combined with those from the literature, indicate that proto-cluster members are more gas-rich than the isolated field (MS or SB) galaxies for $\mathcal{M}\,{<}\,10^{11}$\,M$_{\odot}$; the gas fractions drops by almost a factor of two at larger stellar masses.

The lack of strong CO lines for the other sources at $z\,{\simeq}$1.5 might imply that their molecular gas is depleted. On the other hand, it is possible that we did not detect lines from any other galaxies because they lie outside of our observed spectral window. Indeed, the line of ALMA ID~3 clearly extends beyond the high frequency end of the spectral window. 

It is a valid question to ask whether gravitational galaxy-galaxy lensing can play a role in enhancing the counts in a scenario where extended structures at lower and higher redshift overlap along the line of sight, giving apparent densities above typical proto-cluster measurements. In the current data most ALMA sources are positionally well matched with {\it Spitzer\/} and NIR data and contain no indication of lensing signatures, while only a few sources show offsets (as seen in Fig.~\ref{Fig2}) between the mm and NIR emission (e.g. ALMA IDs~1, 4, and 15).
In terms of statistical arguments, in general the probability for strong galaxy-galaxy lensing is small; for example, \citet{vanderWel+13} estimate one source per 200\,arcmin$^2$ for average counts of strongly lensed sources. However, lensing cannot be completely ruled out, and in particular smaller flux boosts by factors less than 2 could be common, since in our case the counts are enhanced (possibly by a factor of 10 for both the source and the lensed population), 
and the region was selected for high submm surface brightness in the first place. 

\section{Conclusions}\label{sec:conclusions}

Using ALMA in only 24 minutes of on-source time we find 18 individual mm galaxies, showing that follow-up of the {\it Planck\/} high-$z$ sample through targeted pointings of {\it Herschel\/}-SPIRE sources is an efficient use of this telescope. For the first time we are directly resolving the {\it Planck\/} peaks and the {\it Herschel\/}-SPIRE overdensities into individual galaxies at mm wavelengths. The ALMA detections are well matched with {\it Spitzer\/}-IRAC sources in all but one case and mostly show excellent positional agreement (typically $\,{<}\,$0.4\arcsec); three sources that are offset by up to 1\arcsec\ are extended in {\it Spitzer\/}, and may either be blended or have intrinsically more complex structure. 

The surface density of the mm galaxies within the ALMA pointings is 8--30 times higher than the average counts, and we estimate an SFR of ${\simeq}\,$2700\,M$_{\odot}$\,yr$^{-1}$ (of which one third can be attributed to sources consistent with $z\,{\simeq}\,1.5$ and another third to sources at $z\,{\simeq}\,$2.4). Furthermore, the SCUBA-2 data indicate that we have not recovered all of the mm galaxies in this field, possibly not even the brightest, which will require a wider mosaic. Nevertheless, we can conclude that the {\it Planck\/} peak G073.4$-$57.5 consists of a large number of moderately bright mm galaxies, rather than a few extremely bright galaxies. Typical proto-clusters (such as the COSMOS $z\,{=}\,2.47$ structure or the SSA22 $z\,{=}\,3.09$ structure) contain fewer galaxies than we have detected, perhaps because G073.4$-$57.5 is a line-of-sight superposition of two massive structures. The cluster XCS J2215.9$-$1738 at $z\,{=}\,1.46$ also shows a high density of ALMA galaxies, but those are fainter. 

NIR colour diagrams of the ALMA-detected galaxies reveal a \lq{}red sequence\rq{}, a characteristic feature of $z\,{>}\,1.3$ structures that are the progenitors of later massive galaxy clusters. An NIR+FIR photo-$z$ analysis indicates a concentration at $z\,{\simeq}\,1.5$, while a second structure at higher redshift ($z\,{\simeq}\,2.4$) could be present as well, consistent with the interpretation of two line-of-sight structures. In addition to photometric redshifts, we present SFRs, IR luminosities, stellar masses, dust temperatures, dust masses, and gas masses for these galaxies. Three galaxies can clearly be identified as starbursts (i.e. lying a factor of 3 or more above the MS), ID~8 at $z\,{\simeq}\,1.5$ and IDs~1 and 14 at $z\,{\simeq}\,2.4$, while most of the galaxies are within the normal range of SFRs for their stellar masses.

Serendipitous line detections of two galaxies at a common frequency ($\Delta V\,{<}\,$300\,km\,s$^{-1}$) are interpreted as the CO(5--4) transition and can be used to fix the redshift of the main structure to $z\,{=}\,1.5434\,{\pm}\,0.0010$, in agreement with the photo-$z$ estimates. However, this needs to be confirmed with additional spectroscopy.

The CO luminosity of ID~3, combined with the parameters derived from the SED
fitting, indicate that the molecular gas in this source is similar to a
normal star-forming galaxy, but with a smaller SFE, as well as a larger gas
mass and fraction than expected based on the MS relation. The CO
properties of ID~8 are instead inconsistent with the relations observed for
MS or SB galaxies.  Based on the SFR--$\mathcal{M}$ relation, ID~8 is an SB
galaxy with a modest offset from the MS, but its SFE and gas mass are, respectively, 
unusually high and low
with respect to expectations for isolated field galaxies. Interestingly, its ISM
mass is much higher than the CO-derived mass, bringing both its SFE and
gas fraction into agreement with the expectation.  

The overall analysis of stellar masses, SFRs, and gas fractions of the ALMA sources in the two structures at $z\,{=}\,$1.5 and 2.4 suggest that the former contains more evolved galaxies with stellar masses larger than the expected $\mathcal{M}_\ast$, and SFRs consistent with the MS, while the latter contains a mix of sources, some more massive than the expected $\mathcal{M}_\ast$, and with moderate to low SFRs, or with stellar masses consistent with the expected $\mathcal{M}_\ast$, but with SFRs above the MS and in the SB region. These results, derived from a mm/submm selection and to be confirmed with more accurate redshifts, are consistent with a scenario in which the structure at lower redshift is more mature and most of its members have reached the end of their active phase; however, the structure at higher redshift contains some galaxies that are instead still actively growing.

The analysis of the cold gas properties (traced by CO emission or by the mm
continuum) for our ALMA detections and for galaxies in clusters and
proto-clusters at $1\,{<}\,z\,{<}\,3$, have revealed two interesting
results.  Firstly, cluster and proto-cluster galaxies with stellar masses
${<}\,10^{11}\,{\rm M}_{\odot}$ exhibit a broad range of gas fractions,
suggesting that the gas content can change quickly in dense environments. 
Secondly, most cluster and proto-cluster galaxies with stellar masses
${>}\,10^{11}\,{\rm M}_{\odot}$ are gas deficient with respect to field
galaxies.  These results are valid only for normal SFGs, and not for SB
galaxies, but this should be confirmed with unbiased CO samples.

There are several important aspects of our study that should be followed up. First, optical/NIR or mm spectroscopy \citep[see e.g.][]{Casey+17} will allow us to confirm the photo-$z$ estimates and the associations of the individual galaxies with structures in redshift space. Second, it will be helpful to associate all SCUBA-2 sources with their counterparts in ALMA data in order to study those brightest submm peaks in more detail. Third, it is important to address the positional offsets between the NIR and FIR images with future high-resolution data, in particular by searching for elongations or multiple images that would be evidence of strong lensing. And lastly, further imaging and spectroscopy of this {\it Planck\/} peak will enable us to characterise its physical properties in terms of angular and redshift-space morphology and to build a census of its stellar and star-forming properties. Such a detailed study is the only way to determine the nature of these red peaks in the CIB that have been picked out by {\it Planck\/}, which is the most decisive step in determining what exactly they are teaching us about structure formation.

\begin{acknowledgements}
We thank the referee for valuable suggestions that improved the paper. MP acknowledges financial support from Labex OCEVU, CM acknowledges the support provided by FONDECYT postdoctoral research grant no. 3170774, and RH and DS acknowledge support from the Natural Sciences and Engineering Research Council of Canada. This work has been carried out thanks to the support of the OCEVU Labex (ANR-11-LABX-0060) and the A$^*$MIDEX project (ANR-11-IDEX-0001-02) funded by the \lq{}Investissements d'Avenir\rq{} French government programme managed by the ANR. This work is mainly based on the following ALMA data: ADS/JAO.ALMA\# 2013.1.01173.S. ALMA is a partnership of ESO (representing its member states), NSF (USA), and NINS (Japan), together with NRC (Canada), NSC, and ASIAA (Taiwan), and KASI (Republic of Korea), in cooperation with the Republic of Chile. The Joint ALMA Observatory is operated by ESO, AUI/NRAO, and NAOJ. The development of {\it Planck\/} has been supported by: ESA; CNES and CNRS/INSU-IN2P3-INP (France); ASI, CNR, and INAF (Italy); NASA and DoE (USA); STFC and UKSA (UK); CSIC, MICINN, JA, and RES (Spain); Tekes, AoF, and CSC (Finland); DLR and MPG (Germany); CSA (Canada); DTU Space (Denmark); SER/SSO (Switzerland);RCN (Norway); SFI (Ireland); FCT/MCTES (Portugal); and PRACE (EU). {\it Herschel\/} is an ESA space observatory with science instruments provided by European-led Principal Investigator consortia and with important participation from NASA. This work is based in part on observations made with the Spitzer Space Telescope, which is operated by the Jet Propulsion Laboratory, California Institute of Technology under a contract with NASA. This research has made use of the NASA/ IPAC Infrared Science Archive, which is operated by the Jet Propulsion Laboratory, California Institute of Technology, under contract with the National Aeronautics and Space Administration. This publication makes use of data products from the Wide-field Infrared Survey Explorer, which is a joint project of the University of California, Los Angeles, and the Jet Propulsion Laboratory/California Institute of Technology, and NEOWISE, which is a project of the Jet Propulsion Laboratory/California Institute of Technology; WISE and NEOWISE are funded by the National Aeronautics and Space Administration.  The James Clerk Maxwell Telescope has historically been operated by the Joint Astronomy Centre on behalf of the Science and Technology Facilities Council of the United Kingdom, the National Research Council of Canada, and the Netherlands Organisation for Scientific Research. Additional funds for the construction of SCUBA-2 were provided by the Canada Foundation for Innovation. The Pan-STARRS1 Surveys (PS1) and the PS1 public science archive have been made possible through contributions by the Institute for Astronomy, the University of Hawaii, the Pan-STARRS Project Office, the Max-Planck Society and its participating institutes, the Max Planck Institute for Astronomy, Heidelberg and the Max Planck Institute for Extraterrestrial Physics, Garching, The Johns Hopkins University, Durham University, the University of Edinburgh, the Queen's University Belfast, the Harvard-Smithsonian Center for Astrophysics, the Las Cumbres Observatory Global Telescope Network
Incorporated, the National Central University of Taiwan, the Space Telescope Science Institute, the National Aeronautics and Space Administration under Grant No.  NNX08AR22G issued through the Planetary Science Division of the NASA Science Mission Directorate, the National Science Foundation Grant No. AST-1238877, the University of Maryland, Eotvos Lorand University (ELTE), the Los Alamos National Laboratory, and the Gordon and Betty Moore Foundation.\\
\end{acknowledgements}

\bibliographystyle{aa} 

\bibliography{alma_biblio}

\begin{appendix}
\section{Photometric redshift fit results}

The Pan-STARRS, WIRCam, and IRAC data used to estimate photometric redshifts and stellar masses are listed in Table~\ref{phot_data}. In the SED fitting procedure we also used the FIR-mm data listed in Tables~\ref{table:1} and \ref{table:4}. Figure~\ref{bestfit_seds} shows our multi-wavelength flux-density measurements as black circles and the best-fit templates obtained in the second round of {\tt EAZY} \citep{Brammer+2008} fitting using template libraries from \citet{polletta07} and \citet{danielson17} as magenta curves. Downward arrows correspond to 3$\,\sigma$ upper limits, or to the confusion limit (i.e. 5.8\,mJy, 6.3\,mJy, and 6.8\,mJy at 250\,$\mu$m, 350\,$\mu$m, and 500\,$\mu$m, respectively) for {\it Herschel\/} flux densities, and are shown in cases where the measured flux density is below 2$\,\sigma$. The green curve is the best-fit model obtained by fitting the Pan-STARRS-WIRCam-IRAC SED with models of \citet{bc03}. The purple curve represents the best-fit to the far-IR SED obtained using a single-temperature modified black-body model. The dashed red and cyan curves represent, respectively, the highest and lowest temperature modified black-body model consistent with the far-IR SED, within 1$\,\sigma$. For comparison, the blue curve is the best-fit model obtained by fixing the redshift to the precise value of $z\,{=}\,1.54$, assumed from the CO lines found in ALMA IDs~3 and 8. The ALMA ID and photometric redshifts are annotated in the top left corners of each panel. The spectroscopic redshift is annotated in the top right corner of each panel, when available.

\begin{landscape}
\begin{table}[htbp!]
\caption{Pan-STARSS (columns 2--6), WIRCam (columns 7 and 8), and IRAC (columns 9 and 10) data.}
\label{phot_data}.
\begin{tabular}{rccccc rr rr} 
\hline\hline
\noalign{\vskip 3pt}
 ID& $g^{\rm a}$& $r$& $i$& $z$& $y$& $S_J$& $S_{K_{\rm S}}$&$S_{3.6}$& $S_{4.5}$\\
 & [mag]& [mag]& [mag]& [mag]& [mag]& [$\mu$Jy]& [$\mu$Jy]& [$\mu$Jy]& [$\mu$Jy]\\
\hline
\noalign{\vskip 3pt}
  0$^{\rm b}$& $<$23.3& $<$23.2& $<$23.1& $<$22.3& $<$21.4& 24.98$\pm$0.44& 77.15$\pm$0.92& 137.41$\pm$0.82& 114.32$\pm$1.08\\
  1& $<$23.3& $<$23.2& $<$23.1& $<$22.3& $<$21.4& 3.25$\pm$0.69& 4.89$\pm$0.90&   14.37$\pm$0.83& 20.30$\pm$1.11\\
  2& $<$23.3 &$<$23.2& $<$23.1& $<$22.3& $<$21.4& $<$1.32& $<$4.37& 5.47$\pm$0.93& 7.86$\pm$1.27\\
  3& $<$23.3& $<$23.2& $<$23.1& $<$22.3& $<$21.4& 8.61$\pm$0.46& 18.28$\pm$1.47& 45.30$\pm$0.90& 57.37$\pm$1.27\\
  4& $<$23.3& 21.624$\pm$0.058& 20.967$\pm$0.042& 20.658$\pm$0.056& 20.344$\pm$0.153&  50.30$\pm$  0.46&   86.12$\pm$  1.49&   67.00$\pm$0.93&   59.11$\pm$1.26\\
  5& $<$23.3& $<$23.2& $<$23.1& $<$22.3& $<$21.4& 6.21$\pm$0.44& 18.19$\pm$0.92& 39.96$\pm$0.83& 49.79$\pm$1.11\\
  6& $<$23.3& $<$23.2& $<$23.1& $<$22.3& $<$21.4& 2.06$\pm$0.44& 4.60$\pm$0.93&  16.00$\pm$0.84& 22.13$\pm$1.12\\
  7& $<$23.3& $<$23.2& $<$23.1& $<$22.3& $<$21.4& $<$1.32& $<$4.37& 5.72$\pm$0.85& 6.24$\pm$1.11\\
  8& $<$23.3& $<$23.2& $<$23.1& $<$22.3& $<$21.4& 8.40$\pm$0.44& 20.31$\pm$0.92& 49.19$\pm$0.82& 63.65$\pm$1.10\\
  9& $<$23.3& $<$23.2& $<$23.1& $<$22.3& $<$21.4& 2.78$\pm$0.44& 6.66$\pm$0.92&  20.23$\pm$0.83& 23.34$\pm$1.12\\
 10& $<$23.3& $<$23.2& $<$23.1& $<$22.3& $<$21.4& 16.31$\pm$0.44& 34.07$\pm$0.92& 58.24$\pm$2.85& 52.33$\pm$3.03\\
 11& $<$23.3& $<$23.2& $<$23.1& $<$22.3& $<$21.4& $<$1.40& 6.35$\pm$0.96& 14.22$\pm$0.88& 6.78$\pm$1.17\\
 12& $<$23.3& $<$23.2& $<$23.1& $<$22.3& $<$21.4& 2.25$\pm$0.46& 10.30$\pm$0.94& 23.28$\pm$0.87& 32.22$\pm$1.16\\
 13& $<$23.3& $<$23.2& $<$23.1& $<$22.3& $<$21.4& $<$1.71& 4.68$\pm$1.10& 10.40$\pm$1.67& 14.61$\pm$1.92\\
 14& $<$23.3& $<$23.2& $<$23.1& $<$22.3& $<$21.4& $<$1.32& $<$4.37& $<$5.03& $<$5.78\\
 15& $<$23.3& $<$23.2& $<$23.1& $<$22.3& $<$21.4& $<$2.36& 6.38$\pm$1.04& 14.00$\pm$1.67& 20.10$\pm$1.93\\
 16& $<$23.3& $<$23.2& $<$23.1& $<$22.3& $<$21.4& 6.27$\pm$0.54& 13.69$\pm$1.01& 20.54$\pm$1.67& 13.60$\pm$1.92\\
 17& $<$23.3& $<$23.2& $<$23.1& $<$22.3& $<$21.4& 5.48$\pm$0.50& 19.13$\pm$1.00& 40.68$\pm$0.87& 40.01$\pm$1.28\\
\hline
\end{tabular}\\
{{\small $^{\rm a}$ All magnitudes are in the AB system.\\
$^{\rm b}$ ID~0 has a WISE band~4 (22\,$\mu$m) flux density of 0.63$\pm$0.16\,mJy.}}
\end{table}
\end{landscape}

\begin{figure*} [htbp!]
\centering
\includegraphics[width=0.95\linewidth]{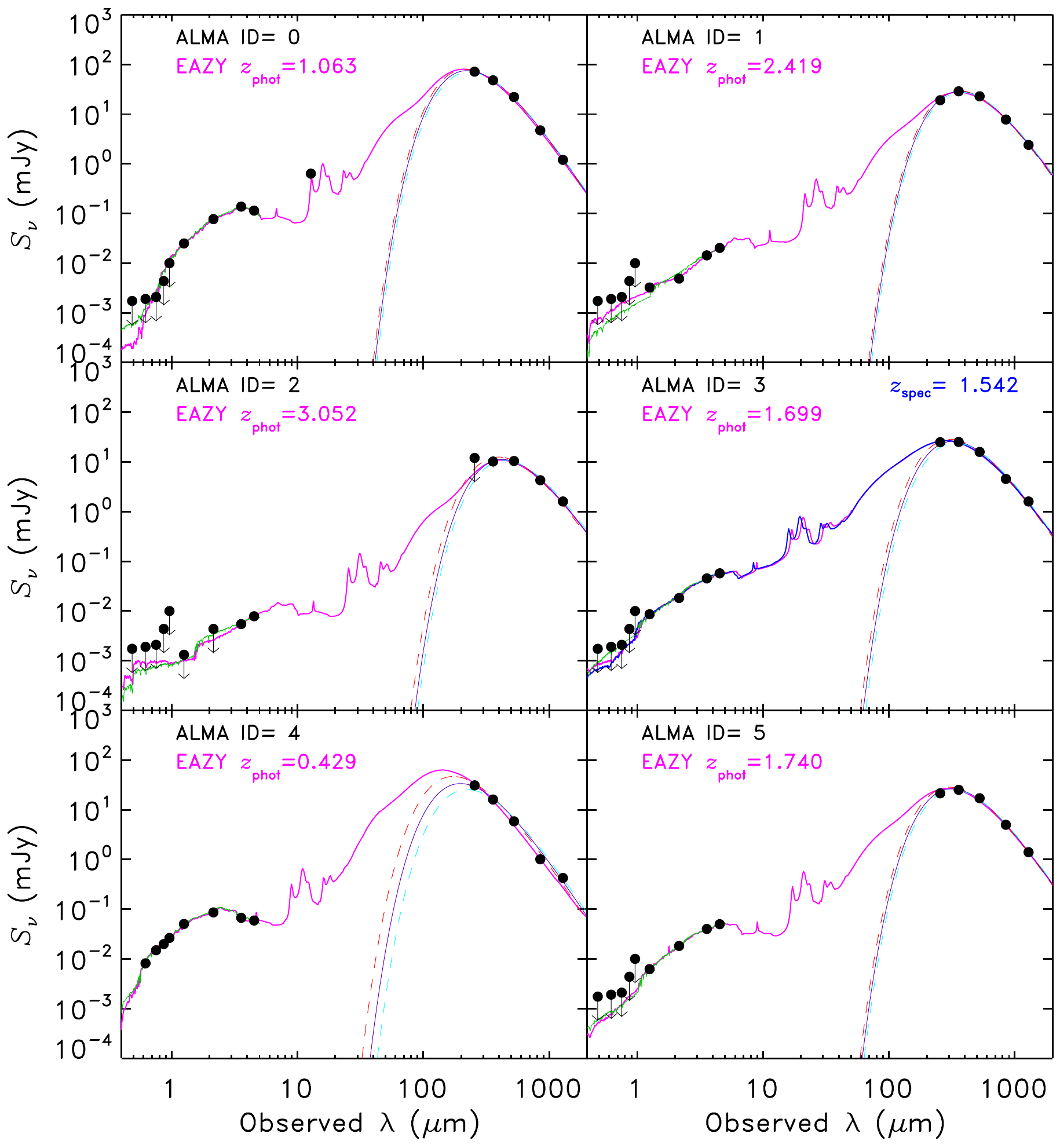}
\caption{Observed SEDs obtained by combining Pan-STARRS, WIRCam, IRAC, SPIRE, SCUBA-2, and ALMA data (filled black circles) and best-fit templates (magenta curves) obtained with {\tt EAZY} \citep{Brammer+2008} at the annotated photometric redshifts. The blue curve is the best-fit template at the spectroscopic redshift, available only for IDs~3 and 8. The green curve is the best-fit template to the Pan-STARRS-WIRCam-IRAC SED obtained using {\tt Hyper-$z$} \citep{bolzonella00} and the stellar population models of \citet{bc03}. Single-temperature modified black-body models that fit the FIR-mm SED to within $\pm1\,\sigma$ are shown as solid purple curves, and dashed red or cyan curves for the warmer and cooler best fits.}
\label{bestfit_seds}
\end{figure*}
\begin{figure*}[htbp!]
\setcounter{figure}{0}
\centering
\includegraphics[width=0.95\linewidth]{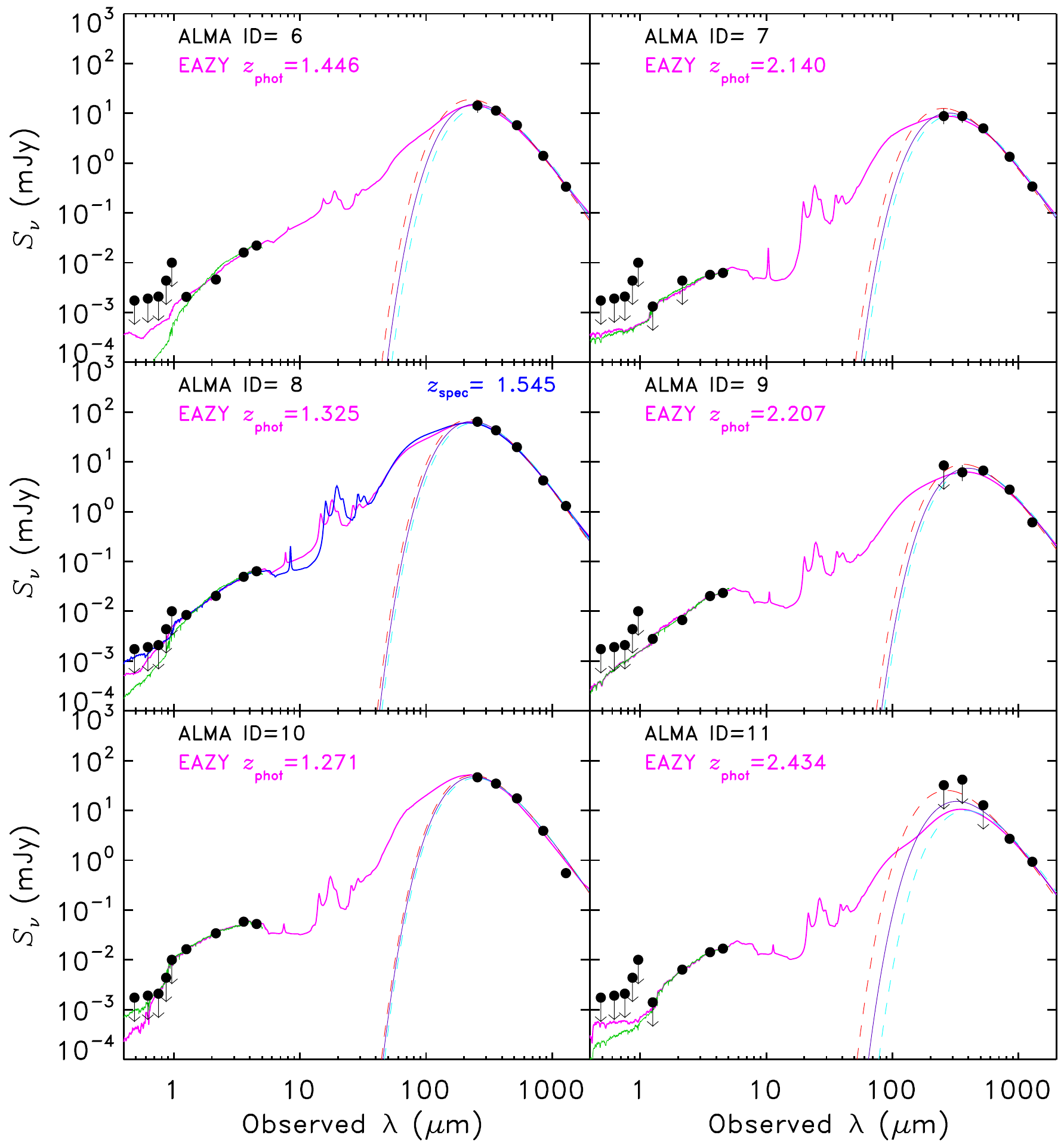}
\caption{{\small Continued.}}
\end{figure*}
\begin{figure*}[htbp!] 
\setcounter{figure}{0}
\centering
\includegraphics[width=0.95\linewidth]{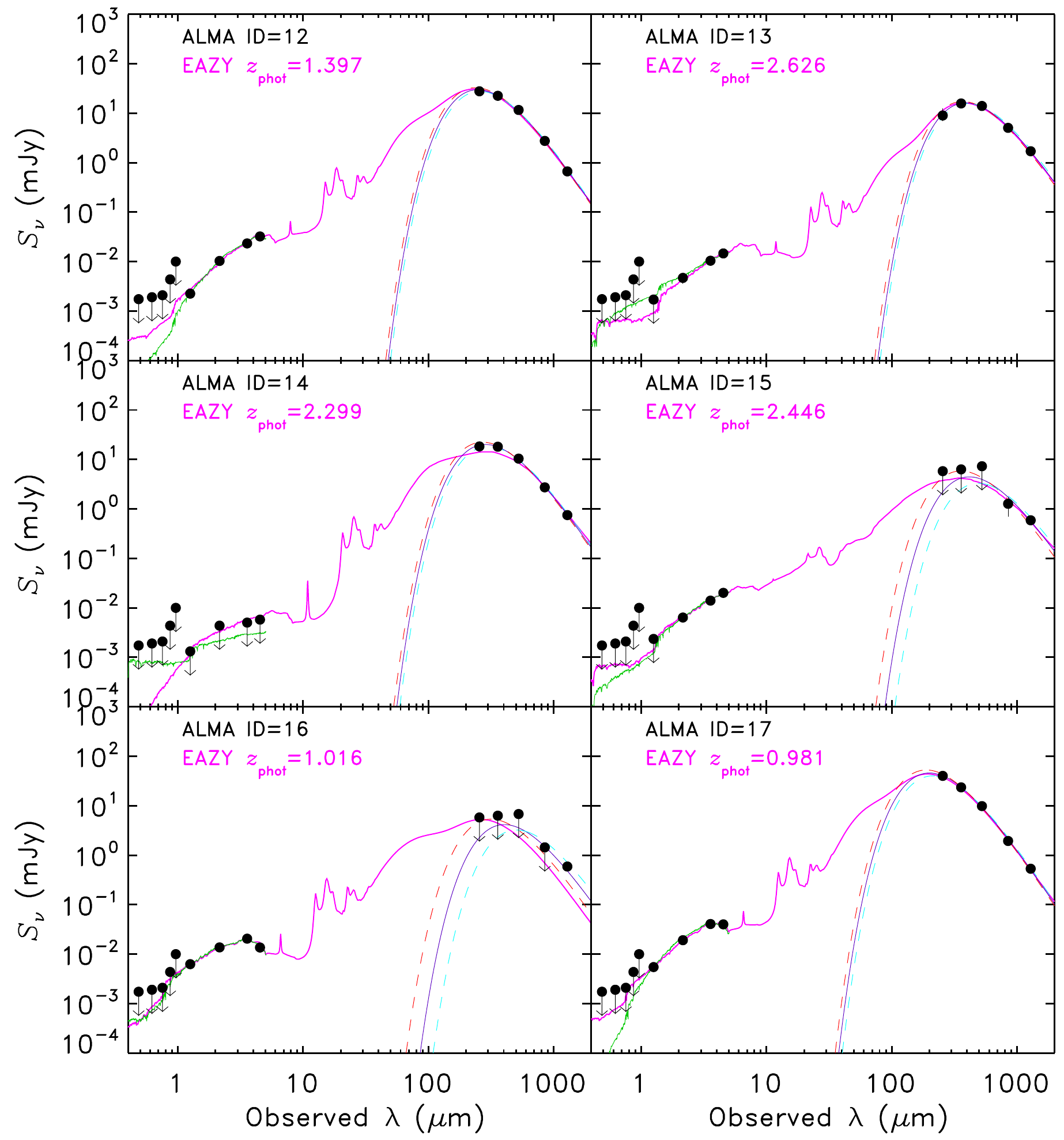}
\caption{{\small Continued.}}
\end{figure*}
\end{appendix}

\end{document}